\documentclass[%
 reprint,
 amsmath,amssymb,
 aps,superscriptaddress,
]{revtex4-1}

\usepackage{color}
\usepackage{amsmath}
\usepackage{natbib}
\usepackage[export]{adjustbox}
\usepackage{graphicx}
\usepackage{dcolumn}
\usepackage{bm}
\usepackage{subfigure}

\begin{document}

\title{Photoelectron and fragmentation dynamics of the H$^{+}$ + H$^{+}$ dissociative channel in NH$_3$ following direct single-photon double ionization}

\author{Kirk A. Larsen}
\email{klarsen@lbl.gov}
\affiliation{%
 Graduate Group in Applied Science and Technology, University of California, Berkeley, CA 94720, USA}
\affiliation{%
 Chemical Sciences Division, Lawrence Berkeley National Laboratory, Berkeley, CA 94720, USA}%

\author{Thomas N. Rescigno}
\email{tnrescigno@lbl.gov}
\affiliation{%
 Chemical Sciences Division, Lawrence Berkeley National Laboratory, Berkeley, CA 94720, USA}%
 
 \author{Travis Severt}
\affiliation{%
 J.R. Macdonald Laboratory, Physics Department, Kansas State University, Manhattan, Kansas 66506, USA}

\author{Zachary L. Streeter}
\affiliation{%
 Chemical Sciences Division, Lawrence Berkeley National Laboratory, Berkeley, CA 94720, USA}%
\affiliation{%
 Department of Chemistry, University of California, Davis, CA 95616, USA}%

\author{Wael Iskandar}
\affiliation{%
 Chemical Sciences Division, Lawrence Berkeley National Laboratory, Berkeley, CA 94720, USA}%
 
\author{Saijoscha Heck}
\affiliation{%
 Chemical Sciences Division, Lawrence Berkeley National Laboratory, Berkeley, CA 94720, USA}%
\affiliation{%
 Max-Planck-Institut f\"{u}r Kernphysik, Saupfercheckweg 1, 69117 Heidelberg, Germany}%
\affiliation{%
 J.W. Goethe Universit\"{a}t, Institut f\"{u}r Kernphysik, Max-von-Laue-Str. 1, 60438 Frankfurt, Germany}%

\author{Averell Gatton}
\affiliation{%
 Chemical Sciences Division, Lawrence Berkeley National Laboratory, Berkeley, CA 94720, USA}%
\affiliation{%
 Department of Physics, Auburn University, Alabama 36849, USA}%
 
\author{Elio G. Champenois}
\affiliation{%
 Graduate Group in Applied Science and Technology, University of California, Berkeley, CA 94720, USA}
\affiliation{%
 Chemical Sciences Division, Lawrence Berkeley National Laboratory, Berkeley, CA 94720, USA}%

 \author{Richard Strom}
\affiliation{%
 Chemical Sciences Division, Lawrence Berkeley National Laboratory, Berkeley, CA 94720, USA}%
\affiliation{%
 Department of Physics, Auburn University, Alabama 36849, USA}%
 
\author{Bethany Jochim}
\affiliation{%
 J.R. Macdonald Laboratory, Physics Department, Kansas State University, Manhattan, Kansas 66506, USA}
 
\author{Dylan Reedy}
\affiliation{%
  Department of Physics, University of Nevada Reno, Reno, Nevada 89557, USA}

\author{Demitri Call}
\affiliation{%
  Department of Physics, University of Nevada Reno, Reno, Nevada 89557, USA}

\author{Robert Moshammer}
\affiliation{%
  Max-Planck-Institut f\"{u}r Kernphysik, Saupfercheckweg 1, 69117 Heidelberg, Germany}
  
\author{Reinhard D\"{o}rner}
\affiliation{%
  J.W. Goethe Universit\"{a}t, Institut f\"{u}r Kernphysik, Max-von-Laue-Str. 1, 60438 Frankfurt, Germany}
 
\author{Allen L. Landers}
\affiliation{%
  Department of Physics, Auburn University, Alabama 36849, USA}
  
\author{Joshua B. Williams}
\affiliation{%
  Department of Physics, University of Nevada Reno, Reno, Nevada 89557, USA}
  
\author{C. William McCurdy}
\affiliation{%
 Chemical Sciences Division, Lawrence Berkeley National Laboratory, Berkeley, CA 94720, USA}%
\affiliation{%
 Department of Chemistry, University of California, Davis, CA 95616, USA}%
 
\author{Robert R. Lucchese}
\affiliation{%
 Chemical Sciences Division, Lawrence Berkeley National Laboratory, Berkeley, CA 94720, USA}%

\author{Itzik Ben-Itzhak}
\affiliation{%
 J.R. Macdonald Laboratory, Physics Department, Kansas State University, Manhattan, Kansas 66506, USA}
 
\author{Daniel S. Slaughter}
\affiliation{%
 Chemical Sciences Division, Lawrence Berkeley National Laboratory, Berkeley, CA 94720, USA}%
 
\author{Thorsten Weber}
\email{tweber@lbl.gov}
\affiliation{%
 Chemical Sciences Division, Lawrence Berkeley National Laboratory, Berkeley, CA 94720, USA}%

\date{\today}

\begin{abstract}
We report measurements on the H$^{+}$ + H$^{+}$ fragmentation channel following direct single-photon double ionization of neutral NH$_{3}$ at 61.5~eV, where the two photoelectrons and two protons are measured in coincidence using 3-D momentum imaging. We identify four dication electronic states that contribute to H$^{+}$ + H$^{+}$ dissociation, based on our multireference configuration-interaction calculations of the dication potential energy surfaces. The extracted branching ratios between these four dication electronic states are presented. Of the four dication electronic states, three dissociate in a concerted process, while the fourth undergoes a sequential fragmentation mechanism. We find evidence that the neutral NH fragment or intermediate NH$^+$ ion is markedly ro-vibrationally excited. We also identify differences in the relative emission angle between the two photoelectrons as a function of their energy sharing for the four different dication states, which bare some similarities to previous observations made on atomic targets. 
\end{abstract}

\pacs{Valid PACS appear here} 

\maketitle

\section{\label{sec:level1}Introduction}

Photo-Double-Ionization (PDI) is a process in which two electrons are ejected from an atom or molecule by absorption of a single photon. The resulting dication can be produced through either an indirect or a direct process. In the indirect process~\cite{Lablanquie, Sann}, the target is first ionized to produce a photoelectron and a singly-charged, excited cation. Subsequently, the cation decays by autoionization to produce a second continuum electron. The secondary electrons in indirect PDI have a unique signature, i.e. often a very narrow kinetic energy distribution and a rather isotropic angular emission pattern, which allows the process to be uniquely identified in a two-electron energy- or momentum-coincidence spectrum. In contrast to the indirect process, direct PDI involves simultaneous projection of two bound electrons to a correlated pair of continuum states. The interaction of the two electrons makes PDI an ideal process for studying electron-electron correlation~\cite{Mergel, Horner, Yip, Weber, Vanroose}.

Because of the repulsive Coulomb interaction between singly charged ions that is active over very large internuclear distances, the vertical double ionization thresholds of small molecules generally lie above the dissociation limits corresponding to formation of singly charged fragments. Since the dissociative electronic states of a polyatomic dication can possess various fragmentation pathways involving different numbers of bodies, distinct fragment species can be measured depending on various factors. Studying the photoelectron pair and various ionic fragments in coincidence can provide information on electron-electron correlation, the features of dication potential energy surfaces, and the nuclear dynamics involved in the dication breakup. The molecular fragmentation that typically follows direct PDI can be broadly described as occurring in a single step (concerted), where all charged and neutral fragments are born simultaneously, or occurring in multiple steps (sequential), where first a portion of the charged and neutral fragments are generated, leading to a metastable intermediate moiety, which then undergoes further dissociation to produce the final set of fragments~\cite{Gaire, ITZAK}.

In sequential fragmentation, the decay of the metastable intermediate(s) can be facilitated by various mechanisms, such as internal conversion or intersystem crossing to a dissociative state. Although spin-orbit coupling is generally weak in low-Z systems, intersystem crossing can in certain instances be the primary decay mechanism of metastable intermediates in a sequential dissociation process. Due to the weak coupling, the rate of intersystem crossing can be low, which leads to a significant period spent in the intermediate, providing time for the metastable fragments to rotate between the two fragmentation steps.

Distinguishing between concerted and sequential fragmentation channels is crucial in certain types of measurements, as concerted fragmentation channels can enable body-fixed frame photoelectron angular distributions to be retrieved, which carry far more information content than laboratory frame angular distributions. These body-fixed frame photoelectron angular distributions can, in most cases, only be reconstructed if the dication dissociates promptly along the relevant internuclear axes relative to rotation of those axes, allowing the molecular orientation at the instant of the PDI to be determined. This requirement is known as the axial recoil approximation~\cite{ZARE}. Since measuring body-frame photoelectron angular distributions following PDI poses a great experimental challenge, there exists only a small body of literature covering this topic, primarily focused on H$_2$ \cite{Weber,Weber1,Weber2,Vanroose,Reddish}. Various experimental methods such as particle coincidence 3-D momentum imaging, including COLd Target Recoil Ion Momentum Spectroscopy (COLTRIMS), allow measurements to be made in the molecular frame, but are predicated on the axial recoil approximation, hence it is useful to first determine which dication states exhibit concerted fragmentation mechanisms. The body-fixed frame electron emission pattern, or Molecular Frame Photoelectron Angular Distributions (MFPADs), can be established if the complete structure of the molecule at the time of dissociation can be reconstructed from the detected heavy ionic fragments. However, if a dissociative channel produces more than two (undetected) neutral fragments, or results in a polyatomic fragment with unknown orientation, only the Recoil Frame Photoelectron Angular Distribution (RFPAD) can be reconstructed. The latter represents the electron emission pattern with respect to a distinguished axis or plane spanned by the (detected) charged fragments. R/MFPADS are particularly sensitive to electron-electron correlation in both the initial and final states.

Various experimental and theoretical studies spanning a few decades have investigated the different dication electronic states and dissociation channels present in NH$_{3}$ following PDI, electron impact double ionization, and double ionization via double-charge-transfer spectroscopy \cite{Winkoun,Stankiewicz,Locht1,Locht2,Eland,Samson,Appell,Cheret,Langford,Griffiths,Locht3,White,Camilloni,Okland,Jennison,Boyd}. Most of these studies have focused on determining the appearance energies of the different fragments and the energetic locations of the dication electronic states. Among these investigations, no study, to our knowledge, has examined the H$^{+}$ + H$^{+}$ fragmentation channels of ammonia.

In this work, we investigate H$^{+}$ + H$^{+}$ dissociation following direct valence PDI of neutral NH$_{3}$ at 61.5~eV, approximately 27 eV above the PDI threshold~\cite{LOCHT}, where both the photoelectron and proton pairs are measured in coincidence using COLTRIMS. Based on Multi-Reference Configuration-Interaction (MRCI) calculations of dication Potential Energy Surfaces (PESs), we identify four dication electronic states that contribute to the H$^{+}$ + H$^{+}$ fragmentation. Our measurement provides the branching ratios between the four involved dication electronic states. As will be detailed below, of these four states, one appears to dissociate via a sequential mechanism and three dissociate in a concerted mechanism. Two of the three concerted dissociative states fragment at geometries near that of the ground state of neutral NH$_3$, where the axial recoil approximation appears valid, while the third state undergoes a significant change in nuclear geometry prior to fragmentation. By measuring the correlated electron and ion fragment momenta, we determine that the neutral NH fragment or charged intermediate NH$^+$ cation is ro-vibrationally excited with considerable internal energy, in some cases more than 2~eV.

\section{\label{sec:level2}Experiment}

The H$^{+}$ + H$^{+}$ fragmentation channel following valence PDI at 61.5~eV was investigated using COLTRIMS \cite{Dorner,Ullrich}, where the two photoelectrons and two protons were collected with full $4\pi$ solid angle, and their 3-D momenta were measured in coincidence, on an event-by-event basis. These four charged particles were guided using weak static parallel electric and magnetic fields, 11.4~V/cm and 10.0~G, respectively, to multi-hit position- and time-sensitive detectors at opposite ends of the spectrometer. Each detector comprised a Multi-Channel Plate (MCP) stack in chevron configuration for time readout, together with a delay-line anode, which decoded the hit position of each particle \cite{Jagutzki}. The electron and ion delay-line detectors were a hex-anode with an 80 mm MCP stack and a quad-anode with a 120~mm MCP stack, respectively. This system encodes a charge particle's 3-D momentum into its hit position on the detector and Time-of-Flight (TOF) relative to each ionizing extreme ultraviolet (XUV) pulse emitted by the synchrotron. These detectors have a small but significant dead-time following each detected particle, therefore they are subject to limited multi-hit capability \cite{Jagutzki}. This problem is most prominent in the electron pair detection, due to the small differences in the electron arrival times and hit positions at the detector. This dead-time effect can influence measured relative electron-electron angular distributions and is thus important to quantify, in order to distinguish real features from those that may emerge due to the detection scheme. We point out that the photoions do not suffer from this dead-time problem to the same degree as the electrons, as they are much more spread out in TOF and hit position on the ion detector. The electron-pair resolution is estimated by simulating the charged particle motion in the spectrometer fields with various sum kinetic energies and in various energy sharing conditions of the electron pair. For each pair of trajectories, the relative hit position and time-of-flight is computed, which is used to determine the fraction of simulated electron-pair events lost due to an estimated detector response, and thus approximate the fraction of actual losses.

The PDI experiment was performed using a tunable monochromatic linearly polarized beam of XUV photons produced at beamline 10.0.1.3. at the Advanced Light Source (ALS) synchrotron located at Lawrence Berkeley National Laboratory. The beamline monochromator was configured to provide 61.5~eV photons to the experiment, with an energy resolution narrower than $\pm$50~meV. The photon energy of 61.5~eV was chosen to be near the maximum of the PDI cross section of NH$_3$, while at the same time providing electron kinetic energies that can be detected with full solid angle and adequate energy resolution (around 1:10). Moreover, it is beneficial to keep the electron sum energy greater than $\sim5$~eV in order to utilize a large region of the 3D electron pair detection phase space, minimizing losses due to the electron detector dead-time (this will be apparent in Fig.~\ref{fig:Ee_Ee} later in the discussion).

A beam of rotationally and vibrationally cold neutral NH$_{3}$ ($\sim$80~K) was produced by an adiabatic expansion of the pressurized target gas ($\sim$35 psi) through a 50~$\mu$m nozzle, and collimated by a pair of downstream skimmers. The first skimmer has a diameter of 0.3~mm and the second skimmer has a diameter of 0.5~mm. The first skimmer is placed 8~mm downstream of the nozzle and in the zone of silence of the supersonic expansion. The second skimmer is 10~mm downstream of the first skimming stage. The resulting supersonic jet of target molecules propagated perpendicular to the photon beam, where the two beams crossed at the interaction region ($\sim0.15 \times 0.15 \times 1.0$~mm$^3$) inside the 3-D momentum imaging spectrometer, where PDI of the neutral ammonia in its ground state occurs at an average rate of less than 0.01 events per XUV pulse, assuring unambiguous coincidence conditions. 

The TOF and hit position of the charge particles produced by PDI were recorded in list mode on an event-by-event basis, enabling relevant events to be selected and examined in a detailed off-line analysis. For each PDI event, the kinetic energies and emission angles of the photoelectrons were determined from the 3-D photoelectron momenta, while the orientation of the recoil frame and the kinetic energy release (KER) of the fragmentation were determined using the measured 3-D momenta of the two protons. We infer the momentum of the center of mass of the remaining neutral NH radical by assuming momentum conservation between it and the two measured protons, treating the fragmentation as three-body breakup (even if the NH diatom fragments to N + H).  

\section{\label{sec:level3}Theory}

\begin{figure*}
\centering
\begin{tabular}{cc}
\includegraphics[width=8.5cm]{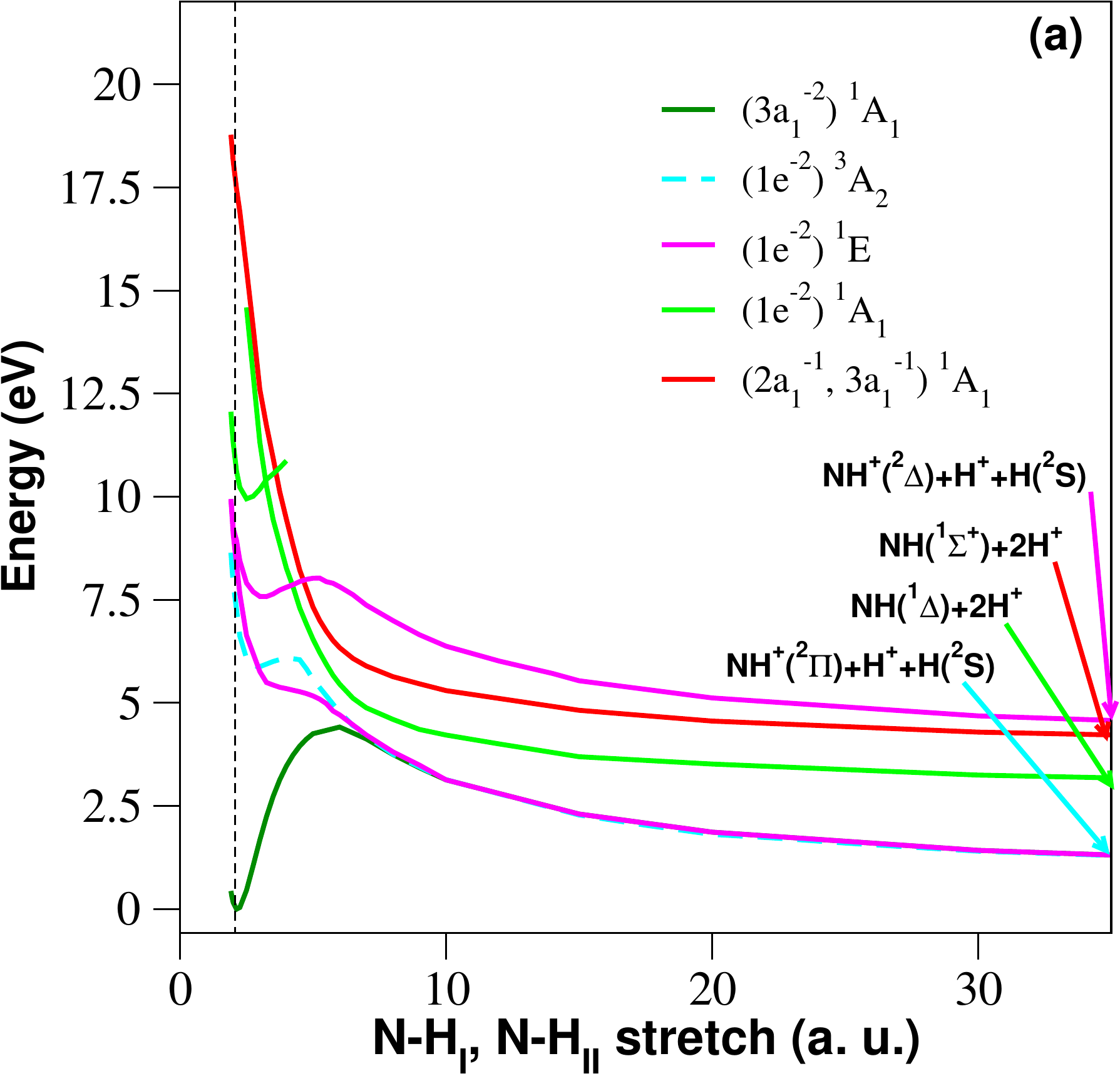}&
\includegraphics[width=8.5cm]{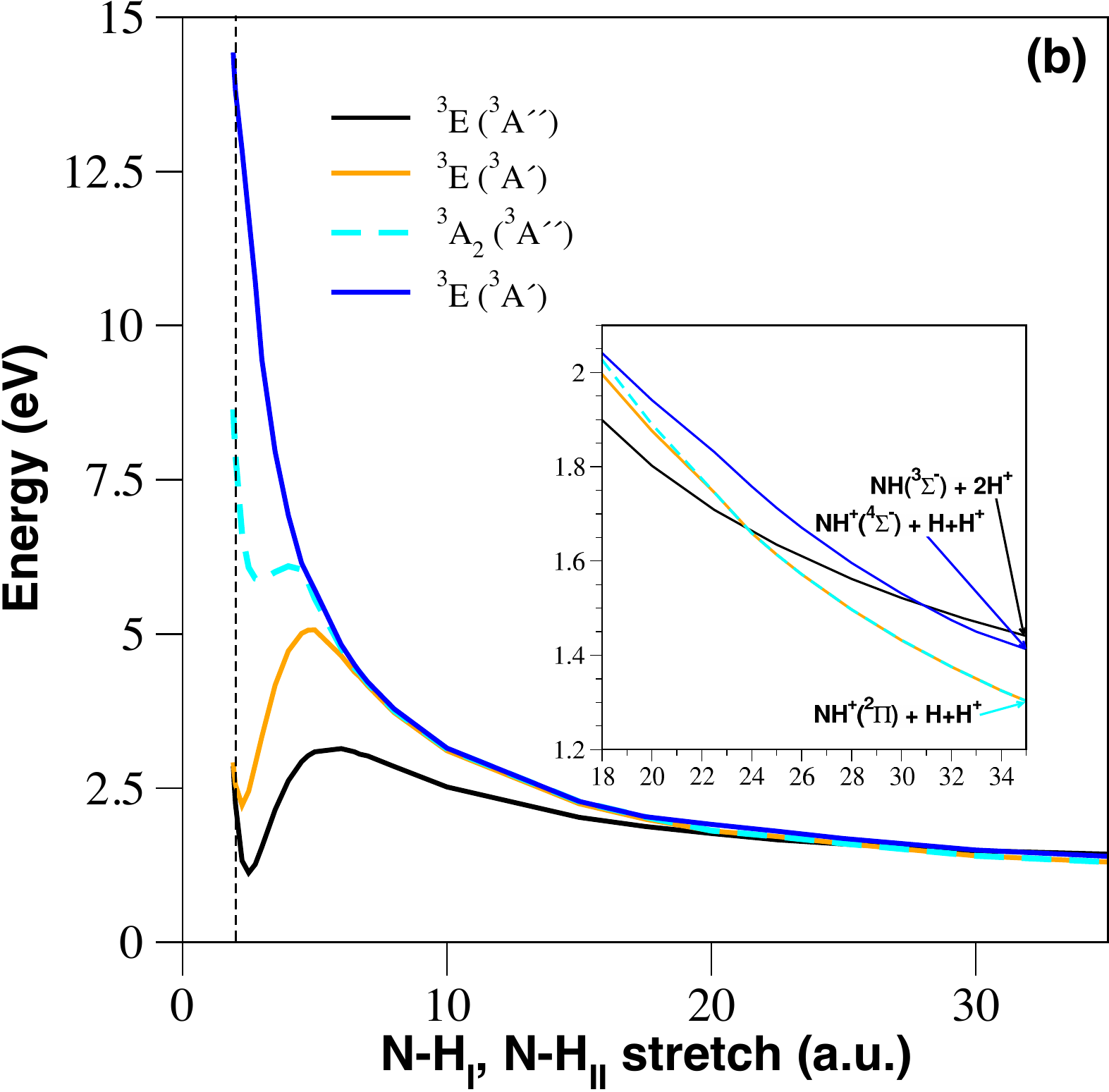}
\end{tabular}
\caption{PES cuts of  the NH$_3$ dication generated from MRCI calculations as described in the text. Here, two protons are symmetrically stretched while the third remains fixed , with all internal angles frozen at the geometry of neutral ammonia. The zero of energy is set at the ground-state ($^1A_1$) of the ammonia dication at the geometry of neutral ammonia, which lies 34.8 eV below the dication~\cite{LOCHT}. On this energy scale, the 61.5~eV photon energy lies at 26.7~eV. The dashed vertical line indicates the equilibrium geometry of neutral ammonia. (a) Cuts of the experimentally identified relevant states;  (b) detail of PES cuts for selected NH$_3$ dication triplet states. The inset indicates a region of large symmetric stretch distances where charge exchange may occur, as discussed in the text.}
\label{fig:PEC_NH3}
\end{figure*}

Most previous work on the ammonia dication have been experimental in nature. Of the earlier theoretical studies, most have focused on computing the vertical double ionization energy of neutral ammonia~\cite{POPE, LOCHT}. Tarantelli {\em et al.}~\cite{CEDERBAUM} computed excited state excitation energies of NH$_3^{2+}$ at the equilibrium geometry of NH$_3$ (see also Table~\ref{table:asymptotes}), but to our knowledge no earlier calculations of NH$_3^{2+}$ potential surfaces have been reported. The electron configuration of NH$_3$ in its ground-state is $(1a_{1})^{2}(2a_{1})^{2}(1e)^{4}(3a_{1})^{2}$. At a photon energy of $61.5$~eV, there are nine dication states which are energetically accessible following a vertical transition. In order to determine which of these states correlate with the three-body NH + H$^+$ + H$^+$ fragmentation channel, we carried out a series of electronic structure calculations. At each molecular geometry considered, we generated a set of molecular orbitals from a two-state, Complete Active Space (CAS) Multi-Configuration Self-Consistent Field (MCSCF) calculation on the lowest triplet ($^3$E) states of the dication. We kept one orbital (N 1s) frozen and included seven orbitals in the active space. We then performed MRCI calculations including all single and double excitations from the CAS reference space to generate 1-D cuts through the PESs. All bond angles were frozen at the equilibrium geometry of neutral ammonia (107$^o$), as was one hydrogen (H$_{\text{III}}$) bond length (1.9138 Bohr), while two hydrogen bonds (H$_{\text{I}}$) and (H$_{\text{II}}$) were symmetrically stretched. The results of the calculations are shown in Fig.~\ref{fig:PEC_NH3} with the electron configuration and state labels of each dication PES cut identified in the legend. The PES cuts were calculated out to a symmetric stretch N-H$_{\text{I}}$/N-H$_{\text{II}}$ distance of 50.0 bohr and extrapolated to infinity under the assumption of a purely repulsive Coulomb interaction between the positively charged fragments. The vertical energies at the neutral NH$_3$ geometry and the energies at the asymptotic limits are given in Table~\ref{table:asymptotes}. Note that here we do not explicitly consider cuts through the dication PESs where only one NH bond is stretched, as that is the subject of a future paper.

\begin{table*}
\centering
\begin{tabular}{  c  c  c  c  } 
 \hline\hline
 State & Vertical Energy (eV) & Asymptote & Adiabatic Limit Energy (eV)\\
 \hline
 (1e$^{-2}$)$^3$A$_2$ (cyan) & 8.64 (8.23)* & NH($^3\Sigma^-$)+H$^+$+H$^+$ & 0.96 \\
 (1e$^{-2}$)$^1$E (magenta) & 9.94 (9.91)* & NH$^+$($^2\Pi$)+H+H$^+$ & 0.52\\
 (1e$^{-2}$)$^1$A$_1$ (green) & 11.94 (11.77)* & NH($^1\Delta$[$^1\Sigma^+$])+H$^+$+H$^+$ & 2.69 [3.74]  \\
 (2a$_1^{-1}$,3a$_1^{-1}$)$^1$A$_1$ (red) & 18.94 (19.33)* & NH($^1\Sigma^+$)+H$^+$+H$^+$ & 3.74 \\
 \hline
\end{tabular}
\caption{Ammonia dication vertical energies at neutral NH$_3$ geometry and asymptotic three-body limits extrapolated from {\em ab initio} calculations at N-H$_{\text{I}}$/N-H$_{\text{II}}$ distances of 50.0 bohr. Note that for the $^1$A$_1$ state (green), two possible asymptotic limits are given (see text). *Values in parentheses are configuration interaction results from Ref.~\cite{CEDERBAUM}.}
\label{table:asymptotes}
\end{table*}

Our calculations reveal that there are only three three-body proton-proton dissociative limits. Of the three-body proton-proton channels, two are singlet states and one is a triplet state. The two singlet states leave the remaining neutral NH molecule in a $^{1}\Delta$ or a $^{1}\Sigma^{+}$ state, while the triplet leaves the neutral NH fragment in a $^{3}\Sigma^{-}$ state. In order to produce the two experimentally observed protons in the fragmentation, the implication is that an excitation must access one of these three dissociative limits, or undergo a four-body fragmentation mechanism that yields two protons, i.e. results in the fragments N + H + H$^+$ + H$^+$.

We identify three relevant singlet states, ($1e^{-2}$) $^1$E, ($1e^{-2}$) $^1$A$_1$, and ($2a_1^{-1}, 3a_1^{-1}$) $^1$A$_1$, shown in Fig.~\ref{fig:PEC_NH3}~(a) as solid curves (magenta, green and red), and a fourth relevant triplet state, ($1e^{-2}$) $^3$A$_2$, shown as a dashed curve (cyan). The curves in Fig.~\ref{fig:PEC_NH3}~(a) are color-coded to be consistent with the experimental features to be discussed in the following section. Since spin-orbit coupling, required for an intersystem crossing, is expected to be weak, the triplet state must dissociate to a triplet fragment state. However, Fig.~\ref{fig:PEC_NH3} shows that the $^3$A$_2$ state (cyan dashed) actually correlates with the NH$^+$($^2\Pi$)+H$^+$+H($^2$S) dissociation channel (cyan dashed in the Fig.~\ref{fig:PEC_NH3}~(b) inset). To reach the NH($^3\Sigma^-$)+2H$^+$ limit (black curve in the inset) requires a charge exchange, which is possible at N-H separations greater than 18~Bohr where the $^3$E ($^3$A$''$) and $^3$A$_2$ ($^3$A$''$) states become nearly degenerate in energy across a range of geometries (see cyan dashed and black curves in Fig.~\ref{fig:PEC_NH3}~(b)). This can result in charge exchange over a large range of distances along the asymmetric stretch coordinate that the dissociating wave packet traverses. We have observed an analogous asymptotic charge-exchange mechanism at such large N-H distances in an earlier study of dissociative electron attachment to ammonia~\cite{Rescigno16}.

For singlet states accessible in the Franck-Condon (FC) region as depicted in Fig.~\ref{fig:PEC_NH3}~(a), there are two different proton-proton limits (red and green curves). The ($1e^{-2}$) $^1$A$_1$ state (green) is seen to cross two other dissociative $^1$A$_1$ states (green and red), which correlate with the products NH ($^1\Delta$) or NH ($^1\Sigma^+$) plus two protons, respectively. Conical intersections (CIs) between the dissociative states and the initially excited $^1$A$_1$ state can result in dissociation to either of the singlet limits. Since the location of the CIs cannot be determined from 1-D energy cuts (although numerous avoided crossing are observed), we must rely on the experimental findings to see which of the singlet limits are populated. 

Previous experimental observations have indicated that PDI to the ($1e^{-2}$) $^1$E state is associated with the NH$^{+}$ + H$^{+}$ + H fragmentation channel \cite{Stankiewicz}. Since the dissociative limit of the ($1e^{-2}$) $^1$E state does not directly yield two protons, excitation to this state must undergo a nonadiabatic transition to either of the two $^1$A$_1$ excited dication states, or the NH$^{+}$ fragment it produces must dissociate to N + H$^{+}$, in order to result in the measured two-proton coincidence.

The ($1e^{-2}$) $^1$E state, doubly degenerate in C$_{3v}$ geometry, splits into A$'$ and A$''$ states when two N-H bonds are symmetrically stretched. Of these two states, the upper state has A$''$ symmetry. Accordingly, internal conversion to either of the $^1$A$'$ states that have limits producing NH + H$^+$ + H$^+$ is unfavorable. Dissociation on the lower curve yields an NH$^+$ fragment in its X$^2\Pi$ ground state. If the NH$^+$ fragment is produced with sufficient internal energy, it can dissociate to N$^+$($^3$P$) +$ H($^2$S) or through intersystem crossing to another NH$^+$ state, to N($^4$S) + H$^+$. In the latter case this results in the production of two protons via a sequential  four-body breakup NH$_3^{++} \rightarrow$ NH$^+$ + H$^+$ + H $\rightarrow$ N + 2H$^+$ + H. This sequential breakup process will be examined in detail below.

\section{\label{sec:level4}Results and Discussion}

Using the insights gained from the calculations on dication electronic states described in the previous section, we provide a detailed discussion of the experimental results below, which has been divided into three sub-sections. In the first sub-section, we present and discuss the energetics of the photoelectrons and photoions, identifying features which correspond with the states outlined in the previous section. In the second sub-section, we address the details of the dissociation dynamics by analyzing the relative emission angle between the two protons in each of these states. Lastly, we present results on the photoelectron dynamics via an analysis of the relative emission angle between the two photoelectrons for the four dication states in different energy sharing conditions of the electron pair.

\subsection{\label{PEPI}Photoelectron and photoion energetics}

\begin{figure}[h!]
        \includegraphics[width=8.5cm]{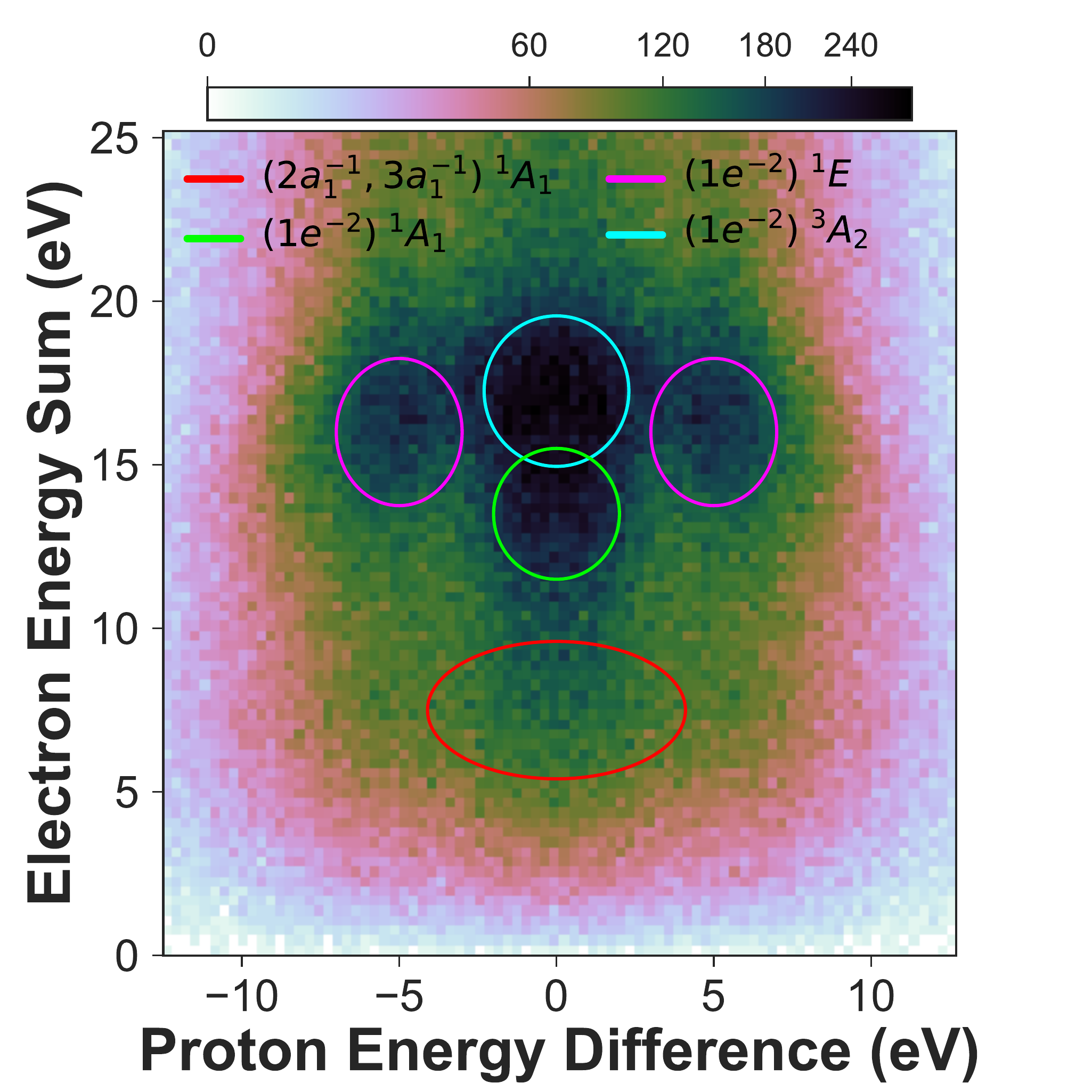}
\caption{The yield of H$^+$ + H$^+$ after valence PDI of NH$_3$ as a function of the energy difference of the proton pair and the energy sum of the photoelectron pair. The four color-coded ellipses guide the eye to the relevant features and dication states discussed in the text. The data has been mirrored about zero proton energy difference, as there is no physical meaning to the order in which the two proton are detected.}
\label{fig:Eesum_Epdiff}
\end{figure}

The H$^{+}$ + H$^{+}$ fragmentation following PDI of NH$_3$ at 61.5~eV, $\sim$27~eV above the PDI threshold, is identified and isolated by selecting the two charged fragments in the time-of-flight spectrum and then in momentum space, and by enforcing that two electrons are measured in coincidence with the two ionic fragments. First, we plot the PDI yield as a function of the energy difference between the two particles of the proton pair and the energy sum of the photoelectron pair. This plot is shown in Fig.~\ref{fig:Eesum_Epdiff}. 

Here we are able to identify four features, which we attribute to the four different dication electronic states calculated and tabulated in the previous section, resulting in photoelectron pairs with energy sums centered around 7.3~eV, 14.1~eV, 16.7~eV, and 17.6~eV. These features are indicated by ellipses to guide the eye and color-coded to be consistent with the calculated values of 7.8, 14.8, 16.8 and 18.1 eV listed in Table~\ref{table:asymptotes}. The measured and calculated values are in excellent agreement and are consistent with the state assignments. Note that the ellipses do not reflect the actual software gates used in the data analysis. In the offline analysis, we choose each of these states by selecting carefully around the center of each feature in Fig.~\ref{fig:Eesum_Epdiff}, while additionally placing constraints on the proton energy sum (which aids in separating the low and high KER features). Enforcing conditions in a multitude of dimensions in this fashion enables us to separate these four features for subsequent analysis. 

Each of these four features possesses a Full Width at Half Maximum (FWHM) in electron energy sum of roughly 6.2~eV, 2.1~eV, 4.2~eV, and 2.4~eV, respectively. The FWHM of the electron energy sum of each dication state roughly indicates the magnitude of the gradient of the PES in the FC region, provided the electron detector energy resolution is smaller than the width of the feature in question. To estimate the expected spread of observed photoelectron energies for the various dication states, we use a variant of the so-called reflection approximation~\cite{Rescigno88}. The range of detectable KERs is determined by the FC envelope of the initial (neutral) vibrational state reflected onto the final dication PESs. We approximate the initial vibrational wavefunction with a harmonic oscillator function $\chi_0$, obtained from a fit of the ground state energy of ammonia as a function of the symmetric stretch coordinate. If we assume that the PDI cross section varies little over the FC region and that the final continuum vibrational wavefunctions can be approximated by delta functions about the classical turning points on the dication PESs~\cite{Vanroose}, then the envelope of the expected photoelectron energies is given by the values of the vertical PDI energies as a function of the symmetric-stretch coordinate, weighted by the square of the symmetric-stretch vibrational wavefunction. We find that $|\chi_0|^2$ reaches half its maximum value at a symmetric-stretch displacement of approximately $\pm$0.11~Bohr from  equilibrium, and we have used these values to calculate the FWHM of the photoelectron distributions. According to this procedure we find widths of 5.1~eV, 1.9~eV, 3.1~eV and 2.2~eV, respectively, which are in good agreement with the measurement (see also Table~\ref{table:Exp_The_Ee_KER}). From this we find that, given our photoelectron spectral resolution of roughly $\Delta E/E \sim 0.1$, the measured FWHM of each state does indeed roughly correspond with the gradient of its PES in the FC region.

\begin{figure}[h!]
          \includegraphics[width=8.5cm]{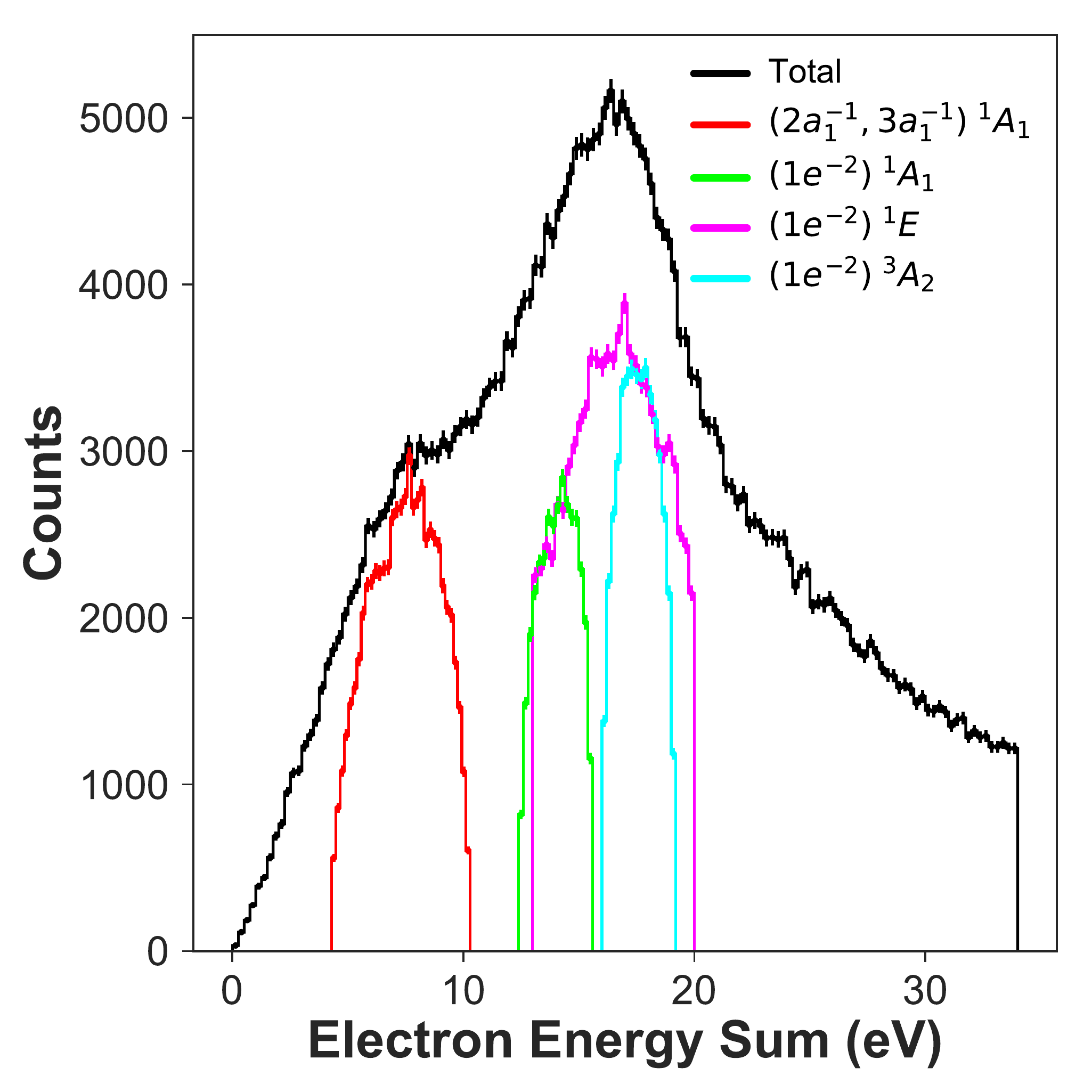}
\caption{The NH$_3$ PDI yield of the H$^{+}$ + H$^{+}$ channel as a function of the photoelectron energy sum integrated over all features (black) as well as for the four color-coded features corresponding with the identified dication states. The electron energy sum distributions for the four features have been scaled by a factor of four, for better visibility.}
\label{fig:Eesum}
\end{figure}

We present the 1-D photoelectron energy sum spectrum in Fig.~\ref{fig:Eesum}, where each feature we identified in Fig.~\ref{fig:Eesum_Epdiff} has been indicated by the color-coded distribution. The peak value of each distribution has been indicated in Table~\ref{table:Exp_The_Ee_KER}, where it is also compared with the theoretically calculated value. We find good agreement between the measurement and calculations. We can clearly identify the feature with a photoelectron energy sum centered near 7.3~eV, while the three higher photoelectron energy features appear clustered together. The branching ratios between the four measured features that correspond with the four dication states are estimated from the relative yield of these four features, and they are presented in Table~\ref{table:branching}. The method for extracting these branching ratios is discussed later.

\begin{figure}[h!]
        \includegraphics[width=8.5cm]{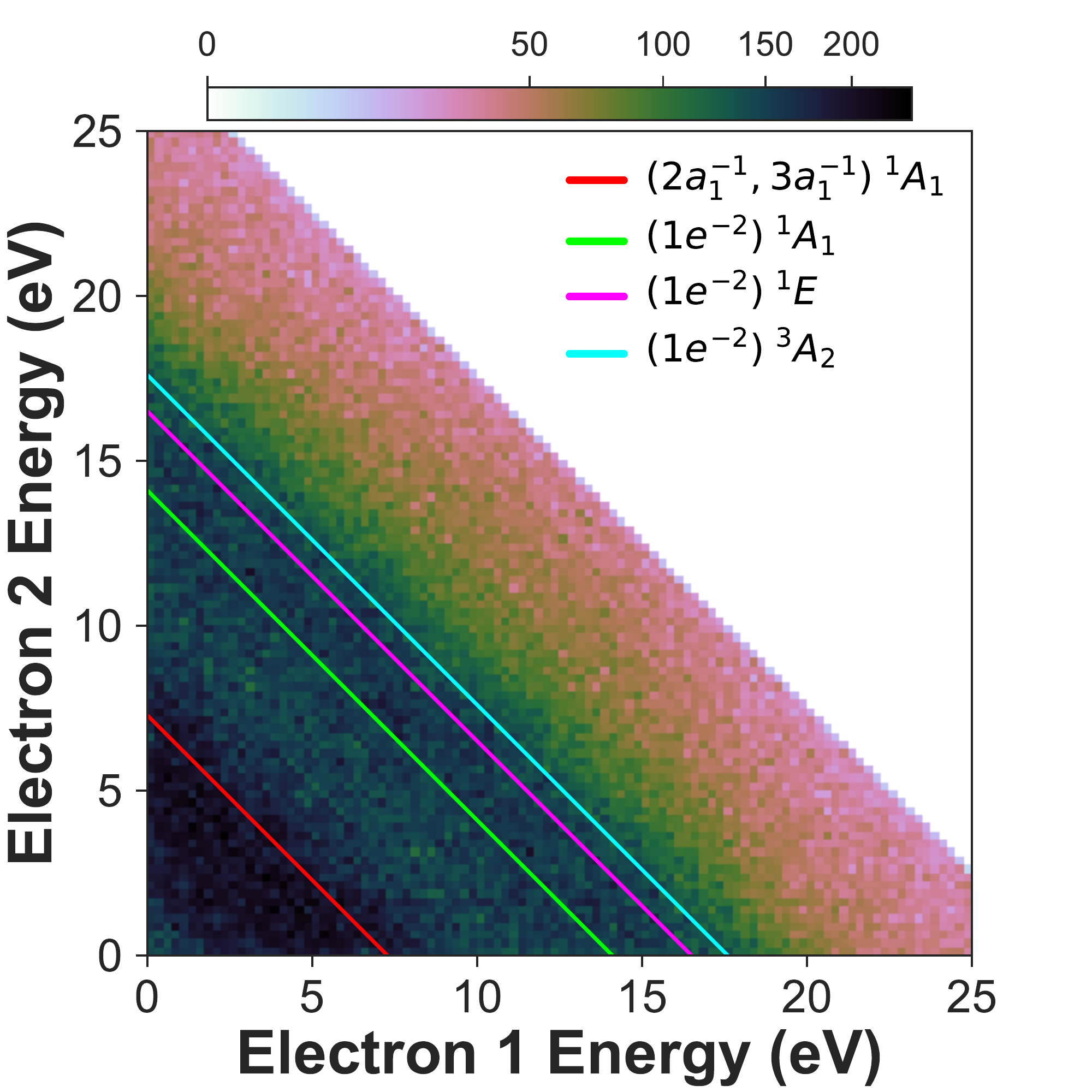}
\caption{Electron-electron energy correlation map for the H$^+$ + H$^+$ channels of the PDI of NH$_3$. The four identified dication states are color-coded and indicated by diagonal lines.}
\label{fig:Ee_Ee}
\end{figure}

The yields of the H$^+$ + H$^+$ channels as a function of the kinetic energy of the first and second detected electron are plotted in the electron-electron energy correlation map shown in Fig.~\ref{fig:Ee_Ee}. Since the two electrons are indistinguishable particles, the labeling (as 1 and 2) is arbitrary and the figure has been symmetrized across the diagonal (the line E$_2$ = E$_1$) to account for this.

The four different features that correspond to the dication electronic states identified in Figs.~\ref{fig:Eesum_Epdiff} and \ref{fig:Eesum} are indicated as color-coded diagonal lines (which take the form E$_2$ = -E$_1$ + E$_{sum}$, where E$_{sum}$ is the photoelectron energy sum corresponding to that feature) in Fig.~\ref{fig:Ee_Ee}. We point out that the red diagonal line appears to be off the center of the diagonal feature, even though this location represents the peak. This is because there are fewer bins along a given constant electron energy sum (i.e. a diagonal of the form E$_2$ = -E$_1$ + E$_{sum}$) as the photoelectron energy sum decreases. Since the length of a constant energy diagonal line scales as $\sqrt{2}E_{e^-}$, the number of available bins that events can populate decreases with decreasing E$_{e^-}$. This leads to the counts at low constant electron energy sums to be concentrated in just a small number of bins, which can render the true location of the peak obscured in this 2-D spectrum, while it is well represented in Fig. \ref{fig:Eesum}.

All four dication states are accessed via direct PDI, as indicated by the uniform diagonal features (taking the form E$_2$ = -E$_1$ + E$_{sum}$) and the absence of any Auger or autoionization lines, which would appear with vertical or horizontal characteristics at very unequal energy sharing due to the autoionization electron possessing a narrow constant (low) energy. The uniformity of the diagonal features in Fig.~\ref{fig:Ee_Ee} indicates that the two photoelectrons do not exhibit a strong preference towards either equal or unequal energy sharing, but rather exhibit roughly constant H$^+$ + H$^+$ yield as a function of the electron energy sharing (see also Fig.~\ref{fig:SDCS}). The photoelectron energy sharing distributions for each of the four states will be presented and discussed in more detail in the final sub-section C. 

\begin{figure}[h!]
        \includegraphics[width=8.5cm]{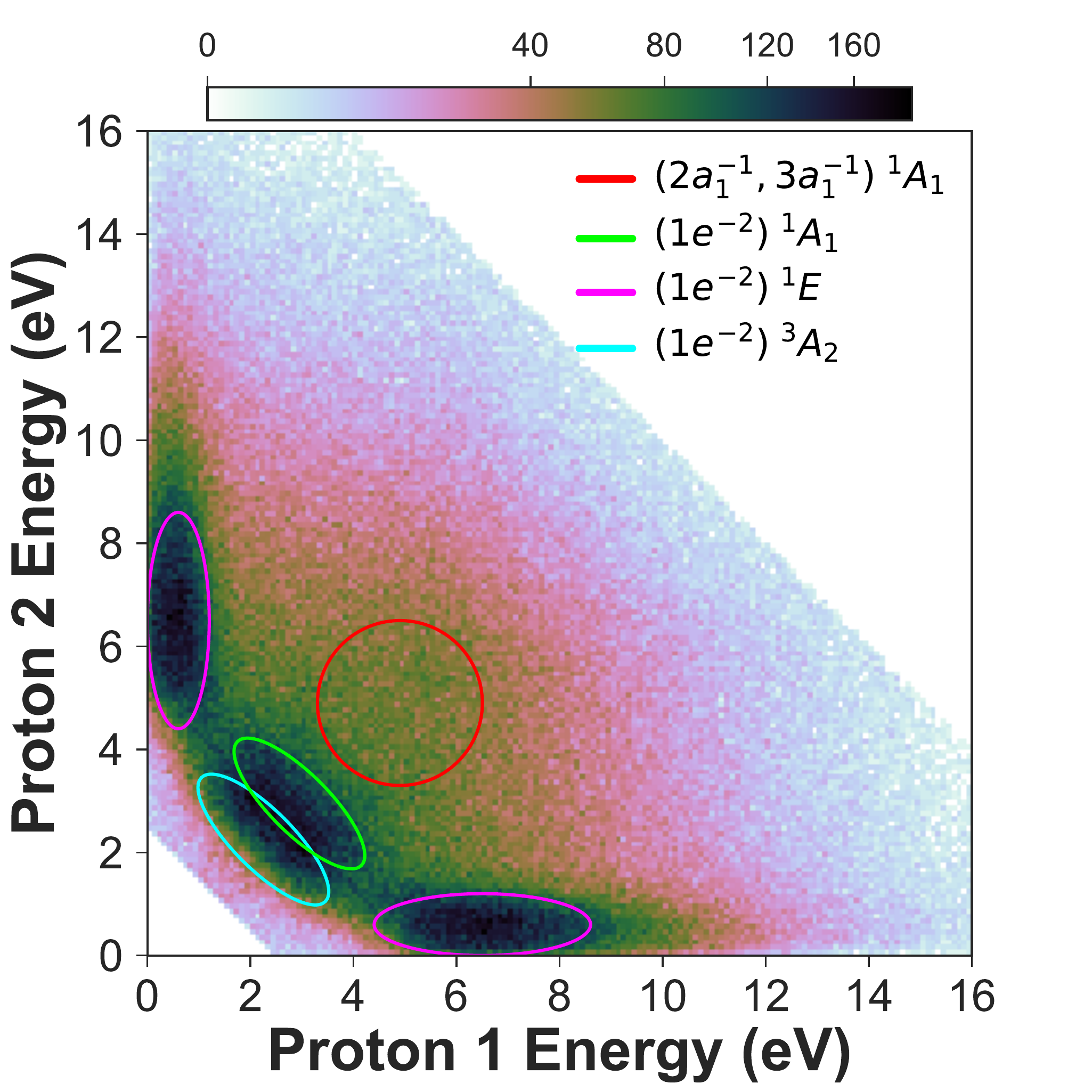} 
\caption{Proton-proton energy correlation map for the H$^+$ +  H$^+$ fragmentation channels of the valence PDI of NH$_3$. The four identified dication states are color-coded and indicated by ellipses to guide the eye.}
\label{fig:Ep_Ep}
\end{figure}

The same four features, corresponding with those seen in Fig.~\ref{fig:Eesum_Epdiff}, are present in the proton-proton energy correlation map given in Fig.~\ref{fig:Ep_Ep}. As in the electron-electron energy correlation map of Fig.~\ref{fig:Ee_Ee}, the two protons are indistinguishable particles, hence the labeling is arbitrary and the figure has been symmetrized across the diagonal (the line E$_2$ = E$_1$). We have removed events that lie in the low energy corner of Fig.~\ref{fig:Ep_Ep}, as the events that lie within this region originate from false coincidences. For each proton-pair we compute the KER by treating the process as a three-body fragmentation and by inferring the momentum of the N-H center of mass via momentum conservation. Each feature seen in Fig.~\ref{fig:Ep_Ep} possesses a different KER distribution centered around 12.7~eV, 5.9~eV, 7.7~eV, and 5.5~eV, each with a FWHM of roughly 6.1~eV, 2.2~eV, 3.0~eV, and 2.0~eV, respectively. These KER distributions are discussed in more detail later. The three KER features we have associated with the $(2a_{1}^{-1}, 3a_{1}^{-1})$ $^{1}$A$_{1}$, $(1e^{-2})$ $^{1}$A$_{1}$ and $^3$A$_2$ states exhibit a tendency towards equal energy sharing between the two protons, consistent with a concerted breakup mechanism. The fourth KER feature, associated with the $^1$E state, exhibits highly unequal energy sharing between the two protons, indicative of a sequential breakup mechanism.

\begin{table*}
\centering
\begin{tabular}{  c  c  c  c  c  } 
 \hline\hline
 State & \multicolumn{2}{c}{Photoelectron Energy Sum (eV)} & \multicolumn{2}{c}{KER (eV)}\\
  & Experiment (FWHM) & Theory (FWHM) & Experiment (FWHM) & Theory* (FWHM)\\
 \hline
 (1e$^{-2}$)$^3$A$_2$ (cyan) & 17.6 (2.4) & 18.1 (2.2) & 5.5 (2.0) & 7.7 (2.2) \\
 (1e$^{-2}$)$^1$E (magenta) & 16.7 (4.2) & 16.8 (3.1) & 7.7 (3.0) & 9.4 (3.1) \\
 (1e$^{-2}$)$^1$A$_1$ (green) & 14.1 (2.1) & 14.8 (1.9) & 5.9 (2.2) & 8.2 (1.9) \\
 (2a$_1^{-1}$,3a$_1^{-1}$)$^1$A$_1$ (red) & 7.3 (6.2) & 7.8 (5.1) & 12.7 (6.1) & 15.2 (5.1) \\
 \hline
\end{tabular}
\caption{The measured and calculated photoelectron energy sum and KER centroids for each of the four identified features from H$^{+}$ + H$^{+}$ fragmentation following PDI of NH$_3$ at 61.5~eV. The asterisk marking the theoretical KER values indicates that these are calculated assuming ro-vibrational ground state fragments, i.e. assuming maximum KER with no energy channeled into internal excitations. The theoretical KER values are all roughly 2~eV higher than the measured values, which is consistent with the dissociation producing fragments possessing approximately 2~eV of ro-vibrational energy (as explained in the text).}
\label{table:Exp_The_Ee_KER}
\end{table*}

Theoretical KER values are obtained by subtracting the asymptotic energies from the associated vertical PDI energies in Table~\ref{table:asymptotes}, while theoretical photoelectron energy sum values are computed by subtracting these vertical PDI energies and the double ionization threshold from the photon energy. These results are displayed in Table~\ref{table:Exp_The_Ee_KER}. For the concerted breakup channels theory gives 15.2, 8.2, and 7.7 eV for the $(2a_{1}^{-1}, 3a_{1}^{-1})$ $^1$A$_1$, $(1e^{-2})$ $^1$A$_1$, and $(1e^{-2})$ $^3$A$_2$ dication states, respectively. (Note that the NH($^1\Sigma^+$) asymptote has been used for both $^1$A$_1$ states.) These values are uniformly higher, by  2.5, 2.3 and 2.2 eV, respectively, than the measured values. This discrepancy is either due to calculated dissociation energies that are all uniformly too small by approximately 2 eV, or can arise if the NH fragment in all three concerted breakup channels is produced with approximately 2 eV of ro-vibrational energy. The energy balance of the sequential breakup is consistent with high internal energy of the NH fragment. For the sequential $^1$E breakup channel, theory gives a KER value of 9.4 eV, which is 1.7 eV higher than the measured value. This corresponds to a four-body breakup mechanism, discussed in detail in Section \ref{sequential}.

The FWHM of the KER distribution associated with each dication state carries similar information as the electron sum energy FWHM (see also Table~\ref{table:Exp_The_Ee_KER}), indicating the steepness of the potential energy surfaces in the FC region, convoluted with the energy resolution of the ion spectrometer (estimated to be on the order of 100~meV). These values are indicated in Table~\ref{table:Exp_The_Ee_KER}.

\begin{figure}[h!]
        \includegraphics[width=8.5cm]{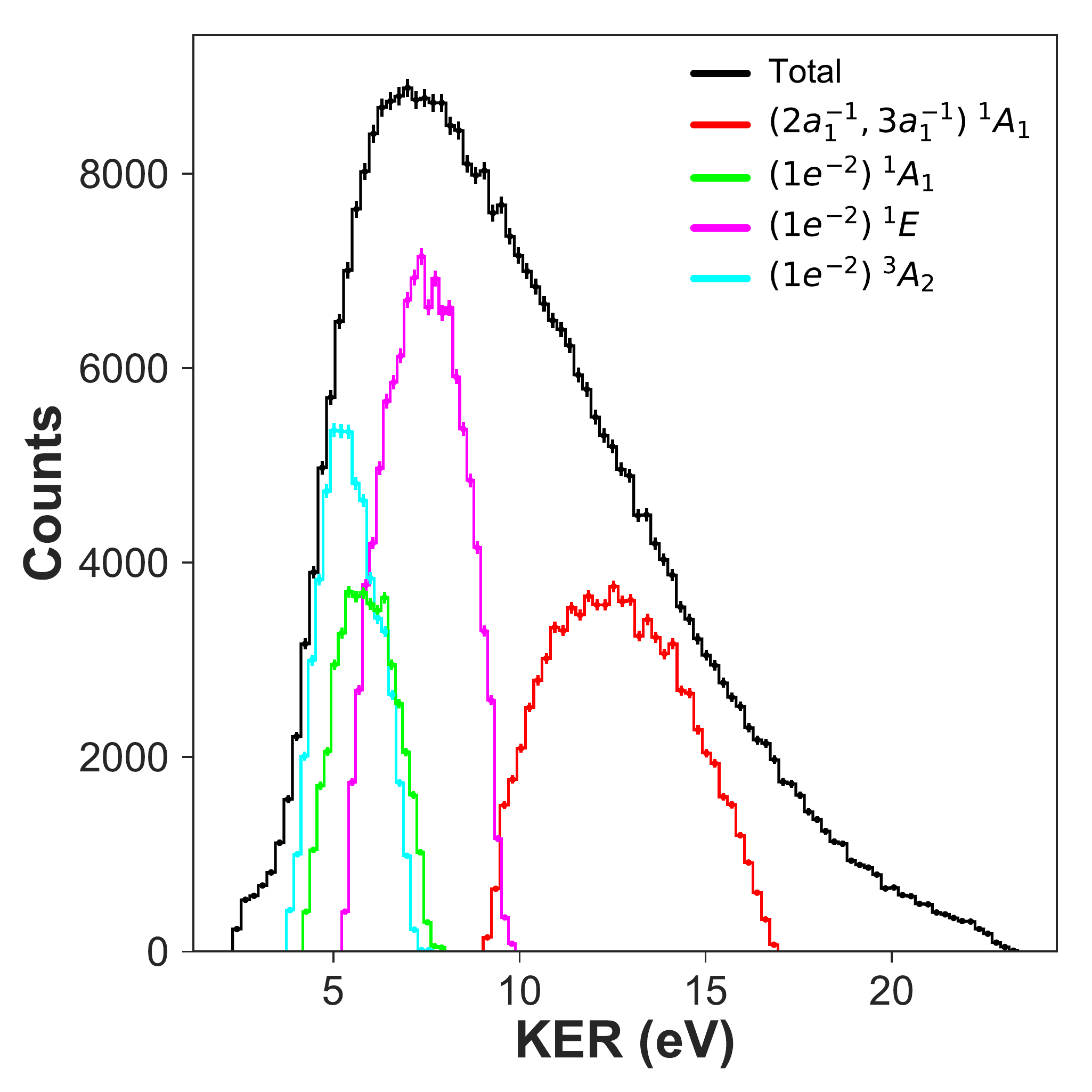} 
\caption{The yield of H$^{+}$ + H$^{+}$ fragmentation channel of the valence PDI of NH$_3$ as a function of KER, shown for the total yield (black), as well as for the four color-coded features corresponding to the identified relevant dication states. The KER distributions for the four features have been scaled by a factor of five, for better visibility.}
\label{fig:KER}
\end{figure}

We show the 1-D KER spectrum in Fig.~\ref{fig:KER}, where each feature we identified in Fig.~\ref{fig:Ep_Ep} has been indicated by the color-coded distribution. The peak value of each distribution is listed in Table~\ref{table:Exp_The_Ee_KER}, where it is also compared with our theoretical results. The differences between the measured and calculated values in Table~\ref{table:Exp_The_Ee_KER} are consistent with the molecular fragments containing roughly 2~eV of internal energy (or the aforementioned four-body breakup mechanism, which is discussed below), not explicitly accounted for in our theory which only considers fragments in their rotational and vibrational ground states.

\begin{figure}[h!]
    {
    {
        \includegraphics[width=8.5cm, trim=0.2cm 0cm 1.8cm 0cm, clip]{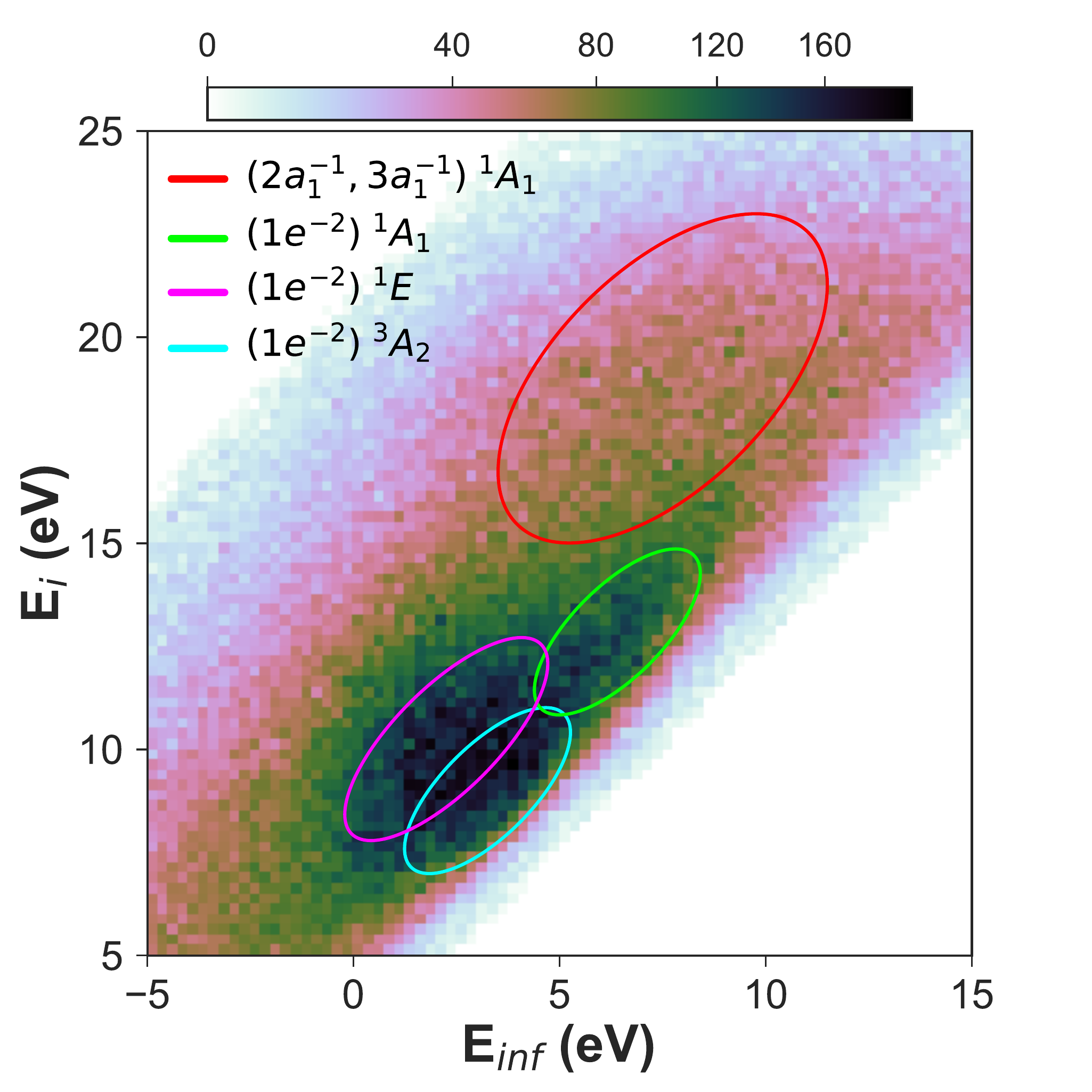}}
    }
\caption{The PDI yield of NH$_3$ at 61.5~eV as a function of the energy above the double ionization threshold at the adiabatic limit following dissociation, E$_{inf}$, and the energy above the double ionization threshold, E$_i$, for each of the four identified relevant dication states from the H$^{+}$ + H$^{+}$ fragmentation channel. The four identified dication states are color-coded and indicated by ellipses to guide the eye.}
\label{fig:EiEinf}
\end{figure}

The estimated branching ratios between these four dication states are displayed in Table~\ref{table:branching}. These branching ratios are approximated by simultaneously fitting each feature in Fig.~\ref{fig:Eesum_Epdiff} with a 2-D Gaussian distribution (although the distributions may not be explicitly Gaussians, this is nonetheless a good approximation). The fitting procedure varied the widths along each dimension independently, while also including a varying constant background offset. Following this fitting procedure, we integrate the fit for each feature individually to estimate its contribution to the total H$^+$ + H$^+$ yield. The main contribution to the uncertainty of the branching ratio is rooted in the aforementioned electron pair deadtime, which influences the detection yield of the electron-ion coincidences for each dication state as a function of the electron sum energy. Applying the simulation mentioned above, we estimate the total possible loss in PDI yield for electron sum energies of 7.3~eV (($2a_1^{-1}, 3a_1^{-1}$) $^1A_1$), 14.1~eV (($1e^{-2}$) $^1A_1$), 16.7~eV (($1e^{-2}$) $^1E$), and 17.6~eV (($1e^{-2}$) $^3A_2$) to be 27.2\%, 10\%, 8.1\%, and 7.5\%, respectively. This translates to an error of up to 5\% in the branching ratio. Errors due to deviations from the assumed Gaussian shape of each feature in the fitting process and the quality of the fit are estimated to be small ($<$1\% and $<$0.3\%, respectively).

\begin{table}
\centering
\begin{tabular}{  c  c  } 
 \hline
 State & Branching Ratio \\
 \hline
 $(2a_{1}^{-1}, 3a_{1}^{-1})$ $^{1}$A$_{1}$ & 14.6\% \\
 $(1e^{-2})$ $^{1}$A$_{1}$ & 4.5\% \\
 $(1e^{-2})$ $^{1}$E & 18.1\% \\
 $(1e^{-2})$ $^{3}$A$_{2}$ & 62.8\% \\
 \hline
\end{tabular}
\caption{The branching ratios for the four dication states contributing to the H$^+$ + H$^+$ dissociation channel following PDI of NH$_3$ at 61.5~eV. The errors on these fractions are estimated to be up to 5\% (see text).}
\label{table:branching}
\end{table}

Last, we plot the H$^+$ + H$^+$ yield as a function of the energy at the adiabatic limit E$_{inf}$ and the energy above the double ionization threshold E$_i$. This plot is shown in Fig.~\ref{fig:EiEinf}, with $E_{i} = \hbar\omega - DIP - (E_{e_1} + E_{e_2})$ and $E_{inf} = \hbar\omega - DIP - (E_{e_1} + E_{e_2} + KER)$, where DIP is the Double Ionization Potential. As a guide to the eye, each of the four identified features have been indicated by ellipses. This plot indicates for each state and its dissociative limit where the NH$_3^{++}$ is excited to upon PDI, relative to the dication ground state. The circled features can be directly compared with the calculated vertical energy and adiabatic energy values shown in Table~\ref{table:asymptotes}, which show good agreement with our theoretical results. As mentioned above, the measured energies E$_{inf}$ are each approximately 2~eV higher than what is theoretically predicted for rotationally and vibrationally cold fragments, whereas the molecular fragments in the experiment can carry away this amount of energy internally, as we think is plausible from our analysis presented in section~\ref{sequential}. 

\subsection{\label{sequential}Photodissociation dynamics: distinguishing concerted and sequential fragmentation}

\begin{figure}[h!]
    {
    {
        \includegraphics[width=4.2cm, trim=0.5cm 1.7cm 0.5cm 0.1cm, clip]{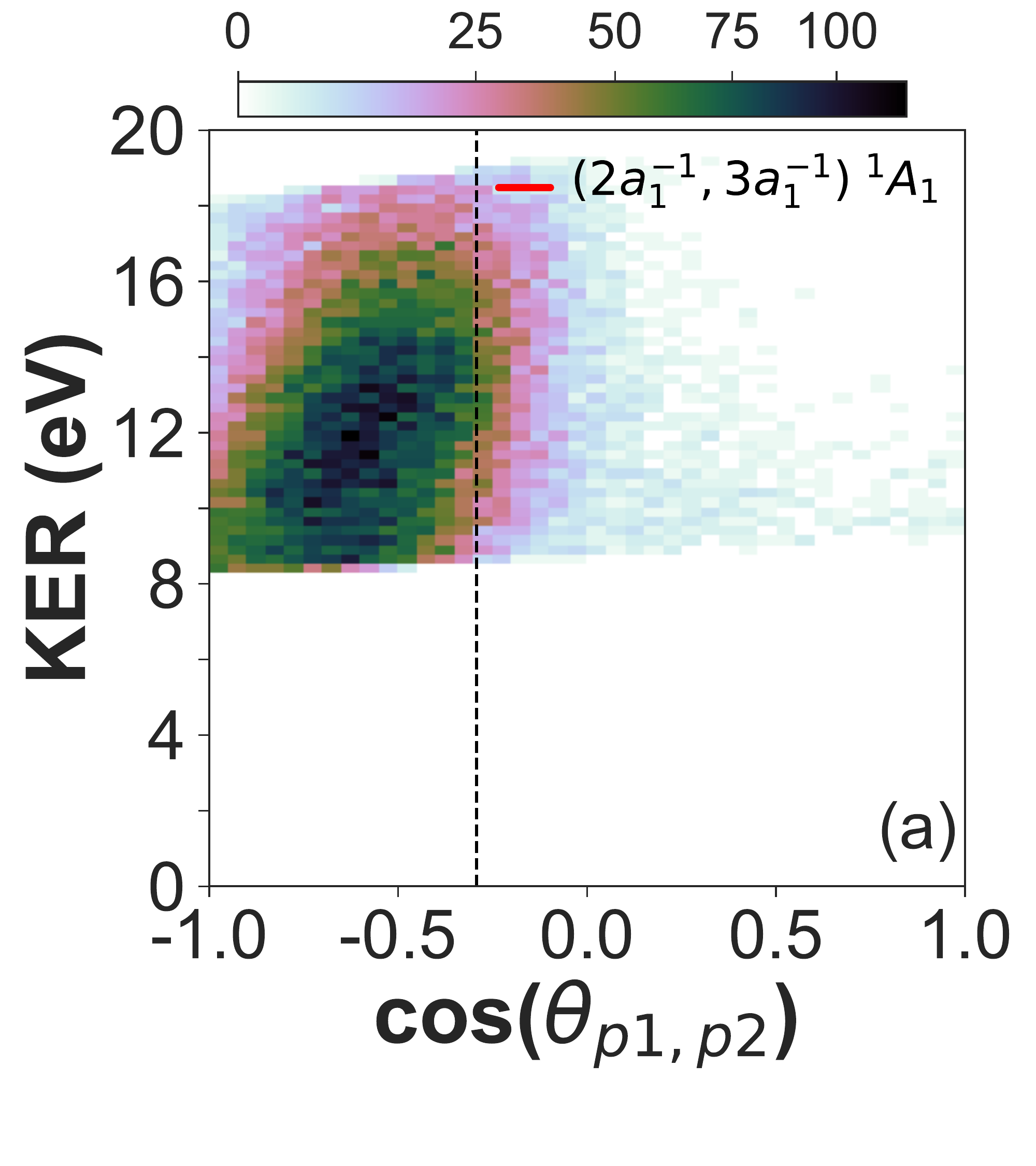}}
    {
        \includegraphics[width=3.85cm, trim=0.5cm 0.95cm 0.5cm 0.0cm, clip]{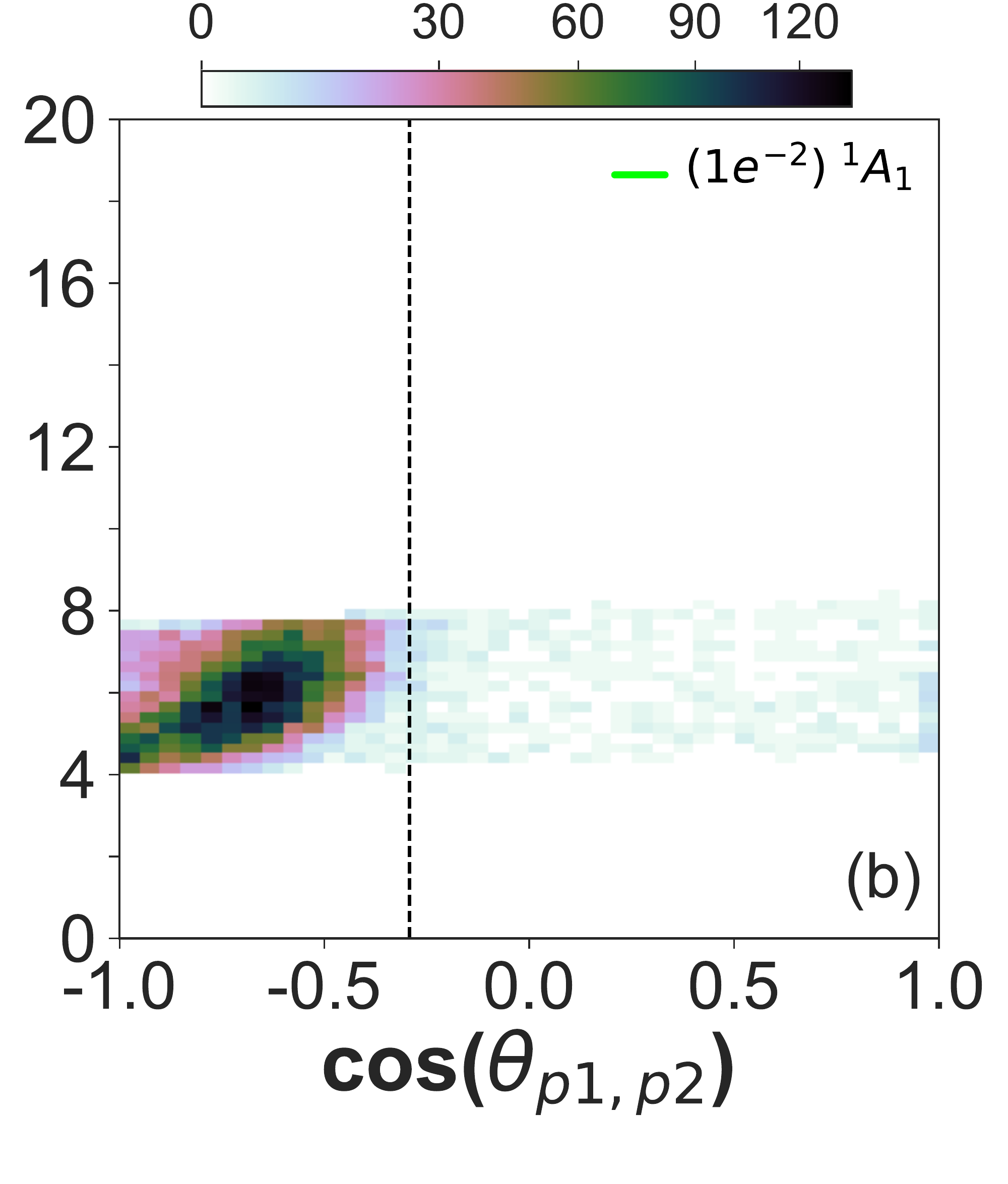}}
    {
        \includegraphics[width=4.2cm, trim=0.5cm 1.7cm 0.5cm 0.1cm, clip]{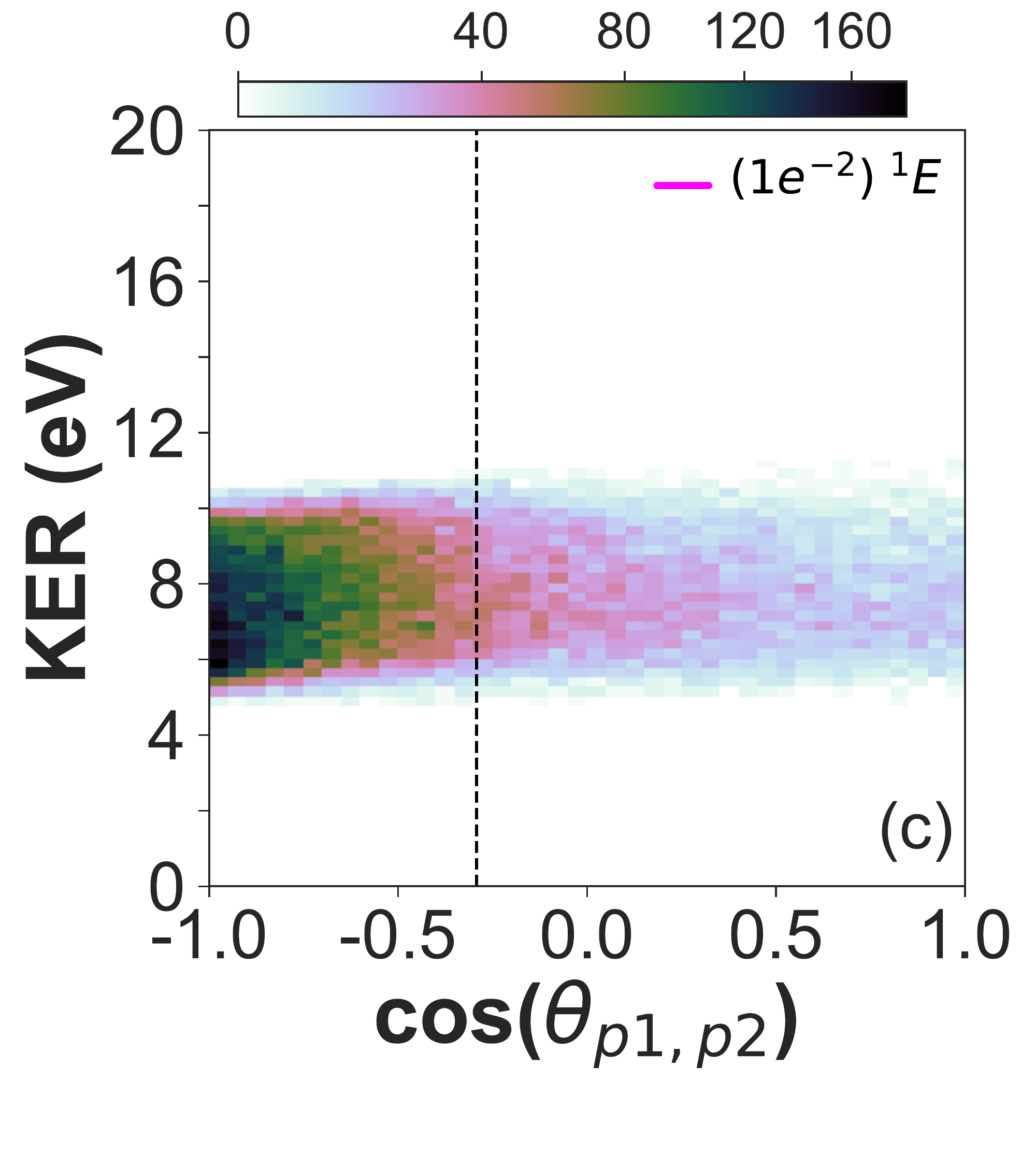}}
    {
        \includegraphics[width=3.85cm, trim=0.5cm 0.95cm 0.5cm 0.0cm, clip]{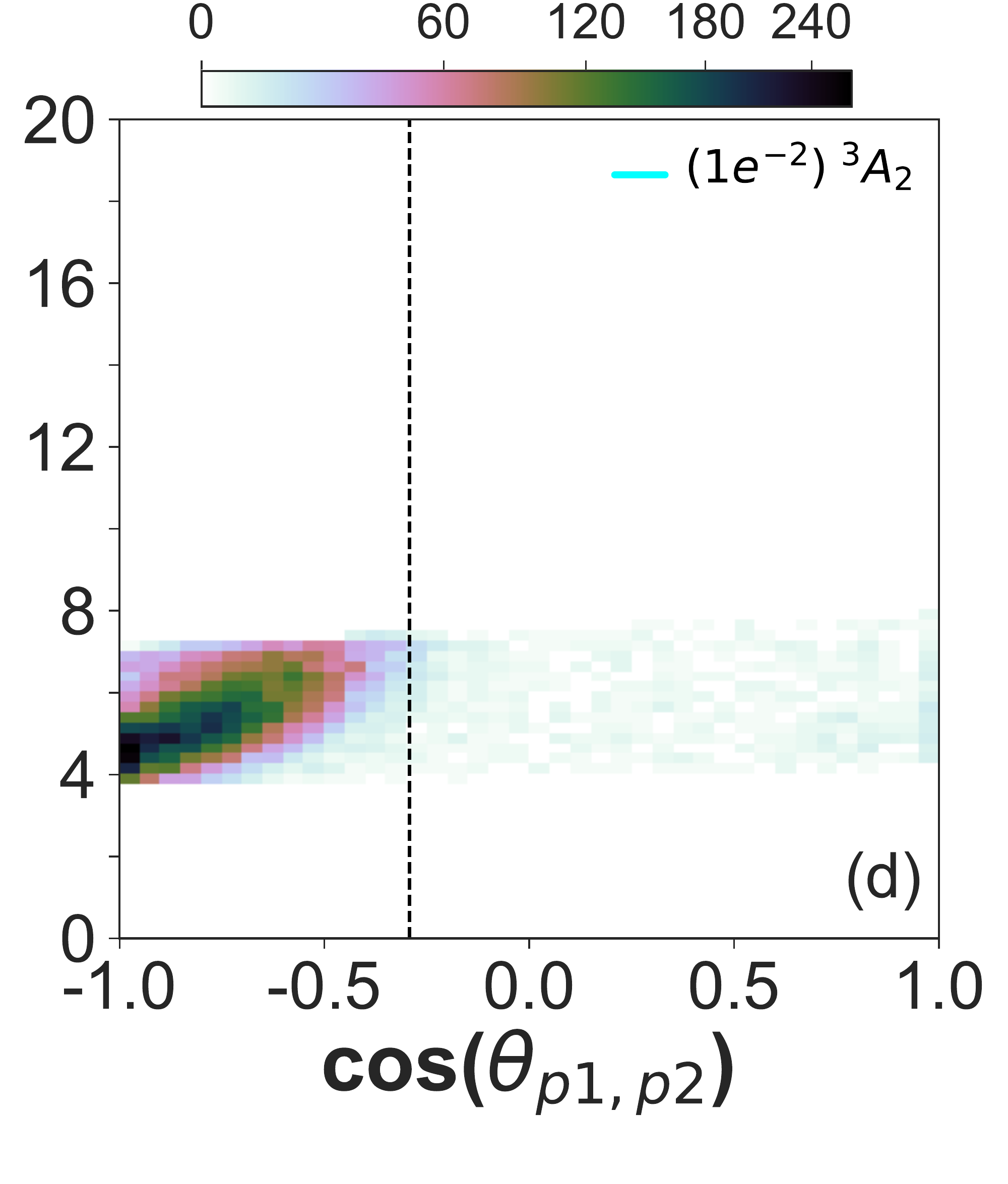}}
    }
\caption{The PDI yield of NH$_3$ as a function of cosine of the measured proton-proton angle, $\cos\theta_{p_1,p_2}$, and KER for each of the four dication states from the H$^{+}$ + H$^{+}$ fragmentation channel at 61.5~eV. The dashed black vertical line indicates the neutral ground state H-N-H angle.}
\label{fig:KER_angle}
\end{figure}

To examine the connection between the measured KER and the molecular geometry in each dication electronic state, we plot the yield as a function of cosine of the measured angle between the momenta of the two protons, $\cos\theta_{p_1,p_2} = \bm{p_1} \cdot \bm{p_2}/|\bm{p_1}||\bm{p_2}|$, and the KER, as shown in Fig.~\ref{fig:KER_angle}. It should be mentioned that due to the Coulomb repulsion between the two photoions, the measured proton-proton angle is an asymptotic dissociation angle, hence its value will be slightly larger than the true angle at which the fragmentation transpires. Although we do not have an exact estimate of how significantly the asymptotic dissociation angles differ from the true bond angles, our analysis carries useful information that differentiates the dissociation dynamics for each of the four features. In Fig.~\ref{fig:KER_angle}, the neutral ground state geometry of NH$_3$ (specifically the H-N-H bond angle) is indicated by the vertical black dashed line. First, we point out that of the four dication states three -  the $(2a_{1}^{-1}3a_{1}^{-1})$ $^{1}$A$_{1}$, $(1e^{-2})$ $^{1}$A$_{1}$, and $(1e^{-2})$ $^{3}$A$_{2}$ states - exhibit decreasing KER with increasing measured dissociation angle between the protons, as seen in Fig.~\ref{fig:KER_angle}. Qualitatively, if the angle between the two protons increases due to nuclear motion in the dication, e.g. the NH$_3$ umbrella opening, their separation increases and their Coulomb repulsion correspondingly decreases, resulting in the negative bivariate correlation between the KER and the proton-proton angle, $\theta_{p_1,p_2}$. Although this type of nuclear motion was not addressed in our calculations (which kept bond angles frozen), we still bring forward this qualitative picture as a possible explanation for the observed correlation. This also gives further support to the notion  that these three dication states dissociate via a concerted mechanism, where the two protons are simultaneously eliminated from the dication.

We also point out that the features associated with the $(2a_{1}^{-1}, 3a_{1}^{-1})$ $^{1}$A$_{1}$ state and the $(1e^{-2})$ $^{1}$A$_{1}$ state dissociate at angles closer to the neutral ground state geometry of the NH$_3$ molecule (H-N-H bond angle $\sim107^\circ$) than the feature associated with the $(1e^{-2})$ $^{3}$A$_{2}$ state, which tends to fragment at angles approaching 180$^{\circ}$. Although the distributions for the $(1e^{-2})$ $^{1}$A$_{1}$ and $(1e^{-2})$ $^{3}$A$_{2}$ states appear similar in shape, each state's fragmentation dynamics can be distinguished as different by the location of their respective peaks in the measured proton-proton angle. This suggests that the $(2a_{1}^{-1}, 3a_{1}^{-1})$ $^{1}$A$_{1}$ and the $(1e^{-2})$ $^{1}$A$_{1}$ states exhibit prompt fragmentation, while the molecular structure in the $(1e^{-2})$ $^{3}$A$_{2}$ state evolves further away from the neutral configuration, driven towards larger bond angles between the two protons, prior to dissociation. This is indeed consistent with the asymptotic charge exchange mechanism, described in Section~\ref{sec:level3}, that couples the $^3$A$_2$ ($^3$A$''$) and $^3$E ($^3$A$''$) states (PES cuts inset in Fig.~\ref{fig:PEC_NH3}(b)). The dissociation on the $(2a_{1}^{-1}, 3a_{1}^{-1})$ $^{1}$A$_{1}$ and $(1e^{-2})$ $^{1}$A$_{1}$ states result in the direct elimination of two protons, which are light and depart fast, providing little time for the molecular structure to evolve away from the neutral equilibrium geometry during the fragmentation. In contrast, the fragmentation on the $(1e^{-2})$ $^{3}$A$_{2}$ state initially involves a heavier NH$^+$ ion preceding the charge exchange mechanism that produces a light proton. Thus the initial dissociation on the $(1e^{-2})$ $^{3}$A$_{2}$ state (prior to the charge exchange) is slower due to the increased mass of one of the charged fragments. 

Although our calculations keep the bond angles frozen, it is known that for molecules of the form AH$_3$, ionization from the $1e$ orbital (as in the case of the $(1e^{-2})$ $^{3}$A$_{2}$ state) drives the molecule towards a planar configuration, i.e. larger H-N-H bond angles (this can be seen in a Walsh diagram, see Ref.~\cite{Higuchi}). The increased fragmentation time leads to an increased likelihood for processes such as the aforementioned charge exchange to take place, as well as more time for the molecular geometry to evolve away from the neutral equilibrium geometry towards larger H-N-H angles, preceding the dissociation. The timescale for a wave packet in the $(1e^{-2})$ $^{3}$A$_{2}$ state to reach the geometry where charge exchange can occur, as well as other details of the dissociation dynamics, precisely explaining the propensity towards fragmentation at H-N-H angles approaching 180$^{\circ}$ (beyond our qualitative description), would need to be addressed in a future study requiring time-dependent calculations that include non-adiabatic coupling.

In contrast to the three states in Fig.~\ref{fig:KER_angle} discussed above, the ($1e^{-2}$) $^1$E state in Fig.~\ref{fig:KER_angle}~(c) displays a band of KER over a wide distribution of $\theta_{p_1,p_2}$ extending all the from zero to 180$^\circ$ and smoothly peaked towards 180$^\circ$. This distribution is consistent with the sequential dissociation mechanism discussed below in detail, namely $\textrm{NH}_3^{++} \rightarrow \textrm{NH}^+ + \textrm{H}^+ + \textrm{H} \rightarrow \textrm{N} + 2\textrm{H}^+ + \textrm{H}$. If prior to the second step of this process the NH$^+$ fragment rotates freely before dissociating via a crossing with another electronic state, the H$^+$ is ejected in a random direction in the body frame of the NH$^+$ molecule. However that is not a random direction in the laboratory frame because the NH$^+$  fragment is translating with a center of mass momentum opposite to the sum of the momenta of the H and H$^+$ atoms produced in the first step, presumably ejected near the directions of the original NH bonds. The diatom's center of mass is therefore moving away from the H$^+$ ion produced in the first step, and consequently the random angular distribution of the proton ejected from the moving NH$^+$ shifted in the direction opposite the direction of the first H$^+$ ion. A similar effect has been seen in dissociation of the water dication following one-photon double ionization, in which a sequential dissociation channel involving dissociation of OH$^+$ is seen \cite{Streeter,Itzik}.

Other evidence also suggests that the different fragmentation dynamics of the $(1e^{-2})$ $^{1}$E state can be specifically attributed to a sequential dissociation mechanism involving four bodies in the final set of fragments. Here, we do not consider the possibility of a sequential dissociation process first resulting in NH$_2^{+}$ + H$^{+}$ fragmentation, with the NH$_2^{+}$ subsequently dissociating to NH + H$^{+}$ or N + H + H$^{+}$. Our interpretation does not include these channels, as we have analyzed the NH$_2^{+}$ + H$^{+}$ dissociation channel (which is the subject of a future paper) and did not observe any electron-ion momentum correlation consistent with shared dication electronic states producing both NH +  H$^{+}$ + H$^{+}$ or NH$_2^{+}$ + H$^{+}$ fragments. However, we cannot totally rule out these possibilities, as the lifetime of the intermediate NH$_2^{+}$ fragment may be too short for these fragments to survive the flight time to the ion detector. However, if intermediate NH$_2^{+}$ fragments dissociate during their flight to the detector, the secondary ion momenta should exhibit a broad spread in momentum. Since this is not observed, we argue in favor of a different sequential dissociation mechanism. 

\begin{figure}[h!]
\includegraphics[width=8.5cm]{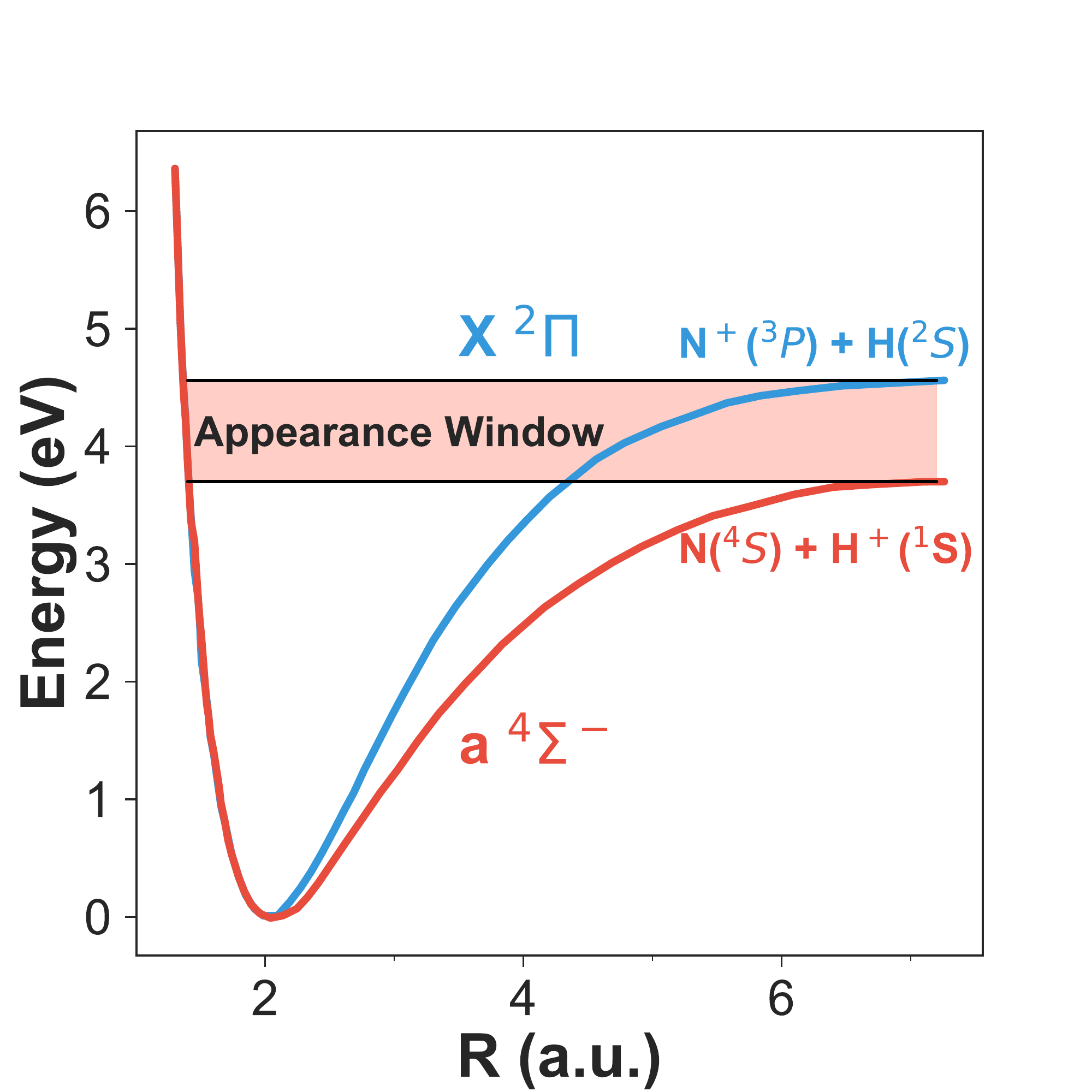}
\caption{The potential energy curves for the X$^2\Pi$ ground state and the a$^4\Sigma^-$ state of NH$^+$, extracted from Ref.~\cite{Amero2}. Population transfer may occur between these states via spin-orbit coupling, where initially bound excitations on the X$^2\Pi$ state can dissociate on the a$^4\Sigma^-$ state. Only diatomic NH$^+$ fragments with internal energy within (or above) the \textit{appearance window} will dissociate.}
\label{fig:PEC_NH}
\end{figure}

Previous measurements have found that PDI to the $(1e^{-2})$ $^{1}$E state produces the fragments NH$^{+}$ + H$^{+}$ + H, where the bound NH$^{+}$ ion is in its ground state, i.e. the X $^{2}\Pi$ state \cite{Stankiewicz}. Although the dissociative limit of the NH$^+$ $^{2}\Pi$ state results in N$^{+}$($^3P$) + H($^2S$) fragmentation, it has been shown that the X $^{2}\Pi$ state crosses the a $^{4}\Sigma^{-}$ state in the FC region and that population transfer between the X and a states can occur via spin-orbit coupling \cite{Liu,Amero2,Amano,Colin,Tarroni}. As seen in Fig.~\ref{fig:PEC_NH}, the NH$^+$  a $^{4}\Sigma^-$  state dissociates to H$^{+}$ + N($^4S$) with a dissociation energy that is roughly 1~eV smaller than the X $^{2}\Pi$ state dissociation energy. Thus, high-lying vibrational states of the NH$^{+}$ fragment that are initially bound in the X $^{2}\Pi$ state can undergo intersystem crossing to the a $^{4}\Sigma^{-}$ state, yielding the final fragments of the reaction NH${_3^{++}} \rightarrow$  N($^4S$) + H($^2S$) + H$^{+}$ + H$^{+}$. In the present context, population transfer can occur along the inner wall of the quasi-degenerate NH$^+$ states when the initial breakup of the ($1e^{-2}$) $^1$E state produces NH$^+$($^2\Pi$) ions with internal energy that lies within the \textit{appearance window} shown in Fig.~\ref{fig:PEC_NH}. We can estimate the location of the four-body limit by first extrapolating the MRCI energy for the $^3$E state (Fig.~\ref{fig:PEC_NH3} blue curve) to infinite separation of the N-H bonds.  This places the NH$^+$($^4\Sigma^-$) + H + H$^+$ limit at 0.63~eV. Adding to this the 3.66~eV dissociation energy of NH$^+$($^4\Sigma^-$) places the four-body limit at 4.29~eV, directly in the center of the \textit{appearance window}. This four-body breakup mechanism also explains why the theoretical KER value of 9.42 eV gleaned from Table~\ref{table:asymptotes} is higher than the measured value of 7.7 eV. From Fig.~\ref{fig:PEC_NH} we see that the NH$^+$ fragment must have a minimum internal energy of 3.7 eV to dissociate to N + H$^+$ at the lower end of the appearance window to produce a fast proton with (9.42--3.7~eV) = 5.72 eV and a zero energy proton. At the upper end of the  \textit{appearance window} we get a fast proton with (9.42--4.5~eV) = 4.92 eV and a slow proton with 1 eV. This interpretation appears to be consistent with the measured particle energy balance and prompts us to believe that each NH fragment in the three concerted dissociation channels was produced with a distribution of ro-vibrational energy around 2~eV, while the NH$^+$ fragment in the sequential dissociation channel was produced with a distribution of ro-vibrational energy that extends well beyond 3.7~eV, enabling the second fragmentation step. These results are also consistent with a previous theoretical treatment of the dissociation of H$_2$O$^{++}$ \cite{Gervais}, where the internal energy distribution of the OH$^+$ fragment in the H$^+$ + OH$^+$ two-body  dissociation channel was observed to span approximately 3--5~eV.
 
Although the initial set of photoions produced via excitation to the $(1e^{-2})$ $^{1}E$ state would not produce the four-particle (two-electron, two-proton) coincidence we measure, highly vibrationally excited ground state NH$^{+}$ fragments (lying within the \textit{appearance window}) can spin-orbit couple to a state where a fragmentation, producing a second proton, is possible, yielding the necessary two-proton coincidence. Since the spin-orbit coupling is weak, and the ensuing dissociation is not instantaneous, the intermediate NH$^{+}$ fragment can rotate prior to coupling to the dissociative state, which results in a proton-proton angular distribution that differs from the other three dication states that involve fewer fragmentation steps. The lifetime of the excited intermediates in the \textit{appearance window} in the X($^{2}\Pi$) state is determined by the strength of the spin-orbit coupling but not deduced in our experiment. It could potentially be measured using a different detection scheme or calculated using a different theoretical approach than the one taken in this study. 

\begin{figure}[h!]
\includegraphics[width=8.5cm]{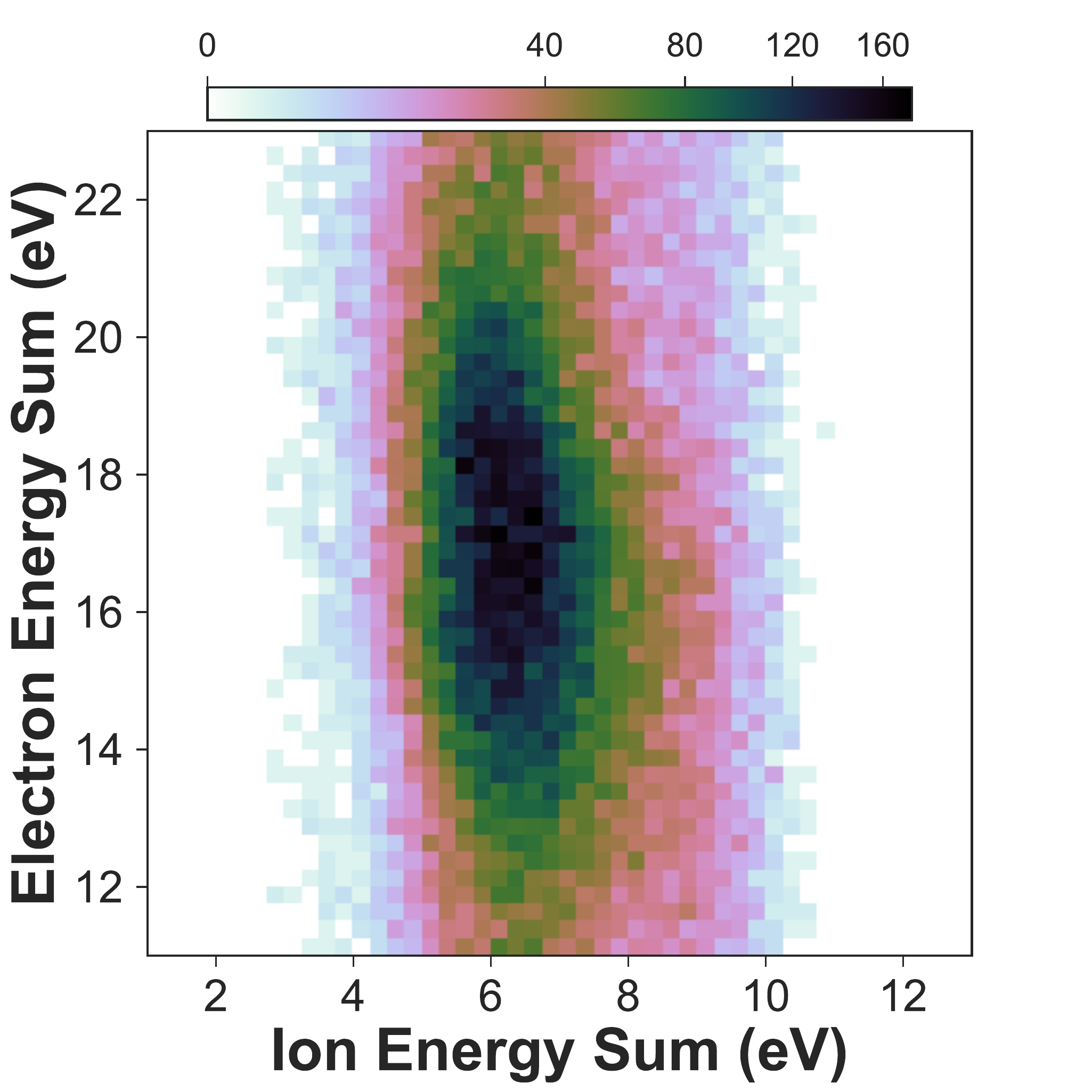}
\caption{The yield of N$^+$ + H$^+$ after valence PDI of NH$_3$ as a function of photoelectron pair energy sum and photoion pair energy sum for the $(1e^{-2})$ $^{1}$E dication state.}
\label{fig:N+H+}
\end{figure}

We discuss the cases of excitations below and above the \textit{appearance window} next. Excitations initially prepared in the X$^{2}\Pi$ state that lie above the \textit{appearance window} directly dissociate to produce N$^+$($^3P$) + H$^{+}$ + H($^2S$) + H($^2S$). Indeed, this is supported by our measurements by analyzing the N$^+$ + H$^{+}$ dissociation channel, which is briefly addressed here. The same procedure used to select the H$^{+}$ + H$^{+}$ dissociation channel and described at the beginning of this sub-section is used to select the N$^+$ + H$^{+}$ channel. We plot the PDI yield of the N$^+$ + H$^{+}$ fragmentation as a function of the photoelectron energy sum and photoion energy sum, shown in Fig.~\ref{fig:N+H+}. In this fragmentation channel we observe a single feature (seen in Fig.~\ref{fig:N+H+}), which we attribute to a single contributing dication electronic state. We argue that this feature corresponds with the magenta color-coded $(1e^{-2})$ $^{1}$E state. This feature possesses an electron energy sum of 16.7~eV, which exactly coincides with the electron energy sum measured for the feature in the H$^+$ + H$^{+}$ dissociation channel corresponding with the $(1e^{-2})$ $^{1}$E state. From this evidence we suggest that the single feature observed in the N$^+$ + H$^{+}$ channel corresponds with the same dication electronic state that contributes to the sequential H$^+$ + H$^{+}$ dissociation mechanism. Comparing the H$^+$ + H$^{+}$ and N$^+$ + H$^{+}$ yields following PDI to the $(1e^{-2})$ $^{1}$E state indicates that roughly the same amount of population ends up above the \textit{appearance window} as compared to within it. As for excitations initially prepared in the X $^{2}\Pi$ state that lie below the \textit{appearance window}, these will remain as bound NH$^+$ fragments. This is also supported by our measurements by analyzing the NH$^+$ + H$^{+}$ dissociation channel (which is the topic of a future paper and thus not presented here). In this dissociation channel we also identify a feature corresponding with the $(1e^{-2})$ $^{1}$E state. These results are entirely consistent with the explanation presented in the paragraph above, where the PDI to the $(1e^{-2})$ $^{1}$E state produces the fragments NH$^{+}$ $^{2}\Pi$ + H$^{+}$ + H for which the excitation in the NH$^{+}$ ion can lie below, within, or above the \textit{appearance window}. All three of these cases are observed in our measurement and illustrate the various levels of complexity in the dissocation dynamics of simple polyatomic molecules that can occur following valence PDI to just a single state.

\begin{figure}[h!]
\includegraphics[width=8.5cm]{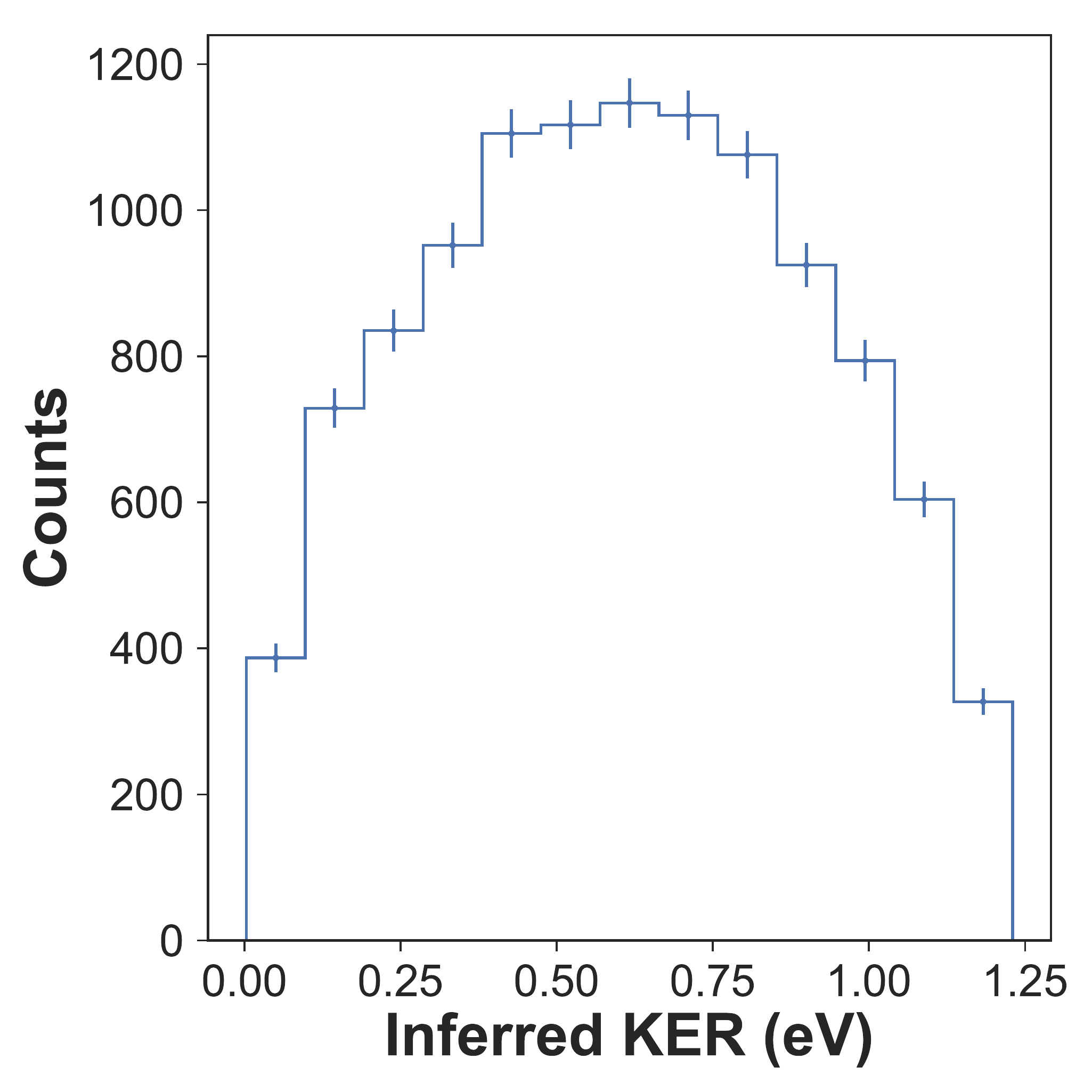}
\caption{The inferred KER from the dissociation of the NH$^{+}$ fragment, involving the measured slow proton and the nitrogen following PDI of NH$_3$ at 61.5~eV to the $(1e^{-2})$ $^{1}$E state, resulting in the four-body fragmentation N + H + H$^+$ + H$^+$. The KER peaks at 0.61~eV, with a FWHM of 0.71~eV.}
\label{fig:inferKER}
\end{figure}

To further support the claim that the H$^+$ + H$^+$ fragmentation on the $(1e^{-2})$ $^{1}$E state occurs via the four-body mechanism discussed above, we analyze the slow proton emerging from the dissociation on the $(1e^{-2})$ $^{1}$E state, using its momentum to infer the KER of the dissociation of the NH$^+$ fragment, shown in Fig.~\ref{fig:inferKER}. Since two neutral particles are left undetected (N and H), and simple conservation of momentum can thus not be applied, this is realized by assuming that the momentum of the undetected neutral N atom is approximately that of the N-H center of mass, inferred from the two proton momenta. We find the inferred KER to peak at 0.61~eV (FWHM 0.71~eV), which lies below the $\sim$1~eV maximum KER permitted by the locations of the two adiabatic limits of the X$^{2}\Pi$ and a$^{4}\Sigma^{-}$ states, i.e. the \textit{appearance window} (see Fig.~\ref{fig:PEC_NH}). This supports the assumption that the slow proton emerges from a dissociation on the a$^{4}\Sigma^{-}$ state. Since our measurement also indicates that the $(1e^{-2})$ $^{1}$E state contributes to the NH$^+$ + H$^+$ + H fragmentation channel (the topic of another manuscript, currently in preparation), which is in agreement with previous measurements \cite{Stankiewicz}, we believe that some small fraction of the NH$^{+}$ fragments of this three-body fragmentation channel can decay through intersystem crossing and feed into the four-body N + H + H$^+$ + H$^+$ fragmentation channel. This conclusion is also corroborated by our analysis of the N$^+$ + H$^{+}$ dissociation channel, which shows that the $(1e^{-2})$ $^{1}$E state also feeds into this four-body fragmentation channel and corresponds with the initial excitations in the NH$^+$ $^{2}\Pi$ ion that lie above the \textit{appearance window}.

\subsection{\label{PEDyn}Photoelectron dynamics}

\begin{figure}[h!]
    {
    \includegraphics[width=8.5cm]{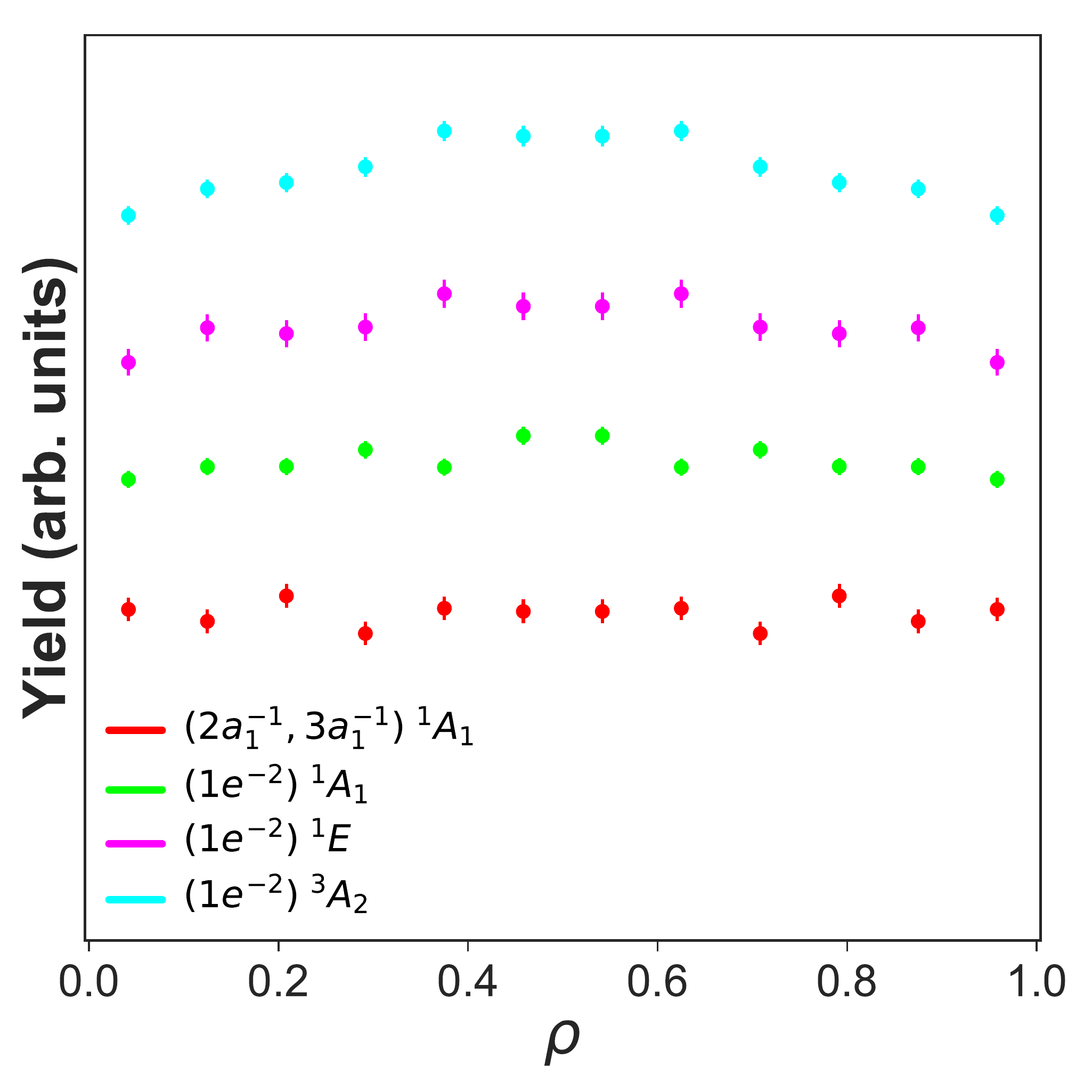}}
\caption{The yield of the H$^+$ + H$^+$ fragmentation after PDI of NH$_3$ at 61.5~eV as a function of photoelectron energy sharing for each of the four relevant dication states. Here the y-axis indicates the PDI yield in arbitrary units on a linear scale. The distributions are not internormalized. They have been staggered in order based on their respective electron energy sum for better visibility (i.e. the states are placed in ascending order with respect to their respective photoelectron energy sum).}
\label{fig:SDCS}
\end{figure}

Next, we display in Fig.~\ref{fig:SDCS} the photoelectron energy sharing distributions for the four dication states. We define the electron energy sharing as:

\begin{equation}
\rho = \frac{E_{e_{1}}}{E_{e_{1}} + E_{e_{2}}},
\end{equation}

where $E_{e_{1}}$ and $E_{e_{2}}$ are the energies of electron 1 and 2, respectively. Values of $\rho$ near 0.5 indicate equal energy sharing between the two photoelectrons, while values near 0 or 1 indicate unequal energy sharing between the two photoelectrons. In all four dication states we do not observe a strong enhancement in yield for any particular values of $\rho$. The distributions are nearly flat. In the absence of autoionization, this is similar to the PDI of atoms and molecules in this excess energy range (see e.g. \cite{Andersson,Reedy}). The exception is the ($1e^{-2}$) $^3$A$_2$ state (cyan) and perhaps the $(1e^{-2})$ $^{1}$E state (magenta), which show some propensity towards increased yield at values of $\rho$ near 0.5. This is surprising since the ($1e^{-2}$) $^3$A$_2$ and the $(1e^{-2})$ $^{1}$E state dication states correspond to the highest electron sum energies (see Fig.~\ref{fig:Eesum}). A maximum PDI yield at equal energy sharing, if any, would be expected for the lowest electron sum energies according the the Wannier threshold law \cite{Wannier}, which favors the emission of two electrons with the same energy and back-to-back close to the PDI threshold. However, the electron pair emission patterns are subject to selection rules that are specific to each dication state and the molecular orientation with respect to the polarization vector; they inherently influence the electron energy sharing to a certain degree. The detailed investigation of this complex problem requires M/RFPADs and is beyond the scope of this work. These distributions have all been normalized to the same value and have been placed in ascending order, based on the corresponding photoelectron energy of the state (the state with the lowest photoelectron energy sum is placed near the bottom and the state with the highest photoelectron energy sum is placed at the top).

\begin{figure}[h!]
    {
    {
        \includegraphics[width=4.2cm, trim=0.4cm 0cm 0.55cm 0cm, clip]{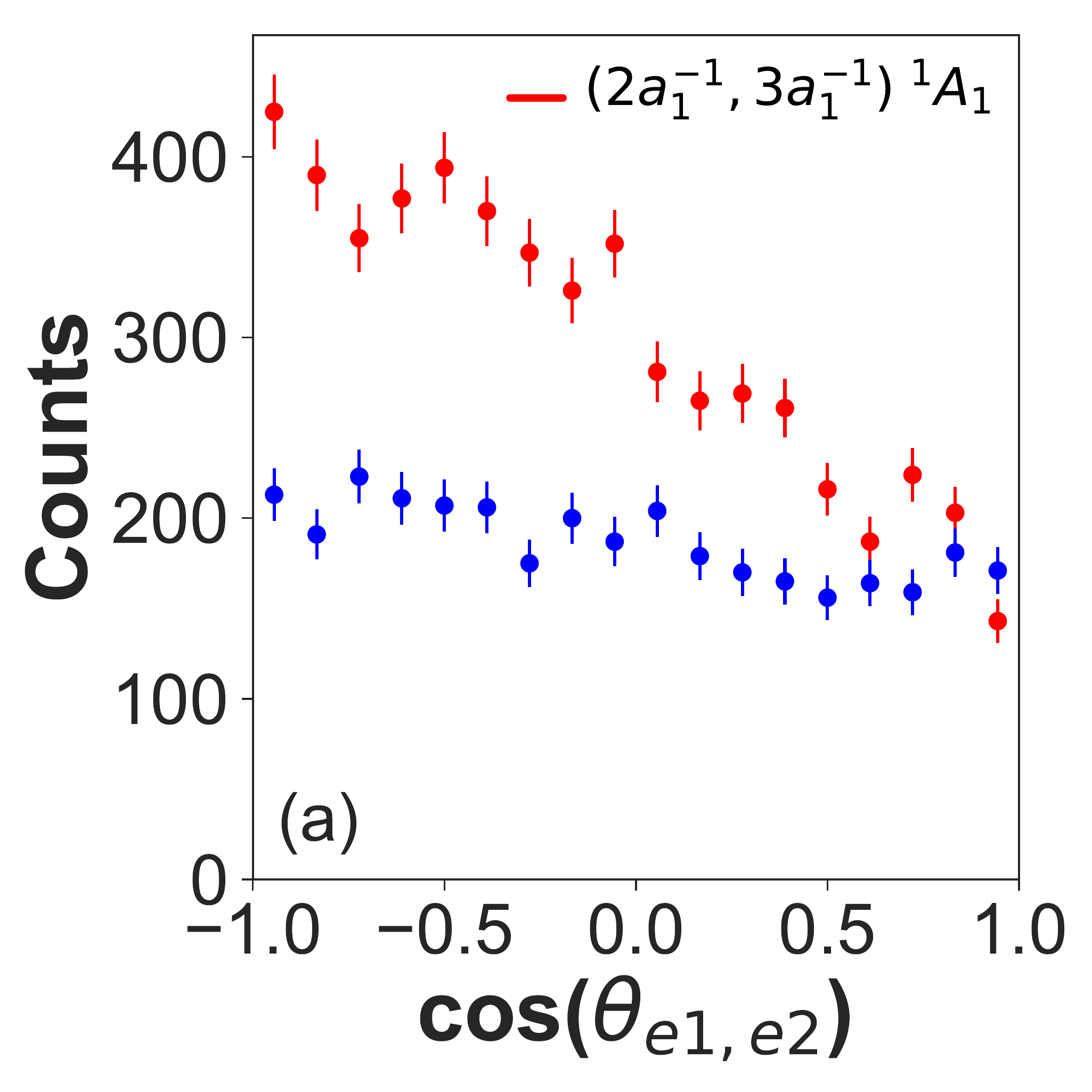}}
    {
        \includegraphics[width=4.2cm, trim=0.5cm 0cm 0.5cm 0cm, clip]{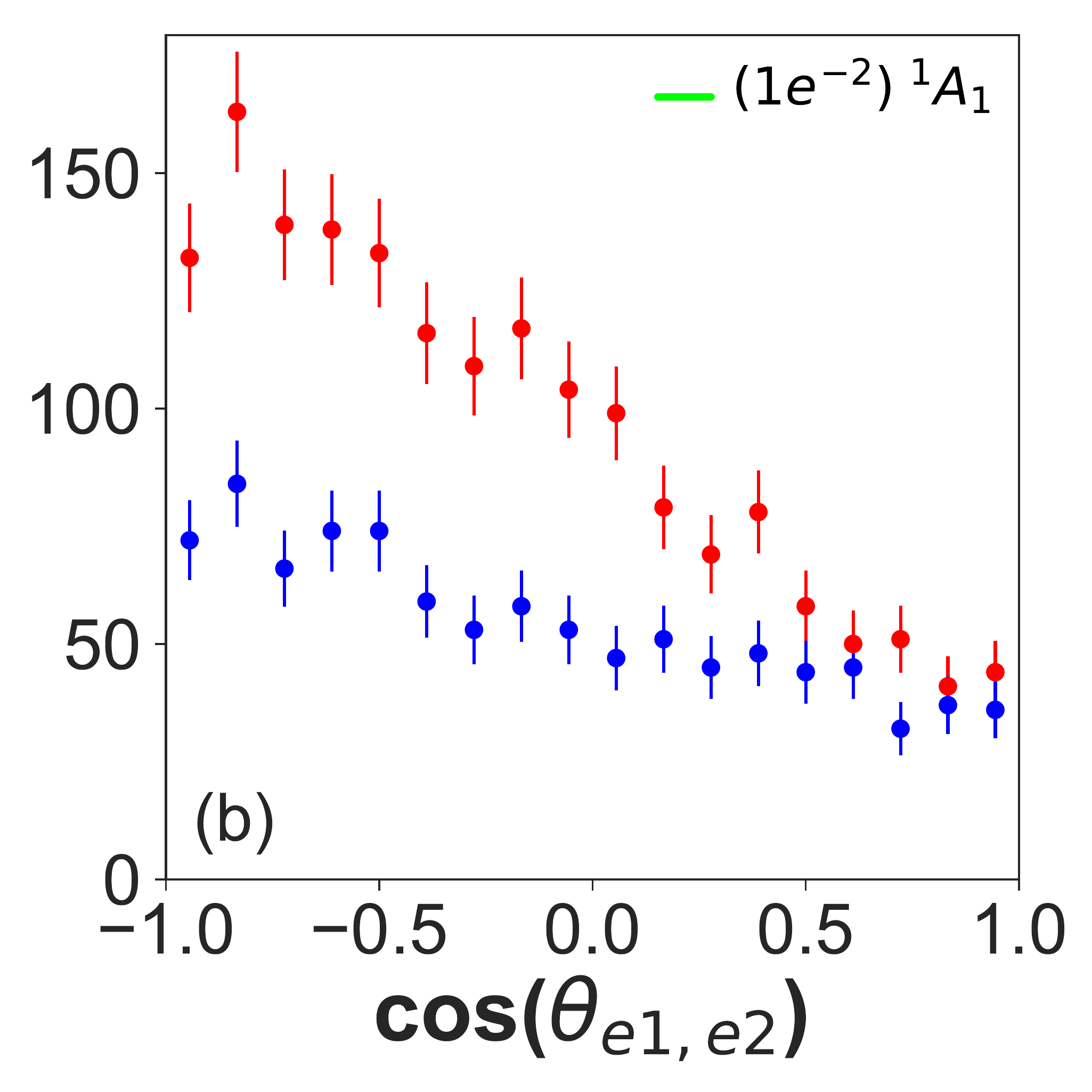}}
    {
        \includegraphics[width=4.2cm, trim=0.4cm 0cm 0.55cm 0cm, clip]{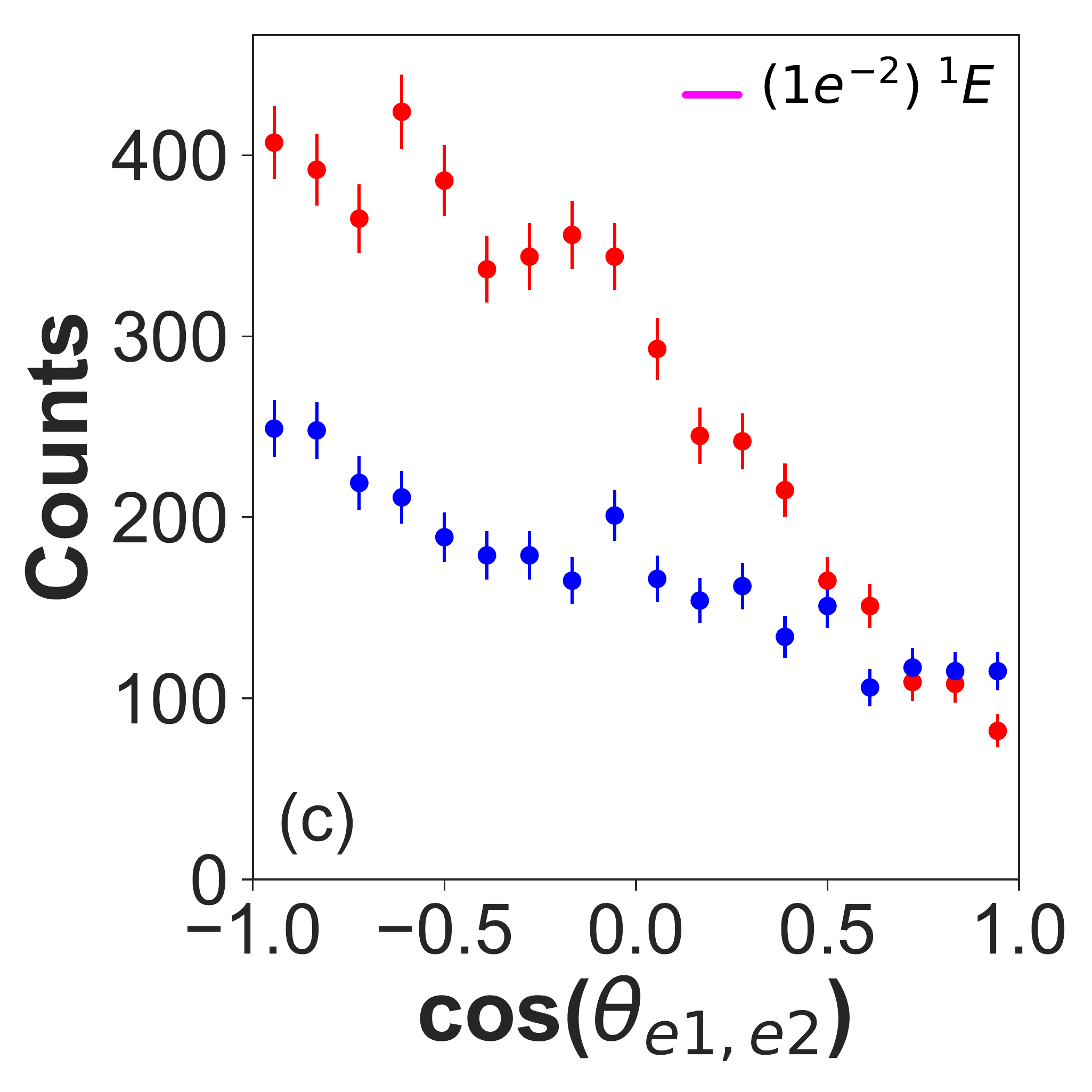}}
    {
        \includegraphics[width=4.2cm, trim=0.5cm 0cm 0.5cm 0cm, clip]{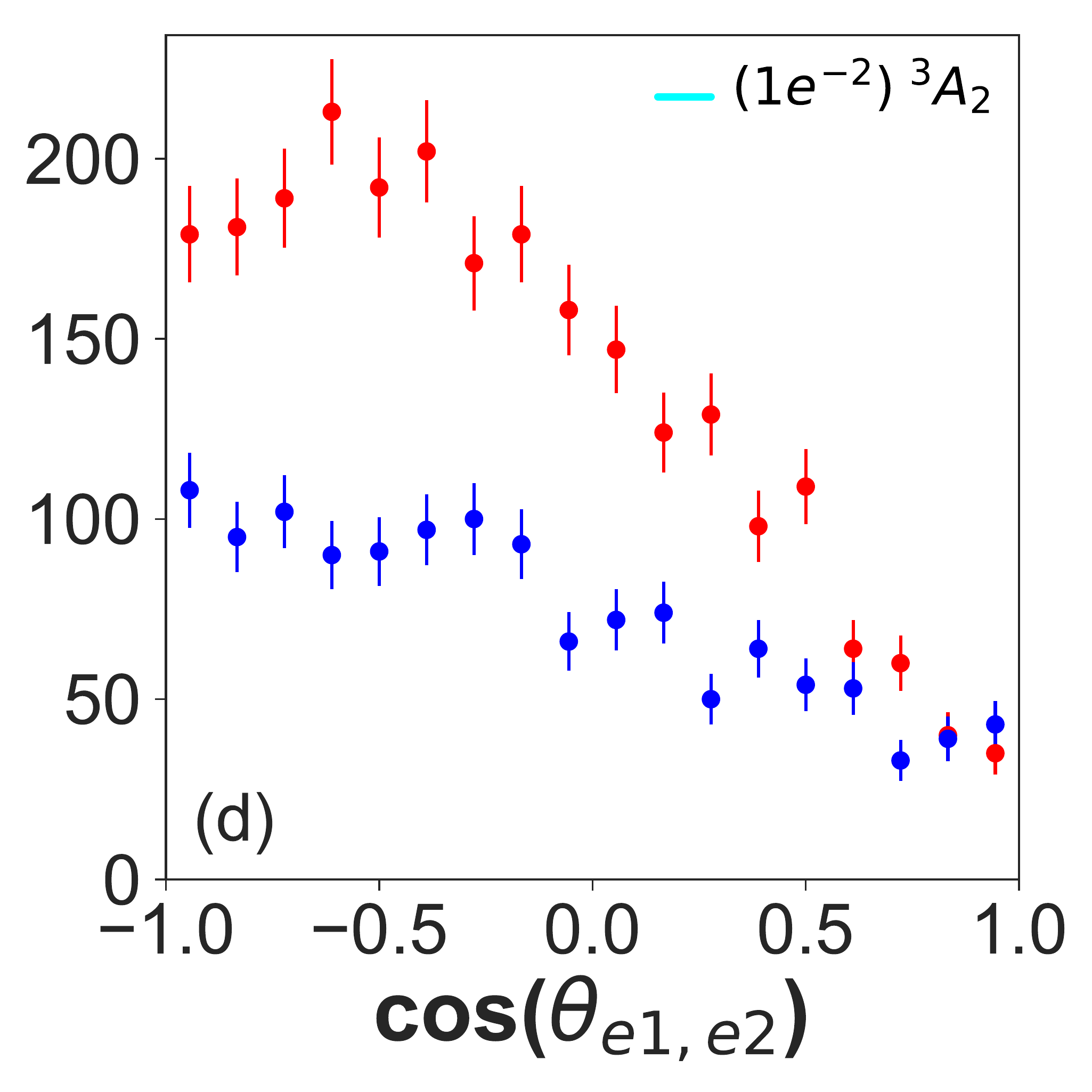}}
    }
\caption{The cosine of the relative emission angle between the two photoelectrons for two different energy sharing conditions for each of the four dication states of NH$_3$ following PDI at 61.5~eV. Electron energy sharing between 0.425 and 0.575 is shown in red, and energy sharing less than 0.05 or greater than 0.95 is shown in blue.}
\label{fig:angle_e}
\end{figure}

Lastly, we plot in Fig.~\ref{fig:angle_e} the yield of the H$^+$ + H$^+$ fragmentation as a function of cosine of the relative emission angle between the two photoelectrons and in different energy sharing conditions of the electron pair for the four dication states, integrated over all molecular orientations relative to the polarization vector of the incoming light and with no restrictions on the emission direction of either electron. The relative electron-electron angles are plotted for $0.425 < \rho < 0.575$ (shown in red) and for $\rho < 0.05$ or $\rho > 0.95$ (shown in blue). We point out that our measurement suffers from some multi-hit detector dead-time effects, which influence the measured yields of the photoelectrons emitted in the same direction with similar kinetic energies. For equal energy sharing between the two emitted electrons and for $\theta_{e_1,e_2} \leq 90^\circ$ (emission into the same hemisphere), we can expect to fail to detect up to $\sim 52\%$ events for the ($1e^{-2}$) $^3A_2$ state, $\sim 27\%$ for the ($1e^{-2}$) $^1E$ state, $\sim 23\%$ for the ($1e^{-2}$) $^1A_1$ state, and $\sim 22\%$ for the ($2a_1^{-1}, 3a_1^{-1}$) $^1A_1$ state. Note that we estimate these losses for the $''$worst case$''$ isotropic relative electron-electron emission, which very well represents autoionization processes that are sequential in nature and are subject to unequal energy sharing between the electrons. The equal energy sharing case on the other hand is dominated by knock-out processes with very few electron pairs emitted into the same hemisphere. The actual losses are expected to be closer to the losses for the case of unequal electron energy sharing reported below.

The relative angles between the two photoelectrons under unequal energy sharing conditions (blue circles in Fig.~\ref{fig:angle_e}) are rather isotropic for all four dication states, where there is a slight propensity towards back-to-back emission (or in other words a lack of events with electrons emitted into the same direction), which we partly attribute to the dead-time problem at relative electron-electrons angles below 90$^\circ$ (emission into the same hemisphere). The simulated losses of events with unequal energy sharing amount to $\sim 26.1\%$ for the ($1e^{-2}$) $^3A_2$ state, $\sim 8.4\%$ for the ($1e^{-2}$) $^1E$ state, $\sim 5.2\%$ for the ($1e^{-2}$) $^1A_1$ state, and $\sim 4.2\%$ for the ($2a_1^{-1}, 3a_1^{-1}$) $^1A_1$ state. Evidently the small anisotropies in the relative angular distributions for the unequal electron energy sharing case (blue), presented in Fig.~\ref{fig:angle_e} for all four dication states, are accounted for by the detector dead-time limitations, and are otherwise consistent with isotropic relative angular distributions. As there is no hint for autoionzation visible in the electron-electron energy correlation map depicted in Fig.~\ref{fig:Ee_Ee}, the unequal electron energy sharing case is likely dominated by knock-out processes, as reasoned below.

In contrast, the photoelectron dynamics for equal energy sharing conditions (red in Fig.~\ref{fig:angle_e}) reveals anisotropic angular distributions that are different for all four dication states and exceed the anisotropy expected from dead-time effects alone. For this case the relative angle between the two photoelectrons producing the ($2a_1^{-1}, 3a_1^{-1}$) $^1$A$_1$ state exhibits a preference towards back-to-back emission. The emission angle between the two photoelectons from the ($1e^{-2}$) $^3$A$_2$ state increases starting at $0^{\circ}$ and peaks at an angle of roughly $125^{\circ}$ before decreasing as the angle approaches $180^{\circ}$. The photoelectrons that produce the ($1e^{-2}$) $^1$E state have relative emission angles that increase starting at $0^{\circ}$, which then begin to level out at $100^{\circ}$, increasing at a slower rate as the angle approaches $180^{\circ}$. Last, the relative electron-electron emission angle of the ($1e^{-2}$) $^1$A$_1$ state increases starting at $0^{\circ}$ and peaks at an angle of roughly $150^{\circ}$ before decreasing as the angle approaches $180^{\circ}$. All four dication states show a non-vanishing PDI yield for small electron-electron angles close to $0^{\circ}$. This contribution is mainly due to the finite angular bin size accepting differences in the relative emission angles of up to $27^{\circ}$ at these values, as well as residual background from random coincidences underneath the features visible in e.g. Fig.~\ref{fig:Eesum_Epdiff}.

These trends in the relative electron-electron angular distributions as a function of the electron energy sharing possess similarities to prior observations made in the PDI of atomic and molecular targets \cite{Brauning,Knapp,Weber,Randazzo}. In the valence PDI experiments for helium~\cite{Brauning}, which is dominated by knock-out processes, rich photoelectron angular distributions emerge due to selection rules and symmetry considerations. It has been seen that for equal energy sharing conditions and the first detected electron fixed along the polarization vector of the ionizing field, the relative emission angle between the photoelectrons can be quite anisotropic and peaked at angles between $90^{\circ}$ and $180^{\circ}$ due to selection rules for dipole allowed transitions, whereas in unequal energy sharing conditions, the relative angle between the electrons can become more isotropic with a smaller peak at $180^{\circ}$. In the atomic case for equal electron energy sharing, there can be a node at a relative electron-electron angle of $180^{\circ}$, regardless of the emission direction of either of the two electrons. This is for instance true for the PDI of He and is due to a selection rule based on parity conservation in one photon transitions. Such a scenario is in general not well pronounced in the PDI of (polyatomic) molecules, and rather resembles the distributions for all cases presented in Fig.~\ref{fig:angle_e}. In addition to the finite angular bin size, again accepting differences in the relative emission angles of up to $27^{\circ}$ at $180^{\circ}$, we attribute this to the fact that we have not enforced any conditions on the molecular orientation or direction of the polarization vector of the XUV field. Integrating over all molecular orientations and the direction of the polarization vector is prone to wash out sharp features in the electron relative angular distribution, since angular momentum can be transferred to the nuclear systems and softens the aforementioned selection rules (as seen and discussed in Refs.~\cite{Weber,Vanroose}), in addition to other features. The limited number of events in the present data set does not allow conditions to be enforced on the molecular orientation or emission direction of one of the photoelectrons with high statistical significance. Future COLTRIMS studies could be directed towards the states that obey the axial recoil approximation to gather appreciable statistics, in order to produce photoelectron angular distributions in the molecular frame, which inter alia would help to study and understand the role of selection rules in the PDI of a symmetric top molecule with respect to the polarization vector of the incoming light.

\section{\label{sec:level5}Conclusion}

We have performed state-selective measurements on the H$^{+}$ + H$^{+}$ dissociation channel of NH$_{3}$ following direct valence PDI at 61.5~eV, where the two photoelectrons and two protons were measured in coincidence on an event-by-event basis using COLTRIMS. With the assistance of theoretical MRCI calculations of dication PES cuts, we identified the four participating dication electronic states that lead to H$^{+}$ + H$^{+}$ fragmentation, which correspond with the four features we observed and have estimated their branching ratios.

The PDI yield as a function of KER and the measured proton-proton angle indicates that three of the four dication states dissociate in a concerted mechanism, while the fourth state, the $(1e^{-2})$ $^{1}$E state, dissociates via a sequential process, with the intermediate ro-vibrationally excited NH$^{+}$ fragment ion decaying through an intersystem crossing that leads to a four-body breakup. Two of the dication states, the $(2a_{1}^{-1}, 3a_{1}^{-1})$ $^{1}$A$_{1}$ and $(1e^{-2})$ $^{1}$A$_{1}$ states, exhibit concerted dissociation mechanisms that fragment near the ground state geometry (axial recoil approximation applies). The third state, the $(1e^{-2})$ $^{3}$A$_{2}$ state, undergoes appreciable evolution in its molecular geometry and an asymptotic electron transfer from H to NH$^+$ at distances greater than 18~Bohr in the dissociating dication, preceding the three-body breakup. Differences between the MRCI calculations and the measured KER suggest that the neutral NH fragment in each of the three-body dissociation channels is highly ro-vibrationally excited. 

The relative emission angle between the two photoelectrons as a function of their energy sharing has some resemblance to prior measurements made on atomic and molecular targets, in spite of integrating over all molecular orientations and emission angles of the first photoelectron, relative to the XUV polarization. While the present study has focused on PDI processes that result in proton-proton breakup channels, we are presently analyzing the two- and three-body PDI breakup channels that produce NH$_2^+$ + H$^+$ and NH$^+$ + H + H$^+$, which is the topic of a future manuscript.

\section{\label{sec:level6}Acknowledgments}

Work at LBNL was performed under the auspices of the US Department of Energy (DOE) under Contract DE-AC02-05CH11231, using the Advanced Light Source and National Energy Research Computing Center, and was supported by the U.S. DOE Office of Basic Energy Sciences, Division of Chemical Sciences. JRML personnel were supported by the same US DOE funding source under Award No. DE-FG02-86ER13491. A.G. was supported by the ALS through a Doctoral Fellowship in Residence. Personnel from the University of Nevada, Reno was supported by the National Science Foundation Grant No. NSF-PHY-1807017. The Frankfurt group acknowledges the support of the Deutsche Akademische Austausch Dienst (DAAD) and the Deutsche Forschungsgemeinschaft (DFG). We thank the staff at the Advanced Light Source for operating the beamline and providing the photon beam. Moreover, we thank the RoentDek GmBH for longtime support with detector hardware and software.

\bibliography{Refs}

\begin{thebibliography}{52}%
\makeatletter
\providecommand \@ifxundefined [1]{%
 \@ifx{#1\undefined}
}%
\providecommand \@ifnum [1]{%
 \ifnum #1\expandafter \@firstoftwo
 \else \expandafter \@secondoftwo
 \fi
}%
\providecommand \@ifx [1]{%
 \ifx #1\expandafter \@firstoftwo
 \else \expandafter \@secondoftwo
 \fi
}%
\providecommand \natexlab [1]{#1}%
\providecommand \enquote  [1]{``#1''}%
\providecommand \bibnamefont  [1]{#1}%
\providecommand \bibfnamefont [1]{#1}%
\providecommand \citenamefont [1]{#1}%
\providecommand \href@noop [0]{\@secondoftwo}%
\providecommand \href [0]{\begingroup \@sanitize@url \@href}%
\providecommand \@href[1]{\@@startlink{#1}\@@href}%
\providecommand \@@href[1]{\endgroup#1\@@endlink}%
\providecommand \@sanitize@url [0]{\catcode `\\12\catcode `\$12\catcode
  `\&12\catcode `\#12\catcode `\^12\catcode `\_12\catcode `\%12\relax}%
\providecommand \@@startlink[1]{}%
\providecommand \@@endlink[0]{}%
\providecommand \url  [0]{\begingroup\@sanitize@url \@url }%
\providecommand \@url [1]{\endgroup\@href {#1}{\urlprefix }}%
\providecommand \urlprefix  [0]{URL }%
\providecommand \Eprint [0]{\href }%
\providecommand \doibase [0]{http://dx.doi.org/}%
\providecommand \selectlanguage [0]{\@gobble}%
\providecommand \bibinfo  [0]{\@secondoftwo}%
\providecommand \bibfield  [0]{\@secondoftwo}%
\providecommand \translation [1]{[#1]}%
\providecommand \BibitemOpen [0]{}%
\providecommand \bibitemStop [0]{}%
\providecommand \bibitemNoStop [0]{.\EOS\space}%
\providecommand \EOS [0]{\spacefactor3000\relax}%
\providecommand \BibitemShut  [1]{\csname bibitem#1\endcsname}%
\let\auto@bib@innerbib\@empty
\bibitem [{\citenamefont {Lablanquie}\ \emph {et~al.}(1989)\citenamefont
  {Lablanquie}, \citenamefont {Delwiche}, \citenamefont {Hubin-Franskin},
  \citenamefont {Nenner}, \citenamefont {Morin}, \citenamefont {Ito},
  \citenamefont {Eland}, \citenamefont {Robbe}, \citenamefont {Gandara},
  \citenamefont {Fournier},\ and\ \citenamefont {Fournier}}]{Lablanquie}%
  \BibitemOpen
  \bibfield  {author} {\bibinfo {author} {\bibfnamefont {P.}~\bibnamefont
  {Lablanquie}}, \bibinfo {author} {\bibfnamefont {J.}~\bibnamefont
  {Delwiche}}, \bibinfo {author} {\bibfnamefont {M.-J.}\ \bibnamefont
  {Hubin-Franskin}}, \bibinfo {author} {\bibfnamefont {I.}~\bibnamefont
  {Nenner}}, \bibinfo {author} {\bibfnamefont {P.}~\bibnamefont {Morin}},
  \bibinfo {author} {\bibfnamefont {K.}~\bibnamefont {Ito}}, \bibinfo {author}
  {\bibfnamefont {J.~H.~D.}\ \bibnamefont {Eland}}, \bibinfo {author}
  {\bibfnamefont {J.-M.}\ \bibnamefont {Robbe}}, \bibinfo {author}
  {\bibfnamefont {G.}~\bibnamefont {Gandara}}, \bibinfo {author} {\bibfnamefont
  {J.}~\bibnamefont {Fournier}}, \ and\ \bibinfo {author} {\bibfnamefont
  {P.~G.}\ \bibnamefont {Fournier}},\ }\href {\doibase
  10.1103/PhysRevA.40.5673} {\bibfield  {journal} {\bibinfo  {journal} {Phys.
  Rev. A}\ }\textbf {\bibinfo {volume} {40}},\ \bibinfo {pages} {5673}
  (\bibinfo {year} {1989})}\BibitemShut {NoStop}%
\bibitem [{\citenamefont {Sann}\ \emph {et~al.}(2011)\citenamefont {Sann},
  \citenamefont {Jahnke}, \citenamefont {Havermeier}, \citenamefont {Kreidi},
  \citenamefont {Stuck}, \citenamefont {Meckel}, \citenamefont {Sch{\"o}ffler},
  \citenamefont {Neumann}, \citenamefont {Wallauer}, \citenamefont {Voss},
  \citenamefont {Czasch}, \citenamefont {Jagutzki}, \citenamefont {Weber},
  \citenamefont {Schmidt-B{\"o}cking}, \citenamefont {Miyabe}, \citenamefont
  {Haxton}, \citenamefont {Orel}, \citenamefont {Rescigno},\ and\ \citenamefont
  {D{\"o}rner}}]{Sann}%
  \BibitemOpen
  \bibfield  {author} {\bibinfo {author} {\bibfnamefont {H.}~\bibnamefont
  {Sann}}, \bibinfo {author} {\bibfnamefont {T.}~\bibnamefont {Jahnke}},
  \bibinfo {author} {\bibfnamefont {T.}~\bibnamefont {Havermeier}}, \bibinfo
  {author} {\bibfnamefont {K.}~\bibnamefont {Kreidi}}, \bibinfo {author}
  {\bibfnamefont {C.}~\bibnamefont {Stuck}}, \bibinfo {author} {\bibfnamefont
  {M.}~\bibnamefont {Meckel}}, \bibinfo {author} {\bibfnamefont {M.~S.}\
  \bibnamefont {Sch{\"o}ffler}}, \bibinfo {author} {\bibfnamefont
  {N.}~\bibnamefont {Neumann}}, \bibinfo {author} {\bibfnamefont
  {R.}~\bibnamefont {Wallauer}}, \bibinfo {author} {\bibfnamefont
  {S.}~\bibnamefont {Voss}}, \bibinfo {author} {\bibfnamefont {A.}~\bibnamefont
  {Czasch}}, \bibinfo {author} {\bibfnamefont {O.}~\bibnamefont {Jagutzki}},
  \bibinfo {author} {\bibfnamefont {T.}~\bibnamefont {Weber}}, \bibinfo
  {author} {\bibfnamefont {H.}~\bibnamefont {Schmidt-B{\"o}cking}}, \bibinfo
  {author} {\bibfnamefont {S.}~\bibnamefont {Miyabe}}, \bibinfo {author}
  {\bibfnamefont {D.~J.}\ \bibnamefont {Haxton}}, \bibinfo {author}
  {\bibfnamefont {A.~E.}\ \bibnamefont {Orel}}, \bibinfo {author}
  {\bibfnamefont {T.~N.}\ \bibnamefont {Rescigno}}, \ and\ \bibinfo {author}
  {\bibfnamefont {R.}~\bibnamefont {D{\"o}rner}},\ }\href {\doibase
  10.1103/PhysRevLett.106.133001} {\bibfield  {journal} {\bibinfo  {journal}
  {Phys. Rev. Lett.}\ }\textbf {\bibinfo {volume} {106}},\ \bibinfo {pages}
  {133001} (\bibinfo {year} {2011})}\BibitemShut {NoStop}%
\bibitem [{\citenamefont {Mergel}\ \emph {et~al.}(1998)\citenamefont {Mergel},
  \citenamefont {Achler}, \citenamefont {D{\"o}rner}, \citenamefont {Khayyat},
  \citenamefont {Kambara}, \citenamefont {Awaya}, \citenamefont {Zoran},
  \citenamefont {Nystr{\"o}m}, \citenamefont {Spielberger}, \citenamefont
  {McGuire}, \citenamefont {Feagin}, \citenamefont {Berakdar}, \citenamefont
  {Azuma},\ and\ \citenamefont {Schmidt-B{\"o}cking}}]{Mergel}%
  \BibitemOpen
  \bibfield  {author} {\bibinfo {author} {\bibfnamefont {V.}~\bibnamefont
  {Mergel}}, \bibinfo {author} {\bibfnamefont {M.}~\bibnamefont {Achler}},
  \bibinfo {author} {\bibfnamefont {R.}~\bibnamefont {D{\"o}rner}}, \bibinfo
  {author} {\bibfnamefont {K.}~\bibnamefont {Khayyat}}, \bibinfo {author}
  {\bibfnamefont {T.}~\bibnamefont {Kambara}}, \bibinfo {author} {\bibfnamefont
  {Y.}~\bibnamefont {Awaya}}, \bibinfo {author} {\bibfnamefont
  {V.}~\bibnamefont {Zoran}}, \bibinfo {author} {\bibfnamefont
  {B.}~\bibnamefont {Nystr{\"o}m}}, \bibinfo {author} {\bibfnamefont
  {L.}~\bibnamefont {Spielberger}}, \bibinfo {author} {\bibfnamefont {J.~H.}\
  \bibnamefont {McGuire}}, \bibinfo {author} {\bibfnamefont {J.}~\bibnamefont
  {Feagin}}, \bibinfo {author} {\bibfnamefont {J.}~\bibnamefont {Berakdar}},
  \bibinfo {author} {\bibfnamefont {Y.}~\bibnamefont {Azuma}}, \ and\ \bibinfo
  {author} {\bibfnamefont {H.}~\bibnamefont {Schmidt-B{\"o}cking}},\ }\href
  {\doibase 10.1103/PhysRevLett.80.5301} {\bibfield  {journal} {\bibinfo
  {journal} {Phys. Rev. Lett.}\ }\textbf {\bibinfo {volume} {80}},\ \bibinfo
  {pages} {5301} (\bibinfo {year} {1998})}\BibitemShut {NoStop}%
\bibitem [{\citenamefont {McCurdy}\ \emph {et~al.}(2004)\citenamefont
  {McCurdy}, \citenamefont {Horner}, \citenamefont {Rescigno},\ and\
  \citenamefont {Mart\'{\i}n}}]{Horner}%
  \BibitemOpen
  \bibfield  {author} {\bibinfo {author} {\bibfnamefont {C.~W.}\ \bibnamefont
  {McCurdy}}, \bibinfo {author} {\bibfnamefont {D.~A.}\ \bibnamefont {Horner}},
  \bibinfo {author} {\bibfnamefont {T.~N.}\ \bibnamefont {Rescigno}}, \ and\
  \bibinfo {author} {\bibfnamefont {F.}~\bibnamefont {Mart\'{\i}n}},\ }\href
  {\doibase 10.1103/PhysRevA.69.032707} {\bibfield  {journal} {\bibinfo
  {journal} {Phys. Rev. A}\ }\textbf {\bibinfo {volume} {69}},\ \bibinfo
  {pages} {032707} (\bibinfo {year} {2004})}\BibitemShut {NoStop}%
\bibitem [{\citenamefont {Yip}\ \emph {et~al.}(2013)\citenamefont {Yip},
  \citenamefont {Rescigno}, \citenamefont {McCurdy},\ and\ \citenamefont
  {Mart\'{\i}n}}]{Yip}%
  \BibitemOpen
  \bibfield  {author} {\bibinfo {author} {\bibfnamefont {F.~L.}\ \bibnamefont
  {Yip}}, \bibinfo {author} {\bibfnamefont {T.~N.}\ \bibnamefont {Rescigno}},
  \bibinfo {author} {\bibfnamefont {C.~W.}\ \bibnamefont {McCurdy}}, \ and\
  \bibinfo {author} {\bibfnamefont {F.}~\bibnamefont {Mart\'{\i}n}},\ }\href
  {\doibase 10.1103/PhysRevLett.110.173001} {\bibfield  {journal} {\bibinfo
  {journal} {Phys. Rev. Lett.}\ }\textbf {\bibinfo {volume} {110}},\ \bibinfo
  {pages} {173001} (\bibinfo {year} {2013})}\BibitemShut {NoStop}%
\bibitem [{\citenamefont {Weber}\ \emph
  {et~al.}(2004{\natexlab{a}})\citenamefont {Weber}, \citenamefont {Czasch},
  \citenamefont {Jagutzki}, \citenamefont {M{\"u}ller}, \citenamefont {Mergel},
  \citenamefont {Kheifets}, \citenamefont {Feagin}, \citenamefont {Rotenberg},
  \citenamefont {Meigs}, \citenamefont {Prior}, \citenamefont {Daveau},
  \citenamefont {Landers}, \citenamefont {Cocke}, \citenamefont {Osipov},
  \citenamefont {Schmidt-B{\"o}cking},\ and\ \citenamefont
  {D{\"o}rner}}]{Weber}%
  \BibitemOpen
  \bibfield  {author} {\bibinfo {author} {\bibfnamefont {T.}~\bibnamefont
  {Weber}}, \bibinfo {author} {\bibfnamefont {A.}~\bibnamefont {Czasch}},
  \bibinfo {author} {\bibfnamefont {O.}~\bibnamefont {Jagutzki}}, \bibinfo
  {author} {\bibfnamefont {A.}~\bibnamefont {M{\"u}ller}}, \bibinfo {author}
  {\bibfnamefont {V.}~\bibnamefont {Mergel}}, \bibinfo {author} {\bibfnamefont
  {A.}~\bibnamefont {Kheifets}}, \bibinfo {author} {\bibfnamefont
  {J.}~\bibnamefont {Feagin}}, \bibinfo {author} {\bibfnamefont
  {E.}~\bibnamefont {Rotenberg}}, \bibinfo {author} {\bibfnamefont
  {G.}~\bibnamefont {Meigs}}, \bibinfo {author} {\bibfnamefont {M.~H.}\
  \bibnamefont {Prior}}, \bibinfo {author} {\bibfnamefont {S.}~\bibnamefont
  {Daveau}}, \bibinfo {author} {\bibfnamefont {A.~L.}\ \bibnamefont {Landers}},
  \bibinfo {author} {\bibfnamefont {C.~L.}\ \bibnamefont {Cocke}}, \bibinfo
  {author} {\bibfnamefont {T.}~\bibnamefont {Osipov}}, \bibinfo {author}
  {\bibfnamefont {H.}~\bibnamefont {Schmidt-B{\"o}cking}}, \ and\ \bibinfo
  {author} {\bibfnamefont {R.}~\bibnamefont {D{\"o}rner}},\ }\href {\doibase
  10.1103/PhysRevLett.92.163001} {\bibfield  {journal} {\bibinfo  {journal}
  {Phys. Rev. Lett.}\ }\textbf {\bibinfo {volume} {92}},\ \bibinfo {pages}
  {163001} (\bibinfo {year} {2004}{\natexlab{a}})}\BibitemShut {NoStop}%
\bibitem [{\citenamefont {Vanroose}\ \emph {et~al.}(2006)\citenamefont
  {Vanroose}, \citenamefont {Horner}, \citenamefont {Mart\'{\i}n},
  \citenamefont {Rescigno},\ and\ \citenamefont {McCurdy}}]{Vanroose}%
  \BibitemOpen
  \bibfield  {author} {\bibinfo {author} {\bibfnamefont {W.}~\bibnamefont
  {Vanroose}}, \bibinfo {author} {\bibfnamefont {D.~A.}\ \bibnamefont
  {Horner}}, \bibinfo {author} {\bibfnamefont {F.}~\bibnamefont {Mart\'{\i}n}},
  \bibinfo {author} {\bibfnamefont {T.~N.}\ \bibnamefont {Rescigno}}, \ and\
  \bibinfo {author} {\bibfnamefont {C.~W.}\ \bibnamefont {McCurdy}},\ }\href
  {\doibase 10.1103/PhysRevA.74.052702} {\bibfield  {journal} {\bibinfo
  {journal} {Phys. Rev. A}\ }\textbf {\bibinfo {volume} {74}},\ \bibinfo
  {pages} {052702} (\bibinfo {year} {2006})}\BibitemShut {NoStop}%
\bibitem [{\citenamefont {Gaire}\ \emph {et~al.}(2014)\citenamefont {Gaire},
  \citenamefont {Lee}, \citenamefont {Haxton}, \citenamefont {Pelz},
  \citenamefont {Bocharova}, \citenamefont {Sturm}, \citenamefont {Gehrken},
  \citenamefont {Honig}, \citenamefont {Pitzer}, \citenamefont {Metz},
  \citenamefont {Kim}, \citenamefont {Sch{\"o}ffler}, \citenamefont
  {D{\"o}rner}, \citenamefont {Gassert}, \citenamefont {Zeller}, \citenamefont
  {Voigtsberger}, \citenamefont {Cao}, \citenamefont {Zohrabi}, \citenamefont
  {Williams}, \citenamefont {Gatton}, \citenamefont {Reedy}, \citenamefont
  {Nook}, \citenamefont {M{\"u}ller}, \citenamefont {Landers}, \citenamefont
  {Cocke}, \citenamefont {Ben-Itzhak}, \citenamefont {Jahnke}, \citenamefont
  {Belkacem},\ and\ \citenamefont {Weber}}]{Gaire}%
  \BibitemOpen
  \bibfield  {author} {\bibinfo {author} {\bibfnamefont {B.}~\bibnamefont
  {Gaire}}, \bibinfo {author} {\bibfnamefont {S.~Y.}\ \bibnamefont {Lee}},
  \bibinfo {author} {\bibfnamefont {D.~J.}\ \bibnamefont {Haxton}}, \bibinfo
  {author} {\bibfnamefont {P.~M.}\ \bibnamefont {Pelz}}, \bibinfo {author}
  {\bibfnamefont {I.}~\bibnamefont {Bocharova}}, \bibinfo {author}
  {\bibfnamefont {F.~P.}\ \bibnamefont {Sturm}}, \bibinfo {author}
  {\bibfnamefont {N.}~\bibnamefont {Gehrken}}, \bibinfo {author} {\bibfnamefont
  {M.}~\bibnamefont {Honig}}, \bibinfo {author} {\bibfnamefont
  {M.}~\bibnamefont {Pitzer}}, \bibinfo {author} {\bibfnamefont
  {D.}~\bibnamefont {Metz}}, \bibinfo {author} {\bibfnamefont {H.-K.}\
  \bibnamefont {Kim}}, \bibinfo {author} {\bibfnamefont {M.}~\bibnamefont
  {Sch{\"o}ffler}}, \bibinfo {author} {\bibfnamefont {R.}~\bibnamefont
  {D{\"o}rner}}, \bibinfo {author} {\bibfnamefont {H.}~\bibnamefont {Gassert}},
  \bibinfo {author} {\bibfnamefont {S.}~\bibnamefont {Zeller}}, \bibinfo
  {author} {\bibfnamefont {J.}~\bibnamefont {Voigtsberger}}, \bibinfo {author}
  {\bibfnamefont {W.}~\bibnamefont {Cao}}, \bibinfo {author} {\bibfnamefont
  {M.}~\bibnamefont {Zohrabi}}, \bibinfo {author} {\bibfnamefont
  {J.}~\bibnamefont {Williams}}, \bibinfo {author} {\bibfnamefont
  {A.}~\bibnamefont {Gatton}}, \bibinfo {author} {\bibfnamefont
  {D.}~\bibnamefont {Reedy}}, \bibinfo {author} {\bibfnamefont
  {C.}~\bibnamefont {Nook}}, \bibinfo {author} {\bibfnamefont {T.}~\bibnamefont
  {M{\"u}ller}}, \bibinfo {author} {\bibfnamefont {A.~L.}\ \bibnamefont
  {Landers}}, \bibinfo {author} {\bibfnamefont {C.~L.}\ \bibnamefont {Cocke}},
  \bibinfo {author} {\bibfnamefont {I.}~\bibnamefont {Ben-Itzhak}}, \bibinfo
  {author} {\bibfnamefont {T.}~\bibnamefont {Jahnke}}, \bibinfo {author}
  {\bibfnamefont {A.}~\bibnamefont {Belkacem}}, \ and\ \bibinfo {author}
  {\bibfnamefont {T.}~\bibnamefont {Weber}},\ }\href {\doibase
  10.1103/PhysRevA.89.013403} {\bibfield  {journal} {\bibinfo  {journal} {Phys.
  Rev. A}\ }\textbf {\bibinfo {volume} {89}},\ \bibinfo {pages} {013403}
  (\bibinfo {year} {2014})}\BibitemShut {NoStop}%
\bibitem [{\citenamefont {Rajput}\ \emph {et~al.}(2018)\citenamefont {Rajput},
  \citenamefont {Severt}, \citenamefont {Berry}, \citenamefont {Jochim},
  \citenamefont {Feizollah}, \citenamefont {Kaderiya}, \citenamefont {Zohrabi},
  \citenamefont {Ablikim}, \citenamefont {Ziaee}, \citenamefont {Raju~P.},
  \citenamefont {Rolles}, \citenamefont {Rudenko}, \citenamefont {Carnes},
  \citenamefont {Esry},\ and\ \citenamefont {Ben-Itzhak}}]{ITZAK}%
  \BibitemOpen
  \bibfield  {author} {\bibinfo {author} {\bibfnamefont {J.}~\bibnamefont
  {Rajput}}, \bibinfo {author} {\bibfnamefont {T.}~\bibnamefont {Severt}},
  \bibinfo {author} {\bibfnamefont {B.}~\bibnamefont {Berry}}, \bibinfo
  {author} {\bibfnamefont {B.}~\bibnamefont {Jochim}}, \bibinfo {author}
  {\bibfnamefont {P.}~\bibnamefont {Feizollah}}, \bibinfo {author}
  {\bibfnamefont {B.}~\bibnamefont {Kaderiya}}, \bibinfo {author}
  {\bibfnamefont {M.}~\bibnamefont {Zohrabi}}, \bibinfo {author} {\bibfnamefont
  {U.}~\bibnamefont {Ablikim}}, \bibinfo {author} {\bibfnamefont
  {F.}~\bibnamefont {Ziaee}}, \bibinfo {author} {\bibfnamefont
  {K.}~\bibnamefont {Raju~P.}}, \bibinfo {author} {\bibfnamefont
  {D.}~\bibnamefont {Rolles}}, \bibinfo {author} {\bibfnamefont
  {A.}~\bibnamefont {Rudenko}}, \bibinfo {author} {\bibfnamefont {K.~D.}\
  \bibnamefont {Carnes}}, \bibinfo {author} {\bibfnamefont {B.~D.}\
  \bibnamefont {Esry}}, \ and\ \bibinfo {author} {\bibfnamefont
  {I.}~\bibnamefont {Ben-Itzhak}},\ }\href {\doibase
  10.1103/PhysRevLett.120.103001} {\bibfield  {journal} {\bibinfo  {journal}
  {Phys. Rev. Lett.}\ }\textbf {\bibinfo {volume} {120}},\ \bibinfo {pages}
  {103001} (\bibinfo {year} {2018})}\BibitemShut {NoStop}%
\bibitem [{\citenamefont {Zare}(1972)}]{ZARE}%
  \BibitemOpen
  \bibfield  {author} {\bibinfo {author} {\bibfnamefont {R.~N.}\ \bibnamefont
  {Zare}},\ }\href@noop {} {\bibfield  {journal} {\bibinfo  {journal}
  {Molecular Photochemistry}\ }\textbf {\bibinfo {volume} {4}},\ \bibinfo
  {pages} {1} (\bibinfo {year} {1972})}\BibitemShut {NoStop}%
\bibitem [{\citenamefont {Weber}\ \emph
  {et~al.}(2004{\natexlab{b}})\citenamefont {Weber}, \citenamefont {Czasch},
  \citenamefont {Jagutzki}, \citenamefont {M{\"u}ller}, \citenamefont {Mergel},
  \citenamefont {Kheifets}, \citenamefont {Rotenberg}, \citenamefont {Meigs},
  \citenamefont {Prior}, \citenamefont {Daveau} \emph {et~al.}}]{Weber1}%
  \BibitemOpen
  \bibfield  {author} {\bibinfo {author} {\bibfnamefont {T.}~\bibnamefont
  {Weber}}, \bibinfo {author} {\bibfnamefont {A.~O.}\ \bibnamefont {Czasch}},
  \bibinfo {author} {\bibfnamefont {O.}~\bibnamefont {Jagutzki}}, \bibinfo
  {author} {\bibfnamefont {A.}~\bibnamefont {M{\"u}ller}}, \bibinfo {author}
  {\bibfnamefont {V.}~\bibnamefont {Mergel}}, \bibinfo {author} {\bibfnamefont
  {A.}~\bibnamefont {Kheifets}}, \bibinfo {author} {\bibfnamefont
  {E.}~\bibnamefont {Rotenberg}}, \bibinfo {author} {\bibfnamefont
  {G.}~\bibnamefont {Meigs}}, \bibinfo {author} {\bibfnamefont {M.~H.}\
  \bibnamefont {Prior}}, \bibinfo {author} {\bibfnamefont {S.}~\bibnamefont
  {Daveau}},  \emph {et~al.},\ }\href@noop {} {\bibfield  {journal} {\bibinfo
  {journal} {Nature}\ }\textbf {\bibinfo {volume} {431}},\ \bibinfo {pages}
  {437} (\bibinfo {year} {2004}{\natexlab{b}})}\BibitemShut {NoStop}%
\bibitem [{\citenamefont {Akoury}\ \emph {et~al.}(2007)\citenamefont {Akoury},
  \citenamefont {Kreidi}, \citenamefont {Jahnke}, \citenamefont {Weber},
  \citenamefont {Staudte}, \citenamefont {Sch{\"o}ffler}, \citenamefont
  {Neumann}, \citenamefont {Titze}, \citenamefont {Schmidt}, \citenamefont
  {Czasch} \emph {et~al.}}]{Weber2}%
  \BibitemOpen
  \bibfield  {author} {\bibinfo {author} {\bibfnamefont {D.}~\bibnamefont
  {Akoury}}, \bibinfo {author} {\bibfnamefont {K.}~\bibnamefont {Kreidi}},
  \bibinfo {author} {\bibfnamefont {T.}~\bibnamefont {Jahnke}}, \bibinfo
  {author} {\bibfnamefont {T.}~\bibnamefont {Weber}}, \bibinfo {author}
  {\bibfnamefont {A.}~\bibnamefont {Staudte}}, \bibinfo {author} {\bibfnamefont
  {M.}~\bibnamefont {Sch{\"o}ffler}}, \bibinfo {author} {\bibfnamefont
  {N.}~\bibnamefont {Neumann}}, \bibinfo {author} {\bibfnamefont
  {J.}~\bibnamefont {Titze}}, \bibinfo {author} {\bibfnamefont {L.~P.~H.}\
  \bibnamefont {Schmidt}}, \bibinfo {author} {\bibfnamefont {A.}~\bibnamefont
  {Czasch}},  \emph {et~al.},\ }\href@noop {} {\bibfield  {journal} {\bibinfo
  {journal} {Science}\ }\textbf {\bibinfo {volume} {318}},\ \bibinfo {pages}
  {949} (\bibinfo {year} {2007})}\BibitemShut {NoStop}%
\bibitem [{\citenamefont {Reddish}\ \emph {et~al.}(2008)\citenamefont
  {Reddish}, \citenamefont {Colgan}, \citenamefont {Bolognesi}, \citenamefont
  {Avaldi}, \citenamefont {Gisselbrecht}, \citenamefont {Lavoll{\'e}e},
  \citenamefont {Pindzola},\ and\ \citenamefont {Huetz}}]{Reddish}%
  \BibitemOpen
  \bibfield  {author} {\bibinfo {author} {\bibfnamefont {T.~J.}\ \bibnamefont
  {Reddish}}, \bibinfo {author} {\bibfnamefont {J.}~\bibnamefont {Colgan}},
  \bibinfo {author} {\bibfnamefont {P.}~\bibnamefont {Bolognesi}}, \bibinfo
  {author} {\bibfnamefont {L.}~\bibnamefont {Avaldi}}, \bibinfo {author}
  {\bibfnamefont {M.}~\bibnamefont {Gisselbrecht}}, \bibinfo {author}
  {\bibfnamefont {M.}~\bibnamefont {Lavoll{\'e}e}}, \bibinfo {author}
  {\bibfnamefont {M.~S.}\ \bibnamefont {Pindzola}}, \ and\ \bibinfo {author}
  {\bibfnamefont {A.}~\bibnamefont {Huetz}},\ }\href {\doibase
  10.1103/PhysRevLett.100.193001} {\bibfield  {journal} {\bibinfo  {journal}
  {Phys. Rev. Lett.}\ }\textbf {\bibinfo {volume} {100}},\ \bibinfo {pages}
  {193001} (\bibinfo {year} {2008})}\BibitemShut {NoStop}%
\bibitem [{\citenamefont {Winkoun}\ and\ \citenamefont
  {Dujardin}(1986)}]{Winkoun}%
  \BibitemOpen
  \bibfield  {author} {\bibinfo {author} {\bibfnamefont {D.}~\bibnamefont
  {Winkoun}}\ and\ \bibinfo {author} {\bibfnamefont {G.}~\bibnamefont
  {Dujardin}},\ }\href {\doibase 10.1007/BF01432498} {\bibfield  {journal}
  {\bibinfo  {journal} {"Zeitschrift f{\"u}r Physik D Atoms, Molecules and
  Clusters"}\ }\textbf {\bibinfo {volume} {4}},\ \bibinfo {pages} {57}
  (\bibinfo {year} {1986})}\BibitemShut {NoStop}%
\bibitem [{\citenamefont {Stankiewicz}\ \emph {et~al.}(1989)\citenamefont
  {Stankiewicz}, \citenamefont {Hatherly}, \citenamefont {Frasinski},
  \citenamefont {Codling},\ and\ \citenamefont {Holland}}]{Stankiewicz}%
  \BibitemOpen
  \bibfield  {author} {\bibinfo {author} {\bibfnamefont {M.}~\bibnamefont
  {Stankiewicz}}, \bibinfo {author} {\bibfnamefont {P.~A.}\ \bibnamefont
  {Hatherly}}, \bibinfo {author} {\bibfnamefont {L.~J.}\ \bibnamefont
  {Frasinski}}, \bibinfo {author} {\bibfnamefont {K.}~\bibnamefont {Codling}},
  \ and\ \bibinfo {author} {\bibfnamefont {D.~M.~P.}\ \bibnamefont {Holland}},\
  }\href {\doibase 10.1088/0953-4075/22/1/006} {\bibfield  {journal} {\bibinfo
  {journal} {Journal of Physics B: Atomic, Molecular and Optical Physics}\
  }\textbf {\bibinfo {volume} {22}},\ \bibinfo {pages} {21} (\bibinfo {year}
  {1989})}\BibitemShut {NoStop}%
\bibitem [{\citenamefont {Locht}\ \emph {et~al.}(1989)\citenamefont {Locht},
  \citenamefont {Davister}, \citenamefont {Denzer}, \citenamefont {Jochims},\
  and\ \citenamefont {Baumg{\"a}rtel}}]{Locht1}%
  \BibitemOpen
  \bibfield  {author} {\bibinfo {author} {\bibfnamefont {R.}~\bibnamefont
  {Locht}}, \bibinfo {author} {\bibfnamefont {M.}~\bibnamefont {Davister}},
  \bibinfo {author} {\bibfnamefont {W.}~\bibnamefont {Denzer}}, \bibinfo
  {author} {\bibfnamefont {H.}~\bibnamefont {Jochims}}, \ and\ \bibinfo
  {author} {\bibfnamefont {H.}~\bibnamefont {Baumg{\"a}rtel}},\ }\href
  {\doibase https://doi.org/10.1016/0301-0104(89)87149-6} {\bibfield  {journal}
  {\bibinfo  {journal} {Chemical Physics}\ }\textbf {\bibinfo {volume} {138}},\
  \bibinfo {pages} {433 } (\bibinfo {year} {1989})}\BibitemShut {NoStop}%
\bibitem [{\citenamefont {Locht}\ and\ \citenamefont
  {Momigny}(1987{\natexlab{a}})}]{Locht2}%
  \BibitemOpen
  \bibfield  {author} {\bibinfo {author} {\bibfnamefont {R.}~\bibnamefont
  {Locht}}\ and\ \bibinfo {author} {\bibfnamefont {J.}~\bibnamefont
  {Momigny}},\ }\href {\doibase https://doi.org/10.1016/0009-2614(87)80527-4}
  {\bibfield  {journal} {\bibinfo  {journal} {Chemical Physics Letters}\
  }\textbf {\bibinfo {volume} {138}},\ \bibinfo {pages} {391 } (\bibinfo {year}
  {1987}{\natexlab{a}})}\BibitemShut {NoStop}%
\bibitem [{\citenamefont {Eland}(2006)}]{Eland}%
  \BibitemOpen
  \bibfield  {author} {\bibinfo {author} {\bibfnamefont {J.~H.}\ \bibnamefont
  {Eland}},\ }\href {\doibase https://doi.org/10.1016/j.chemphys.2005.09.047}
  {\bibfield  {journal} {\bibinfo  {journal} {Chemical Physics}\ }\textbf
  {\bibinfo {volume} {323}},\ \bibinfo {pages} {391 } (\bibinfo {year}
  {2006})}\BibitemShut {NoStop}%
\bibitem [{\citenamefont {Samson}\ \emph {et~al.}(1987)\citenamefont {Samson},
  \citenamefont {Haddad},\ and\ \citenamefont {Kilcoyne}}]{Samson}%
  \BibitemOpen
  \bibfield  {author} {\bibinfo {author} {\bibfnamefont {J.~A.}\ \bibnamefont
  {Samson}}, \bibinfo {author} {\bibfnamefont {G.}~\bibnamefont {Haddad}}, \
  and\ \bibinfo {author} {\bibfnamefont {L.}~\bibnamefont {Kilcoyne}},\
  }\href@noop {} {\bibfield  {journal} {\bibinfo  {journal} {The Journal of
  chemical physics}\ }\textbf {\bibinfo {volume} {87}},\ \bibinfo {pages}
  {6416} (\bibinfo {year} {1987})}\BibitemShut {NoStop}%
\bibitem [{\citenamefont {Appell}\ and\ \citenamefont
  {Horsley}(1974)}]{Appell}%
  \BibitemOpen
  \bibfield  {author} {\bibinfo {author} {\bibfnamefont {J.}~\bibnamefont
  {Appell}}\ and\ \bibinfo {author} {\bibfnamefont {J.}~\bibnamefont
  {Horsley}},\ }\href@noop {} {\bibfield  {journal} {\bibinfo  {journal} {The
  Journal of Chemical Physics}\ }\textbf {\bibinfo {volume} {60}},\ \bibinfo
  {pages} {3445} (\bibinfo {year} {1974})}\BibitemShut {NoStop}%
\bibitem [{\citenamefont {M{\"a}rk}\ \emph {et~al.}(1977)\citenamefont
  {M{\"a}rk}, \citenamefont {Egger},\ and\ \citenamefont {Cheret}}]{Cheret}%
  \BibitemOpen
  \bibfield  {author} {\bibinfo {author} {\bibfnamefont {T.}~\bibnamefont
  {M{\"a}rk}}, \bibinfo {author} {\bibfnamefont {F.}~\bibnamefont {Egger}}, \
  and\ \bibinfo {author} {\bibfnamefont {M.}~\bibnamefont {Cheret}},\
  }\href@noop {} {\bibfield  {journal} {\bibinfo  {journal} {The Journal of
  Chemical Physics}\ }\textbf {\bibinfo {volume} {67}},\ \bibinfo {pages}
  {3795} (\bibinfo {year} {1977})}\BibitemShut {NoStop}%
\bibitem [{\citenamefont {Langford}\ \emph {et~al.}(1992)\citenamefont
  {Langford}, \citenamefont {Harris}, \citenamefont {Fournier},\ and\
  \citenamefont {Fournier}}]{Langford}%
  \BibitemOpen
  \bibfield  {author} {\bibinfo {author} {\bibfnamefont {M.}~\bibnamefont
  {Langford}}, \bibinfo {author} {\bibfnamefont {F.}~\bibnamefont {Harris}},
  \bibinfo {author} {\bibfnamefont {P.}~\bibnamefont {Fournier}}, \ and\
  \bibinfo {author} {\bibfnamefont {J.}~\bibnamefont {Fournier}},\ }\href@noop
  {} {\bibfield  {journal} {\bibinfo  {journal} {International journal of mass
  spectrometry and ion processes}\ }\textbf {\bibinfo {volume} {116}},\
  \bibinfo {pages} {53} (\bibinfo {year} {1992})}\BibitemShut {NoStop}%
\bibitem [{\citenamefont {Griffiths}\ and\ \citenamefont
  {Harris}(1990)}]{Griffiths}%
  \BibitemOpen
  \bibfield  {author} {\bibinfo {author} {\bibfnamefont {W.}~\bibnamefont
  {Griffiths}}\ and\ \bibinfo {author} {\bibfnamefont {F.}~\bibnamefont
  {Harris}},\ }\href@noop {} {\bibfield  {journal} {\bibinfo  {journal} {Rapid
  Communications in Mass Spectrometry}\ }\textbf {\bibinfo {volume} {4}},\
  \bibinfo {pages} {366} (\bibinfo {year} {1990})}\BibitemShut {NoStop}%
\bibitem [{\citenamefont {Locht}\ \emph {et~al.}(1988)\citenamefont {Locht},
  \citenamefont {Servais}, \citenamefont {Ligot}, \citenamefont {Derwa},\ and\
  \citenamefont {Momigny}}]{Locht3}%
  \BibitemOpen
  \bibfield  {author} {\bibinfo {author} {\bibfnamefont {R.}~\bibnamefont
  {Locht}}, \bibinfo {author} {\bibfnamefont {C.}~\bibnamefont {Servais}},
  \bibinfo {author} {\bibfnamefont {M.}~\bibnamefont {Ligot}}, \bibinfo
  {author} {\bibfnamefont {F.}~\bibnamefont {Derwa}}, \ and\ \bibinfo {author}
  {\bibfnamefont {J.}~\bibnamefont {Momigny}},\ }\href@noop {} {\bibfield
  {journal} {\bibinfo  {journal} {Chemical physics}\ }\textbf {\bibinfo
  {volume} {123}},\ \bibinfo {pages} {443} (\bibinfo {year}
  {1988})}\BibitemShut {NoStop}%
\bibitem [{\citenamefont {White}\ \emph {et~al.}(1977)\citenamefont {White},
  \citenamefont {Rye},\ and\ \citenamefont {Houston}}]{White}%
  \BibitemOpen
  \bibfield  {author} {\bibinfo {author} {\bibfnamefont {J.}~\bibnamefont
  {White}}, \bibinfo {author} {\bibfnamefont {R.}~\bibnamefont {Rye}}, \ and\
  \bibinfo {author} {\bibfnamefont {J.}~\bibnamefont {Houston}},\ }\href@noop
  {} {\bibfield  {journal} {\bibinfo  {journal} {Chemical Physics Letters}\
  }\textbf {\bibinfo {volume} {46}},\ \bibinfo {pages} {146} (\bibinfo {year}
  {1977})}\BibitemShut {NoStop}%
\bibitem [{\citenamefont {Camilloni}\ \emph {et~al.}(1977)\citenamefont
  {Camilloni}, \citenamefont {Stefani},\ and\ \citenamefont
  {Giardini-Guidoni}}]{Camilloni}%
  \BibitemOpen
  \bibfield  {author} {\bibinfo {author} {\bibfnamefont {R.}~\bibnamefont
  {Camilloni}}, \bibinfo {author} {\bibfnamefont {G.}~\bibnamefont {Stefani}},
  \ and\ \bibinfo {author} {\bibfnamefont {A.}~\bibnamefont
  {Giardini-Guidoni}},\ }\href@noop {} {\bibfield  {journal} {\bibinfo
  {journal} {Chemical Physics Letters}\ }\textbf {\bibinfo {volume} {50}},\
  \bibinfo {pages} {213} (\bibinfo {year} {1977})}\BibitemShut {NoStop}%
\bibitem [{\citenamefont {{\O}kland}\ \emph {et~al.}(1976)\citenamefont
  {{\O}kland}, \citenamefont {F{\ae}gri~Jr},\ and\ \citenamefont
  {Manne}}]{Okland}%
  \BibitemOpen
  \bibfield  {author} {\bibinfo {author} {\bibfnamefont {M.~T.}\ \bibnamefont
  {{\O}kland}}, \bibinfo {author} {\bibfnamefont {K.}~\bibnamefont
  {F{\ae}gri~Jr}}, \ and\ \bibinfo {author} {\bibfnamefont {R.}~\bibnamefont
  {Manne}},\ }\href@noop {} {\bibfield  {journal} {\bibinfo  {journal}
  {Chemical Physics Letters}\ }\textbf {\bibinfo {volume} {40}},\ \bibinfo
  {pages} {185} (\bibinfo {year} {1976})}\BibitemShut {NoStop}%
\bibitem [{\citenamefont {Jennison}(1981)}]{Jennison}%
  \BibitemOpen
  \bibfield  {author} {\bibinfo {author} {\bibfnamefont {D.~R.}\ \bibnamefont
  {Jennison}},\ }\href@noop {} {\bibfield  {journal} {\bibinfo  {journal}
  {Physical Review A}\ }\textbf {\bibinfo {volume} {23}},\ \bibinfo {pages}
  {1215} (\bibinfo {year} {1981})}\BibitemShut {NoStop}%
\bibitem [{\citenamefont {Boyd}\ \emph {et~al.}(1985)\citenamefont {Boyd},
  \citenamefont {Singh},\ and\ \citenamefont {Beynon}}]{Boyd}%
  \BibitemOpen
  \bibfield  {author} {\bibinfo {author} {\bibfnamefont {R.}~\bibnamefont
  {Boyd}}, \bibinfo {author} {\bibfnamefont {S.}~\bibnamefont {Singh}}, \ and\
  \bibinfo {author} {\bibfnamefont {J.}~\bibnamefont {Beynon}},\ }\href@noop {}
  {\bibfield  {journal} {\bibinfo  {journal} {Chemical physics}\ }\textbf
  {\bibinfo {volume} {100}},\ \bibinfo {pages} {297} (\bibinfo {year}
  {1985})}\BibitemShut {NoStop}%
\bibitem [{\citenamefont {Locht}\ and\ \citenamefont
  {Momigny}(1987{\natexlab{b}})}]{LOCHT}%
  \BibitemOpen
  \bibfield  {author} {\bibinfo {author} {\bibfnamefont {R.}~\bibnamefont
  {Locht}}\ and\ \bibinfo {author} {\bibfnamefont {J.}~\bibnamefont
  {Momigny}},\ }\href {\doibase https://doi.org/10.1016/0009-2614(87)80527-4}
  {\bibfield  {journal} {\bibinfo  {journal} {Chemical Physics Letters}\
  }\textbf {\bibinfo {volume} {138}},\ \bibinfo {pages} {391 } (\bibinfo {year}
  {1987}{\natexlab{b}})}\BibitemShut {NoStop}%
\bibitem [{\citenamefont {D{\"o}rner}\ \emph {et~al.}(2000)\citenamefont
  {D{\"o}rner}, \citenamefont {Mergel}, \citenamefont {Jagutzki}, \citenamefont
  {Spielberger}, \citenamefont {Ullrich}, \citenamefont {Moshammer},\ and\
  \citenamefont {Schmidt-B{\"o}cking}}]{Dorner}%
  \BibitemOpen
  \bibfield  {author} {\bibinfo {author} {\bibfnamefont {R.}~\bibnamefont
  {D{\"o}rner}}, \bibinfo {author} {\bibfnamefont {V.}~\bibnamefont {Mergel}},
  \bibinfo {author} {\bibfnamefont {O.}~\bibnamefont {Jagutzki}}, \bibinfo
  {author} {\bibfnamefont {L.}~\bibnamefont {Spielberger}}, \bibinfo {author}
  {\bibfnamefont {J.}~\bibnamefont {Ullrich}}, \bibinfo {author} {\bibfnamefont
  {R.}~\bibnamefont {Moshammer}}, \ and\ \bibinfo {author} {\bibfnamefont
  {H.}~\bibnamefont {Schmidt-B{\"o}cking}},\ }\href {\doibase doi: DOI:
  10.1016/S0370-1573(99)00109-X} {\bibfield  {journal} {\bibinfo  {journal}
  {Physics Reports}\ }\textbf {\bibinfo {volume} {330}},\ \bibinfo {pages} {95}
  (\bibinfo {year} {2000})}\BibitemShut {NoStop}%
\bibitem [{\citenamefont {Ullrich}\ \emph {et~al.}(2003)\citenamefont
  {Ullrich}, \citenamefont {Moshammer}, \citenamefont {Dorn}, \citenamefont
  {D{\"o}rner}, \citenamefont {Schmidt},\ and\ \citenamefont
  {Schmidt-B{\"o}cking}}]{Ullrich}%
  \BibitemOpen
  \bibfield  {author} {\bibinfo {author} {\bibfnamefont {J.}~\bibnamefont
  {Ullrich}}, \bibinfo {author} {\bibfnamefont {R.}~\bibnamefont {Moshammer}},
  \bibinfo {author} {\bibfnamefont {A.}~\bibnamefont {Dorn}}, \bibinfo {author}
  {\bibfnamefont {R.}~\bibnamefont {D{\"o}rner}}, \bibinfo {author}
  {\bibfnamefont {L.~P.~H.}\ \bibnamefont {Schmidt}}, \ and\ \bibinfo {author}
  {\bibfnamefont {H.}~\bibnamefont {Schmidt-B{\"o}cking}},\ }\href@noop {}
  {\bibfield  {journal} {\bibinfo  {journal} {Reports on Progress in Physics}\
  }\textbf {\bibinfo {volume} {66}},\ \bibinfo {pages} {1463} (\bibinfo {year}
  {2003})}\BibitemShut {NoStop}%
\bibitem [{\citenamefont {Jagutzki}\ \emph {et~al.}(2002)\citenamefont
  {Jagutzki}, \citenamefont {Cerezo}, \citenamefont {Czasch}, \citenamefont
  {Dorner}, \citenamefont {Hattas}, \citenamefont {Huang}, \citenamefont
  {Mergel}, \citenamefont {Spillmann}, \citenamefont {Ullmann-Pfleger},
  \citenamefont {Weber} \emph {et~al.}}]{Jagutzki}%
  \BibitemOpen
  \bibfield  {author} {\bibinfo {author} {\bibfnamefont {O.}~\bibnamefont
  {Jagutzki}}, \bibinfo {author} {\bibfnamefont {A.}~\bibnamefont {Cerezo}},
  \bibinfo {author} {\bibfnamefont {A.}~\bibnamefont {Czasch}}, \bibinfo
  {author} {\bibfnamefont {R.}~\bibnamefont {Dorner}}, \bibinfo {author}
  {\bibfnamefont {M.}~\bibnamefont {Hattas}}, \bibinfo {author} {\bibfnamefont
  {M.}~\bibnamefont {Huang}}, \bibinfo {author} {\bibfnamefont
  {V.}~\bibnamefont {Mergel}}, \bibinfo {author} {\bibfnamefont
  {U.}~\bibnamefont {Spillmann}}, \bibinfo {author} {\bibfnamefont
  {K.}~\bibnamefont {Ullmann-Pfleger}}, \bibinfo {author} {\bibfnamefont
  {T.}~\bibnamefont {Weber}},  \emph {et~al.},\ }\href@noop {} {\bibfield
  {journal} {\bibinfo  {journal} {IEEE Transactions on Nuclear Science}\
  }\textbf {\bibinfo {volume} {49}},\ \bibinfo {pages} {2477} (\bibinfo {year}
  {2002})}\BibitemShut {NoStop}%
\bibitem [{\citenamefont {Pope}\ \emph {et~al.}(1983)\citenamefont {Pope},
  \citenamefont {Hillier}, \citenamefont {Guest},\ and\ \citenamefont
  {Kendric}}]{POPE}%
  \BibitemOpen
  \bibfield  {author} {\bibinfo {author} {\bibfnamefont {S.~A.}\ \bibnamefont
  {Pope}}, \bibinfo {author} {\bibfnamefont {I.~H.}\ \bibnamefont {Hillier}},
  \bibinfo {author} {\bibfnamefont {M.~F.}\ \bibnamefont {Guest}}, \ and\
  \bibinfo {author} {\bibfnamefont {J.}~\bibnamefont {Kendric}},\ }\href@noop
  {} {\bibfield  {journal} {\bibinfo  {journal} {Chem. Phys. Lett.}\ }\textbf
  {\bibinfo {volume} {95}},\ \bibinfo {pages} {247} (\bibinfo {year}
  {1983})}\BibitemShut {NoStop}%
\bibitem [{\citenamefont {Tarantelli}\ \emph {et~al.}(1985)\citenamefont
  {Tarantelli}, \citenamefont {Tarantelli}, \citenamefont {Sgamellotti},
  \citenamefont {Schirmer},\ and\ \citenamefont {Cederbaum}}]{CEDERBAUM}%
  \BibitemOpen
  \bibfield  {author} {\bibinfo {author} {\bibfnamefont {F.}~\bibnamefont
  {Tarantelli}}, \bibinfo {author} {\bibfnamefont {A.}~\bibnamefont
  {Tarantelli}}, \bibinfo {author} {\bibfnamefont {A.}~\bibnamefont
  {Sgamellotti}}, \bibinfo {author} {\bibfnamefont {J.}~\bibnamefont
  {Schirmer}}, \ and\ \bibinfo {author} {\bibfnamefont {L.~S.}\ \bibnamefont
  {Cederbaum}},\ }\href@noop {} {\bibfield  {journal} {\bibinfo  {journal}
  {Chem. Phys. Lett.}\ }\textbf {\bibinfo {volume} {117}},\ \bibinfo {pages}
  {577} (\bibinfo {year} {1985})}\BibitemShut {NoStop}%
\bibitem [{\citenamefont {Rescigno}\ \emph {et~al.}(2016)\citenamefont
  {Rescigno}, \citenamefont {Trevisan}, \citenamefont {Orel}, \citenamefont
  {Slaughter}, \citenamefont {Adaniya}, \citenamefont {Belkacem}, \citenamefont
  {Weyland}, \citenamefont {Dorn},\ and\ \citenamefont {McCurdy}}]{Rescigno16}%
  \BibitemOpen
  \bibfield  {author} {\bibinfo {author} {\bibfnamefont {T.~N.}\ \bibnamefont
  {Rescigno}}, \bibinfo {author} {\bibfnamefont {C.~S.}\ \bibnamefont
  {Trevisan}}, \bibinfo {author} {\bibfnamefont {A.~E.}\ \bibnamefont {Orel}},
  \bibinfo {author} {\bibfnamefont {D.~S.}\ \bibnamefont {Slaughter}}, \bibinfo
  {author} {\bibfnamefont {H.}~\bibnamefont {Adaniya}}, \bibinfo {author}
  {\bibfnamefont {A.}~\bibnamefont {Belkacem}}, \bibinfo {author}
  {\bibfnamefont {M.}~\bibnamefont {Weyland}}, \bibinfo {author} {\bibfnamefont
  {A.}~\bibnamefont {Dorn}}, \ and\ \bibinfo {author} {\bibfnamefont {C.~W.}\
  \bibnamefont {McCurdy}},\ }\href {\doibase 10.1103/PhysRevA.93.052704}
  {\bibfield  {journal} {\bibinfo  {journal} {Phys. Rev. A}\ }\textbf {\bibinfo
  {volume} {93}},\ \bibinfo {pages} {052704} (\bibinfo {year}
  {2016})}\BibitemShut {NoStop}%
\bibitem [{\citenamefont {Rescigno}\ and\ \citenamefont
  {Schneider}(1988)}]{Rescigno88}%
  \BibitemOpen
  \bibfield  {author} {\bibinfo {author} {\bibfnamefont {T.~N.}\ \bibnamefont
  {Rescigno}}\ and\ \bibinfo {author} {\bibfnamefont {B.~I.}\ \bibnamefont
  {Schneider}},\ }\href@noop {} {\bibfield  {journal} {\bibinfo  {journal} {J.
  Phys. B: At. Mol. Opt. Phys.}\ }\textbf {\bibinfo {volume} {21}},\ \bibinfo
  {pages} {L691} (\bibinfo {year} {1988})}\BibitemShut {NoStop}%
\bibitem [{\citenamefont {Higuchi}(1956)}]{Higuchi}%
  \BibitemOpen
  \bibfield  {author} {\bibinfo {author} {\bibfnamefont {J.}~\bibnamefont
  {Higuchi}},\ }\href@noop {} {\bibfield  {journal} {\bibinfo  {journal} {The
  Journal of Chemical Physics}\ }\textbf {\bibinfo {volume} {24}},\ \bibinfo
  {pages} {535} (\bibinfo {year} {1956})}\BibitemShut {NoStop}%
\bibitem [{\citenamefont {Streeter}\ \emph {et~al.}(2018)\citenamefont
  {Streeter}, \citenamefont {Yip}, \citenamefont {Lucchese}, \citenamefont
  {Gervais}, \citenamefont {Rescigno},\ and\ \citenamefont
  {McCurdy}}]{Streeter}%
  \BibitemOpen
  \bibfield  {author} {\bibinfo {author} {\bibfnamefont {Z.~L.}\ \bibnamefont
  {Streeter}}, \bibinfo {author} {\bibfnamefont {F.~L.}\ \bibnamefont {Yip}},
  \bibinfo {author} {\bibfnamefont {R.~R.}\ \bibnamefont {Lucchese}}, \bibinfo
  {author} {\bibfnamefont {B.}~\bibnamefont {Gervais}}, \bibinfo {author}
  {\bibfnamefont {T.~N.}\ \bibnamefont {Rescigno}}, \ and\ \bibinfo {author}
  {\bibfnamefont {C.~W.}\ \bibnamefont {McCurdy}},\ }\href@noop {} {\bibfield
  {journal} {\bibinfo  {journal} {Physical Review A}\ }\textbf {\bibinfo
  {volume} {98}},\ \bibinfo {pages} {053429} (\bibinfo {year}
  {2018})}\BibitemShut {NoStop}%
\bibitem [{\citenamefont {Ben-Itzhak}\ and\ \citenamefont
  {Severt}(2020)}]{Itzik}%
  \BibitemOpen
  \bibfield  {author} {\bibinfo {author} {\bibfnamefont {I.}~\bibnamefont
  {Ben-Itzhak}}\ and\ \bibinfo {author} {\bibfnamefont {T.}~\bibnamefont
  {Severt}},\ }\href@noop {} {}\bibinfo {howpublished} {private communication}
  (\bibinfo {year} {2020})\BibitemShut {NoStop}%
\bibitem [{\citenamefont {Liu}\ and\ \citenamefont {Verhaegen}(1970)}]{Liu}%
  \BibitemOpen
  \bibfield  {author} {\bibinfo {author} {\bibfnamefont {H.}~\bibnamefont
  {Liu}}\ and\ \bibinfo {author} {\bibfnamefont {G.}~\bibnamefont
  {Verhaegen}},\ }\href@noop {} {\bibfield  {journal} {\bibinfo  {journal} {The
  Journal of Chemical Physics}\ }\textbf {\bibinfo {volume} {53}},\ \bibinfo
  {pages} {735} (\bibinfo {year} {1970})}\BibitemShut {NoStop}%
\bibitem [{Ame()}]{Amero2}%
  \BibitemOpen
  \href@noop {} {}\bibinfo {note} {J. M . Amero, and G. J. V{\'a}zquez,
  International Journal of Quantum Chemistry {\bf 99}, 353 (2004); {\it ibid}
  {\bf 101}, 396 (2005)}\BibitemShut {NoStop}%
\bibitem [{\citenamefont {Kawaguchi}\ and\ \citenamefont
  {Amano}(1988)}]{Amano}%
  \BibitemOpen
  \bibfield  {author} {\bibinfo {author} {\bibfnamefont {K.}~\bibnamefont
  {Kawaguchi}}\ and\ \bibinfo {author} {\bibfnamefont {T.}~\bibnamefont
  {Amano}},\ }\href@noop {} {\bibfield  {journal} {\bibinfo  {journal} {The
  Journal of chemical physics}\ }\textbf {\bibinfo {volume} {88}},\ \bibinfo
  {pages} {4584} (\bibinfo {year} {1988})}\BibitemShut {NoStop}%
\bibitem [{\citenamefont {Colin}\ and\ \citenamefont {Douglas}(1968)}]{Colin}%
  \BibitemOpen
  \bibfield  {author} {\bibinfo {author} {\bibfnamefont {R.}~\bibnamefont
  {Colin}}\ and\ \bibinfo {author} {\bibfnamefont {A.}~\bibnamefont
  {Douglas}},\ }\href@noop {} {\bibfield  {journal} {\bibinfo  {journal}
  {Canadian Journal of Physics}\ }\textbf {\bibinfo {volume} {46}},\ \bibinfo
  {pages} {61} (\bibinfo {year} {1968})}\BibitemShut {NoStop}%
\bibitem [{\citenamefont {Tarroni}\ \emph {et~al.}(1997)\citenamefont
  {Tarroni}, \citenamefont {Palmieri}, \citenamefont {Mitrushenkov},
  \citenamefont {Tosi},\ and\ \citenamefont {Bassi}}]{Tarroni}%
  \BibitemOpen
  \bibfield  {author} {\bibinfo {author} {\bibfnamefont {R.}~\bibnamefont
  {Tarroni}}, \bibinfo {author} {\bibfnamefont {P.}~\bibnamefont {Palmieri}},
  \bibinfo {author} {\bibfnamefont {A.}~\bibnamefont {Mitrushenkov}}, \bibinfo
  {author} {\bibfnamefont {P.}~\bibnamefont {Tosi}}, \ and\ \bibinfo {author}
  {\bibfnamefont {D.}~\bibnamefont {Bassi}},\ }\href@noop {} {\bibfield
  {journal} {\bibinfo  {journal} {The Journal of chemical physics}\ }\textbf
  {\bibinfo {volume} {106}},\ \bibinfo {pages} {10265} (\bibinfo {year}
  {1997})}\BibitemShut {NoStop}%
\bibitem [{\citenamefont {Gervais}\ \emph {et~al.}(2009)\citenamefont
  {Gervais}, \citenamefont {Giglio}, \citenamefont {Adoui}, \citenamefont
  {Cassimi}, \citenamefont {Duflot},\ and\ \citenamefont {Galassi}}]{Gervais}%
  \BibitemOpen
  \bibfield  {author} {\bibinfo {author} {\bibfnamefont {B.}~\bibnamefont
  {Gervais}}, \bibinfo {author} {\bibfnamefont {E.}~\bibnamefont {Giglio}},
  \bibinfo {author} {\bibfnamefont {L.}~\bibnamefont {Adoui}}, \bibinfo
  {author} {\bibfnamefont {A.}~\bibnamefont {Cassimi}}, \bibinfo {author}
  {\bibfnamefont {D.}~\bibnamefont {Duflot}}, \ and\ \bibinfo {author}
  {\bibfnamefont {M.}~\bibnamefont {Galassi}},\ }\href@noop {} {\bibfield
  {journal} {\bibinfo  {journal} {The Journal of chemical physics}\ }\textbf
  {\bibinfo {volume} {131}},\ \bibinfo {pages} {024302} (\bibinfo {year}
  {2009})}\BibitemShut {NoStop}%
\bibitem [{\citenamefont {Andersson}\ \emph {et~al.}(2019)\citenamefont
  {Andersson}, \citenamefont {Zagorodskikh}, \citenamefont {Roos},
  \citenamefont {Talaee}, \citenamefont {Squibb}, \citenamefont {Koulentianos},
  \citenamefont {Wallner}, \citenamefont {Zhaunerchyk}, \citenamefont {Singh},
  \citenamefont {Eland} \emph {et~al.}}]{Andersson}%
  \BibitemOpen
  \bibfield  {author} {\bibinfo {author} {\bibfnamefont {J.}~\bibnamefont
  {Andersson}}, \bibinfo {author} {\bibfnamefont {S.}~\bibnamefont
  {Zagorodskikh}}, \bibinfo {author} {\bibfnamefont {A.~H.}\ \bibnamefont
  {Roos}}, \bibinfo {author} {\bibfnamefont {O.}~\bibnamefont {Talaee}},
  \bibinfo {author} {\bibfnamefont {R.}~\bibnamefont {Squibb}}, \bibinfo
  {author} {\bibfnamefont {D.}~\bibnamefont {Koulentianos}}, \bibinfo {author}
  {\bibfnamefont {M.}~\bibnamefont {Wallner}}, \bibinfo {author} {\bibfnamefont
  {V.}~\bibnamefont {Zhaunerchyk}}, \bibinfo {author} {\bibfnamefont
  {R.}~\bibnamefont {Singh}}, \bibinfo {author} {\bibfnamefont
  {J.}~\bibnamefont {Eland}},  \emph {et~al.},\ }\href@noop {} {\bibfield
  {journal} {\bibinfo  {journal} {Scientific Reports}\ }\textbf {\bibinfo
  {volume} {9}},\ \bibinfo {pages} {1} (\bibinfo {year} {2019})}\BibitemShut
  {NoStop}%
\bibitem [{\citenamefont {Reedy}\ \emph {et~al.}(2018)\citenamefont {Reedy},
  \citenamefont {Williams}, \citenamefont {Gaire}, \citenamefont {Gatton},
  \citenamefont {Weller}, \citenamefont {Menssen}, \citenamefont {Bauer},
  \citenamefont {Henrichs}, \citenamefont {Burzynski}, \citenamefont {Berry}
  \emph {et~al.}}]{Reedy}%
  \BibitemOpen
  \bibfield  {author} {\bibinfo {author} {\bibfnamefont {D.}~\bibnamefont
  {Reedy}}, \bibinfo {author} {\bibfnamefont {J.}~\bibnamefont {Williams}},
  \bibinfo {author} {\bibfnamefont {B.}~\bibnamefont {Gaire}}, \bibinfo
  {author} {\bibfnamefont {A.}~\bibnamefont {Gatton}}, \bibinfo {author}
  {\bibfnamefont {M.}~\bibnamefont {Weller}}, \bibinfo {author} {\bibfnamefont
  {A.}~\bibnamefont {Menssen}}, \bibinfo {author} {\bibfnamefont
  {T.}~\bibnamefont {Bauer}}, \bibinfo {author} {\bibfnamefont
  {K.}~\bibnamefont {Henrichs}}, \bibinfo {author} {\bibfnamefont
  {P.}~\bibnamefont {Burzynski}}, \bibinfo {author} {\bibfnamefont
  {B.}~\bibnamefont {Berry}},  \emph {et~al.},\ }\href@noop {} {\bibfield
  {journal} {\bibinfo  {journal} {Physical Review A}\ }\textbf {\bibinfo
  {volume} {98}},\ \bibinfo {pages} {053430} (\bibinfo {year}
  {2018})}\BibitemShut {NoStop}%
\bibitem [{\citenamefont {Wannier}(1953)}]{Wannier}%
  \BibitemOpen
  \bibfield  {author} {\bibinfo {author} {\bibfnamefont {G.~H.}\ \bibnamefont
  {Wannier}},\ }\href@noop {} {\bibfield  {journal} {\bibinfo  {journal}
  {Physical Review}\ }\textbf {\bibinfo {volume} {90}},\ \bibinfo {pages} {817}
  (\bibinfo {year} {1953})}\BibitemShut {NoStop}%
\bibitem [{\citenamefont {Br{\"a}uning}\ \emph {et~al.}(1998)\citenamefont
  {Br{\"a}uning}, \citenamefont {D{\"o}rner}, \citenamefont {Cocke},
  \citenamefont {Prior}, \citenamefont {Kr{\"a}ssig}, \citenamefont {Kheifets},
  \citenamefont {Bray}, \citenamefont {Br{\"a}uning-Demian}, \citenamefont
  {Carnes}, \citenamefont {Dreuil}, \citenamefont {Mergel}, \citenamefont
  {Richard}, \citenamefont {Ullrich},\ and\ \citenamefont
  {Schmidt-B{\"o}cking}}]{Brauning}%
  \BibitemOpen
  \bibfield  {author} {\bibinfo {author} {\bibfnamefont {H.}~\bibnamefont
  {Br{\"a}uning}}, \bibinfo {author} {\bibfnamefont {R.}~\bibnamefont
  {D{\"o}rner}}, \bibinfo {author} {\bibfnamefont {C.~L.}\ \bibnamefont
  {Cocke}}, \bibinfo {author} {\bibfnamefont {M.~H.}\ \bibnamefont {Prior}},
  \bibinfo {author} {\bibfnamefont {B.}~\bibnamefont {Kr{\"a}ssig}}, \bibinfo
  {author} {\bibfnamefont {A.~S.}\ \bibnamefont {Kheifets}}, \bibinfo {author}
  {\bibfnamefont {I.}~\bibnamefont {Bray}}, \bibinfo {author} {\bibfnamefont
  {A.}~\bibnamefont {Br{\"a}uning-Demian}}, \bibinfo {author} {\bibfnamefont
  {K.}~\bibnamefont {Carnes}}, \bibinfo {author} {\bibfnamefont
  {S.}~\bibnamefont {Dreuil}}, \bibinfo {author} {\bibfnamefont
  {V.}~\bibnamefont {Mergel}}, \bibinfo {author} {\bibfnamefont
  {P.}~\bibnamefont {Richard}}, \bibinfo {author} {\bibfnamefont
  {J.}~\bibnamefont {Ullrich}}, \ and\ \bibinfo {author} {\bibfnamefont
  {H.}~\bibnamefont {Schmidt-B{\"o}cking}},\ }\href {\doibase
  10.1088/0953-4075/31/23/012} {\bibfield  {journal} {\bibinfo  {journal}
  {Journal of Physics B: Atomic, Molecular and Optical Physics}\ }\textbf
  {\bibinfo {volume} {31}},\ \bibinfo {pages} {5149} (\bibinfo {year}
  {1998})}\BibitemShut {NoStop}%
\bibitem [{\citenamefont {Knapp}\ \emph {et~al.}(2005)\citenamefont {Knapp},
  \citenamefont {Kr{\"a}ssig}, \citenamefont {Kheifets}, \citenamefont {Bray},
  \citenamefont {Weber}, \citenamefont {Landers}, \citenamefont
  {Sch{\"o}ssler}, \citenamefont {Jahnke}, \citenamefont {Nickles},
  \citenamefont {Kammer}, \citenamefont {Jagutzki}, \citenamefont {Schmidt},
  \citenamefont {Sch{\"o}ffler}, \citenamefont {Osipov}, \citenamefont {Prior},
  \citenamefont {Schmidt-B{\"o}cking}, \citenamefont {Cocke},\ and\
  \citenamefont {D{\"o}rner}}]{Knapp}%
  \BibitemOpen
  \bibfield  {author} {\bibinfo {author} {\bibfnamefont {A.}~\bibnamefont
  {Knapp}}, \bibinfo {author} {\bibfnamefont {B.}~\bibnamefont {Kr{\"a}ssig}},
  \bibinfo {author} {\bibfnamefont {A.}~\bibnamefont {Kheifets}}, \bibinfo
  {author} {\bibfnamefont {I.}~\bibnamefont {Bray}}, \bibinfo {author}
  {\bibfnamefont {T.}~\bibnamefont {Weber}}, \bibinfo {author} {\bibfnamefont
  {A.~L.}\ \bibnamefont {Landers}}, \bibinfo {author} {\bibfnamefont
  {S.}~\bibnamefont {Sch{\"o}ssler}}, \bibinfo {author} {\bibfnamefont
  {T.}~\bibnamefont {Jahnke}}, \bibinfo {author} {\bibfnamefont
  {J.}~\bibnamefont {Nickles}}, \bibinfo {author} {\bibfnamefont
  {S.}~\bibnamefont {Kammer}}, \bibinfo {author} {\bibfnamefont
  {O.}~\bibnamefont {Jagutzki}}, \bibinfo {author} {\bibfnamefont {L.~P.~H.}\
  \bibnamefont {Schmidt}}, \bibinfo {author} {\bibfnamefont {M.}~\bibnamefont
  {Sch{\"o}ffler}}, \bibinfo {author} {\bibfnamefont {T.}~\bibnamefont
  {Osipov}}, \bibinfo {author} {\bibfnamefont {M.~H.}\ \bibnamefont {Prior}},
  \bibinfo {author} {\bibfnamefont {H.}~\bibnamefont {Schmidt-B{\"o}cking}},
  \bibinfo {author} {\bibfnamefont {C.~L.}\ \bibnamefont {Cocke}}, \ and\
  \bibinfo {author} {\bibfnamefont {R.}~\bibnamefont {D{\"o}rner}},\ }\href
  {\doibase 10.1088/0953-4075/38/6/003} {\bibfield  {journal} {\bibinfo
  {journal} {Journal of Physics B: Atomic, Molecular and Optical Physics}\
  }\textbf {\bibinfo {volume} {38}},\ \bibinfo {pages} {645} (\bibinfo {year}
  {2005})}\BibitemShut {NoStop}%
\bibitem [{\citenamefont {Randazzo}\ \emph {et~al.}(2020)\citenamefont
  {Randazzo}, \citenamefont {Turri}, \citenamefont {Bolognesi}, \citenamefont
  {Mathis}, \citenamefont {Ancarani},\ and\ \citenamefont {Avaldi}}]{Randazzo}%
  \BibitemOpen
  \bibfield  {author} {\bibinfo {author} {\bibfnamefont {J.~M.}\ \bibnamefont
  {Randazzo}}, \bibinfo {author} {\bibfnamefont {G.}~\bibnamefont {Turri}},
  \bibinfo {author} {\bibfnamefont {P.}~\bibnamefont {Bolognesi}}, \bibinfo
  {author} {\bibfnamefont {J.}~\bibnamefont {Mathis}}, \bibinfo {author}
  {\bibfnamefont {L.~U.}\ \bibnamefont {Ancarani}}, \ and\ \bibinfo {author}
  {\bibfnamefont {L.}~\bibnamefont {Avaldi}},\ }\href {\doibase
  10.1103/PhysRevA.101.033407} {\bibfield  {journal} {\bibinfo  {journal}
  {Phys. Rev. A}\ }\textbf {\bibinfo {volume} {101}},\ \bibinfo {pages}
  {033407} (\bibinfo {year} {2020})}\BibitemShut {NoStop}%
\end{thebibliography}%

\end{document}